\documentclass[aps,prl,twocolumn,superscriptaddress,longbibliography]{revtex4-2}
\usepackage[T1]{fontenc}
\usepackage{amsmath,amsthm,amssymb}
\usepackage{graphicx}
\usepackage{appendix}
\usepackage{xcolor}
\usepackage{scalerel}
\definecolor{blueviolet}{rgb}{0.2, 0.2, 0.6}
\definecolor{webgreen}{rgb}{0,.5,0}
\definecolor{webbrown}{rgb}{.6,0,0}
\definecolor{urlColor}{HTML}{0BCB9A}
\usepackage[bookmarks=false,colorlinks=true,linkcolor=blueviolet,citecolor=webgreen,urlcolor=urlColor]{hyperref}

\usepackage{cleveref}
\usepackage{orcidlink}

\newtheorem{theorem}{Theorem}\crefname{theorem}{Theorem}{Theorems}

\newtheorem*{theorem*}{Theorem}
\newtheorem*{corollary*}{Corollary}

\crefname{lemma}{Lemma}{Lemmas}
\crefname{proposition}{Proposition}{Propositions}
\crefname{corollary}{Corollary}{Corollaries}
\crefname{definition}{Definition}{Definitions}

\usepackage{times}

\begin{document}
\title{Generalized Schrieffer--Wolff transformation}

\author{Tie-Cheng Guo\orcidlink{0000-0001-8252-7970}}
\email{gtc16@tsinghua.org.cn}
\affiliation{International Research Centre MagTop, Institute of Physics, Polish Academy of Sciences, Aleja Lotnikow 32/46, PL-02668 Warsaw, Poland}

\date{\today}

\begin{abstract}

The Schrieffer--Wolff transformation (SWT) is a cornerstone of effective-Hamiltonian theory. 
However, the SWT is predominantly used for weak-coupling problems, while a systematic counterpart for strongly coupled regimes remains largely unexplored.
A general closed-form SWT generator and a clear physical principle for gauge fixing remain elusive, obscuring the transformation's underlying structure.
We introduce the dual SWT to deal with the strong-coupling problems.
The generalized SWT (gSWT) is formulated for any Hermitian operator and block partition.
We establish the geodesic principle for its canonical gauge.
Furthermore, we derive an exact closed-form solution of the gSWT generator at every order as a finite nested-commutator sum whose only input is the active gap set.
Both the geometric and algebraic aspects of the SWT physics are revealed.
The gSWT framework is illustrated in two quantum many-body models, the collective spin model and the transverse-field Ising model.

\end{abstract}

\maketitle

The Schrieffer--Wolff transformation (SWT) is an important method for deriving effective Hamiltonians~\cite{schrieffer1966relation,bravyi2011schrieffer}. It is closely related to the Foldy--Wouthuysen transformation and quasidegenerate perturbation theory~\cite{foldy1950dirac,lowdin1951note,klein1974degenerate}.
Initially introduced to connect the Anderson and Kondo models~\cite{schrieffer1966relation}, the SWT now has applications in strongly correlated electron systems~\cite{macdonald1988expansion}, semiconductor spin qubits~\cite{golovach2006electric}, quantum optics and light--matter interactions~\cite{zhu2013dispersive,hwang2015quantum,garbe2020critical}, superconducting quantum circuits and gates~\cite{magesan2020effective,li2022nonperturbative}, quantum simulation~\cite{bravyi2008quantum,kale2022schrieffer}, and Floquet engineering~\cite{bukov2016schrieffer}.

Constructing the SWT generator is the key step. Analytical and recursive methods have been developed for specific systems~\cite{golovach2006electric,li2022nonperturbative},
automated tools support symbolic generator construction~\cite{diotallevi2024sympt,arayaday2025pymablock},
and variational methods provide an alternative route to many-body subspace decoupling~\cite{wurtz2020variational}.
The SWT has also been generalized to time-dependent and periodically driven systems~\cite{xiao2022perturbative,bukov2016schrieffer}, dissipative systems~\cite{kessler2012generalized}, and non-Hermitian Hamiltonians~\cite{starkov2023schrieffer}.

Despite these advances, several challenges remain.
First, no generally usable closed-form expression for the generator is known.
Second, the SWT has so far primarily addressed weak-coupling systems, while strong-coupling and complementary limits have no systematic method. Only a few systems have been studied case by case:
ultrarelativistic Dirac particles~\cite{cini1958relativistic}, the polaron system~\cite{lang1963kinetic}, and the generalized rotating-wave approximation~\cite{twyeffort2007generalized}.
Third, a more fundamental issue concerns the principle determining the generator's gauge: least action (direct rotation) and block-off-diagonality of the generator coincide only for two blocks, but generally differ from third order onward for three or more blocks~\cite{cederbaum1989block,mankodi2024perturbative}.

\begin{figure}[htb]
	\includegraphics[width=\columnwidth]{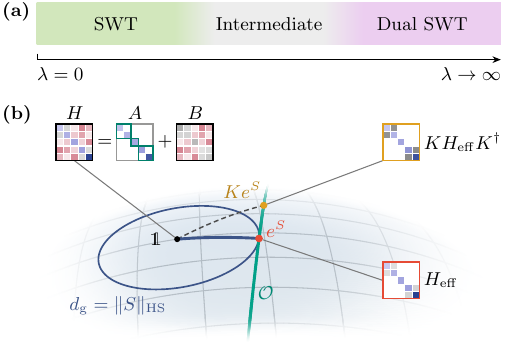}
	\caption{\textbf{Generalized Schrieffer--Wolff transformation}. (a)~Regimes of $H=H_1+\lambda H_2$ along the $\lambda$ axis: the SWT regime $(\lambda\rightarrow 0)$, the dual SWT regime $(\lambda\rightarrow \infty)$, and the intermediate regime, interpolated by a two-point Pad\'e approximant.
	(b)~Schematic of the gSWT. Icons are matrices in the eigenbasis of $A$, with teal frames marking the blocks of the partition $\Pi$: $H$ at $U=\openone$, and $UHU^\dagger$ at $U=e^{S}$ and $Ke^{S}$. On the surface, a schematic two-dimensional slice of the unitary group, the block-diagonalizing unitaries form the gauge orbit $\mathcal O=\{Ke^{S}\}$ (teal), with $K$ unitary within the blocks, and the canonical gauge $\mathcal P_\Pi S=0$ selects the point $e^{S}$ (red) closest to $\openone$ in geodesic distance $d_g$.
	}
	\label{fig:gswt}
\end{figure}

In this Letter, we solve these challenges and present the generalized SWT (gSWT).
We introduce the dual SWT for strong-coupling systems and prove its structural properties.
The gSWT is formulated for an arbitrary Hermitian operator $H$ with two parts, the dominant operator and the perturbation operator, and a block partition.
Furthermore, we present a geodesic principle for the canonical gauge condition.
Crucially, we derive an exact solution for the generator at every perturbative order, providing a constructive procedure for the gSWT.
Two quantum many-body models, the collective spin model and the transverse-field Ising model, are used to illustrate the gSWT.

\textit{Dual Schrieffer--Wolff transformation.---}%
Consider a Hamiltonian with a $\mathbb Z_2$ grading structure,
\begin{equation}\label{eq:H}
	H=H_0+V,\qquad [\Gamma,H_0]=0,\quad \{\Gamma,V\}=0,
\end{equation}
where $\Gamma=\Gamma^\dagger=\Gamma^{-1}$ is the $\mathbb Z_2$ grading operator. Its eigenvalues $\pm1$ label two sectors: $H_0$ acts within the sectors and $V$ couples them.
For example, $\Gamma$ can be a parity or sublattice operator.

When $H_0$ dominates and $V$ is a perturbation, the SWT~\cite{schrieffer1966relation,bravyi2011schrieffer} removes the coupling between the eigenspaces of $H_0$ order by order in $V$: $H_{\mathrm{eff}}=e^{S}He^{-S}$ is block diagonal with respect to the eigenspaces of $H_0$. The second-order effective Hamiltonian and the first-order generator $S^{(1)}$ are
\begin{equation}\label{eq:swt}
	H_{\mathrm{eff}}^{[2]}=H_0+V^{\parallel}+\tfrac12\mathcal{P}_{H_0}[S^{(1)},V^{\perp}],\quad
	\operatorname{ad}_{H_0}S^{(1)}=V^{\perp},
\end{equation}
where the superscript $[2]$ denotes truncation at second order and $\operatorname{ad}_XY=[X,Y]$.
Here $\mathcal{P}_{H_0}$ (and $\mathcal{P}_{V}$ below) is the block-diagonal projector, and $\parallel$ and $\perp$ denote the \emph{secular} and \emph{nonsecular} parts, which act within and between the eigenspaces of the dominant operator.

When $V$ dominates, the strong-coupling problem appears.
The dual SWT exchanges the roles of $H_0$ and $V$: it removes the coupling between the eigenspaces of $V$ order by order in $H_0$, so that $\tilde H_{\mathrm{eff}}=e^{\tilde S}He^{-\tilde S}$ is block diagonal in the eigenspaces of $V$. The second-order effective Hamiltonian and the first-order generator are
\begin{equation}\label{eq:dual}
	\tilde H_{\mathrm{eff}}^{[2]}=V+H_0^{\parallel}+\tfrac12\mathcal{P}_{V}[\tilde S^{(1)},H_0^{\perp}],\quad
	\operatorname{ad}_{V}\tilde S^{(1)}=H_0^{\perp}.
\end{equation}
The dual SWT is an expansion in the inverse coupling and gives a strong-coupling effective Hamiltonian.
Its structural properties and its relation to the conventional SWT are given in the End Matter and the Supplemental Material (SM)~\cite{SM}.

Both are special cases of the gSWT defined below, with (dominant operator, perturbation operator) $=(H_0,V)$ and $(V,H_0)$, respectively, in the finest partition.
They are compared in Table~\ref{tab:compare} (End Matter), and the canonical gauges are $\mathcal{P}_{H_0}S=0$ and $\mathcal{P}_{V}\tilde S=0$. 

The duality can be extended beyond Eq.~\eqref{eq:H} to general systems. A Hamiltonian $H=H_1+\lambda H_2$ has three regimes [Fig.~\ref{fig:gswt}(a)]: the SWT regime ($\lambda\to0$), the dual SWT regime ($\lambda\to\infty$), and the intermediate regime, where a two-point Pad\'e approximant constrained by the SWT and dual-SWT series interpolates observables.

\textit{Generalized Schrieffer--Wolff transformation.---}%
We now generalize the SWT and the dual SWT beyond the $\Gamma$-even/$\Gamma$-odd structure to an arbitrary Hermitian operator $H=A+B$,
with $A$ the dominant operator and $B$ the perturbation operator, and to a general block partition.
The spectral decomposition of $A$ is $A=\sum_aE_aQ_a$, with $E_a$ its distinct eigenvalues and $Q_a$ the spectral projectors, $Q_aQ_b=\delta_{ab}Q_a$, $\sum_aQ_a=\openone$. Let $\Pi=\{P_\alpha\}$ be a partition compatible with $A$: $P_\alpha=\sum_{a\in I_\alpha}Q_a$, with disjoint index sets $I_\alpha$ exhausting the spectrum, so that $A=\sum_\alpha P_\alpha AP_\alpha$.
Let $\mathcal{A}_\Pi$ be the algebra of block-diagonal operators
and $\mathcal{P}_\Pi$ the block-diagonal projector, which maps any operator $X$ onto its block-diagonal part, $\mathcal{P}_\Pi X=\sum_\alpha P_\alpha XP_\alpha$.
The complement $\mathcal{Q}_\Pi=\mathcal{I}-\mathcal{P}_\Pi$ is the block-off-diagonal projector, with $\mathcal I$ the identity superoperator.
The secular part $X^{\parallel}=\mathcal{P}_\Pi X$ of $X$ acts inside the blocks, and $X^{\perp}=\mathcal{Q}_\Pi X$ is its nonsecular part.
In the \emph{finest partition} every block is a single eigenspace $Q_a$ of $A$, 
so that $\mathcal{P}_\Pi=\mathcal{P}_A$ projects onto the operators commuting with $A$. In a \emph{coarse-grained partition} some blocks unite several eigenspaces. Unless stated otherwise, the finest partition is used.
The sector gap of the partition is
$g_\Pi=\min\bigl\{\,|E_a-E_b|:\ a\in I_\alpha,\ b\in I_\beta,\ \alpha\neq\beta\,\bigr\}$.

The gSWT of $H$ is the unitary transformation $H_{\mathrm{eff}}=UHU^\dagger$ with $U=e^{S}$, $S^\dagger=-S$, constrained by two conditions. The first one is the block-diagonal condition,
\begin{equation}\label{eq:gswt1}
	U\Pi_\alpha U^\dagger=P_\alpha\ \ \forall\alpha,
\end{equation}
where $\Pi_\alpha$ is the spectral projector of $H$ onto the eigenvalues continuously connected to those of $A$ in block $\alpha$.
The second is the canonical gauge condition,
\begin{equation}\label{eq:gswt2}
	\mathcal{P}_\Pi S=0.
\end{equation}

The canonical gauge fixes the expansion $S=\sum_{n\ge1}S^{(n)}$, $H_{\mathrm{eff}}=A+\sum_{n\ge1}H^{(n)}$ uniquely at every order $n$ in $B$, with $\|S^{(n)}\|=O(x^n)$ and $\|H^{(n)}\|=O(g_\Pi x^n)$.
The expansion parameter is $x\equiv\|B\|/g_\Pi$, with $\|\cdot\|$ the operator norm.
Two sufficient conditions play different roles. The first guarantees that the gSWT exists: for $x<\tfrac12$ the levels of $A+tB$ evolving from different blocks remain separated for $0\le t\le1$, so the $\Pi_\alpha$ are well defined, unitaries obeying Eq.~\eqref{eq:gswt1} exist, and each block of $H_{\mathrm{eff}}$ carries the exact levels continued from that block of $A$. The second guarantees that the expansion converges: for $\|B\|_{\mathrm{HS}}<cg_\Pi$, with $\|\cdot\|_{\mathrm{HS}}$ the Hilbert--Schmidt norm and $c=e^{-1/2}-\tfrac12\approx0.107$, it converges absolutely to a unitary $U=e^{S}$ obeying both Eqs.~\eqref{eq:gswt1} and~\eqref{eq:gswt2} (see SM~\cite{SM}). Since $\|B\|\le\|B\|_{\mathrm{HS}}$, the second condition implies the first.

Order by order in $B$, the expansion of $e^{S}He^{-S}$ gives $H^{(n)}+\operatorname{ad}_AS^{(n)}=R^{(n)}$, where the source term $R^{(n)}$ contains only $B$ and the generators of lower orders (End Matter and SM~\cite{SM}). By Eqs.~\eqref{eq:gswt1} and~\eqref{eq:gswt2}, $H^{(n)}$ is block diagonal, while $S^{(n)}$ and $\operatorname{ad}_AS^{(n)}$ are block off-diagonal, $\mathcal{P}_\Pi S^{(n)}=0$. We thus obtain the $n$th-order effective Hamiltonian,
\begin{equation}\label{eq:Hn}
	H^{(n)}=\mathcal{P}_\Pi R^{(n)},
\end{equation}
and the generator is determined by
\begin{equation}\label{eq:Sn}
	\operatorname{ad}_A S^{(n)}=\mathcal{Q}_\Pi R^{(n)}.
\end{equation}

The projector $\mathcal{P}_\Pi$ cannot in general be omitted. 
The cases in which $\mathcal{P}_\Pi$ reduces to a selection rule (two blocks; a single active gap at second order; or a Lie algebra closing on the transition energies $0,\pm g$), and its evaluation through transition energies or block labels, are given in the SM~\cite{SM}.

\textit{Geodesic principle.---}%
The unitaries that satisfy Eq.~\eqref{eq:gswt1} achieve matrix block diagonalization, but they are not unique: they form the \emph{gauge orbit} $\mathcal{O}=\{KU:K\in\mathcal{A}_\Pi\ \text{unitary}\}$ [Fig.~\ref{fig:gswt}(b)], where $K$ performs the rotation inside the blocks and $H_{\mathrm{eff}}\mapsto KH_{\mathrm{eff}}K^\dagger$.
The canonical gauge, Eq.~\eqref{eq:gswt2}, is used to fix the gauge in the SWT. Here we give its geodesic principle.

For $U=e^S$ with $S^\dagger=-S$ and $\|S\|<\pi$, the geodesic distance from the identity $\openone$ to $U$, measured
with the Hilbert--Schmidt metric on the unitary group, is
\begin{equation}
	d_g(U)=\|S\|_{\mathrm{HS}}=\Bigl(\sum_{j=1}^{N}\theta_j^2\Bigr)^{1/2},
\end{equation}
where $S=\log U$ is the principal anti-Hermitian logarithm
and $\theta_j\in(-\pi,\pi)$ are the eigenphases of $U$.

\begin{theorem}[Geodesic principle]\label{thm:extremal}
	Let $x<\tfrac12$ and $U=e^{S}\in\mathcal{O}$ with $S=\log U$, $\|S\|<\pi$. (i)~$U$ is a stationary point of $d_g^2$ on $\mathcal{O}$ if and only if $\mathcal{P}_\Pi S=0$. (ii)~The stationary point is a strict local minimum of $d_g$ if $\max_j\theta_j-\min_j\theta_j<\pi$. (iii)~A stationary point with $d_g(U)<\pi/(2\sqrt2)$ is the unique global minimizer of $d_g$ on $\mathcal{O}$.
	Such a $U$ exists whenever $\|B\|_{\mathrm{HS}}<(e^{-1/2}-\tfrac12)g_\Pi$, and more generally under a principal-angle condition.
\end{theorem}
The proof is given in the SM~\cite{SM}.
The canonical gauge of Eq.~\eqref{eq:gswt2} is thus equivalent to the stationarity condition of the geodesic distance along the gauge orbit, and under the condition of (iii) it chooses the block-diagonalizing unitary closest to the identity.
Theorem~\ref{thm:extremal} gives gauge fixing a geometric principle.

The choice of distance is important. Minimizing instead the chordal distance $d_c(U)=\|U-\openone\|_{\mathrm{HS}}$ yields the least-action transformation~\cite{cederbaum1989block}, which satisfies the canonical gauge for two blocks~\cite{davis1970rotation,bravyi2011schrieffer} but deviates from it at third order for three or more blocks~\cite{mankodi2024perturbative} (End Matter).

\textit{Exact solution of the gSWT.---}%
At every order the generator obeys the same Sylvester equation, Eq.~\eqref{eq:Sn}: $\operatorname{ad}_AG=R^{\perp}$, where the block-off-diagonal operator $R^{\perp}=\mathcal{Q}_\Pi R^{(n)}$ is given and the block-off-diagonal $G=S^{(n)}$ is to be determined.
At first order $R^{\perp}=B^{\perp}$.

The physical observation is that commutation with $A$ multiplies each transition operator by its energy difference, $\operatorname{ad}_A^k(Q_aXQ_b)=(E_a-E_b)^kQ_aXQ_b$, so the solution depends on $A$ only through the energy differences that $R^{\perp}$ actually couples. The \emph{active gap set} is defined by
\begin{equation}\label{eq:active}
	\mathcal{G}_A(R^{\perp})=\{\,|E_a-E_b|:\ Q_aR^{\perp}Q_b\neq0\,\}.
\end{equation}
Energy gaps that $R^{\perp}$ does not couple and gaps inside a block are inactive.

\begin{theorem}[Nested-commutator solution]\label{thm:exactsolution}
	For the gSWT problem with $(A,B,\Pi)$, let $R^{\perp}$ be block off-diagonal, with active gap set $\mathcal{G}_A(R^{\perp})=\{g_1,\dots,g_m\}$. The equation $\operatorname{ad}_AG=R^{\perp}$ has the unique block-off-diagonal solution $G=p(\operatorname{ad}_A)\,R^{\perp}$ with $p(\omega)=\omega^{-1}\bigl[1-\prod_{i=1}^{m}(1-\omega^2/g_i^2)\bigr]$. Explicitly,
	\begin{equation}\label{eq:ek}
		G=\sum_{k=1}^{m}(-1)^{k+1}e_k\bigl(g_1^{-2},\dots,g_m^{-2}\bigr)\,\operatorname{ad}_A^{2k-1}R^{\perp},
	\end{equation}
	with $e_k$ the elementary symmetric polynomials, and $p(\omega)$ is the unique odd polynomial of degree $\le2m-1$ with $p(\pm g_i)=\pm1/g_i$.
\end{theorem}

The proof is given in the SM~\cite{SM}. By Lagrange--Sylvester interpolation~\cite{higham2008functions} of $1/\omega$ on the active gaps, any polynomial $q$ satisfying $q(\pm g_i)=\pm1/g_i$ yields the exact solution $G=q(\operatorname{ad}_A)R^\perp$, because $\operatorname{ad}_A^{-1}$ acts as $\pm1/g_i$ on the components of $R^{\perp}$ with transition energy $\pm g_i$.
Nodes at which $R^{\perp}$ has no component impose no condition: \emph{supersets of the active gap set preserve exactness}.
Any finite set $\mathcal{G}'\supseteq\mathcal{G}_A(R^{\perp})$, e.g., all inter-block gaps of $A$, yields the same generator at the cost of a larger commutator depth. 

For a single active gap, $m=1$, the nested commutator collapses to a single one, $G=[A,R^{\perp}]/g^2=(R_{+g}-R_{-g})/g$, with $R_{\pm g}=\tfrac12(R^{\perp}\pm[A,R^{\perp}]/g)$ the parts of $R^{\perp}$ that raise and lower $A$ by $g$. At first order, with $B_\pm$ the corresponding parts of $B^{\perp}$, we have
\begin{equation}\label{eq:single}
	S^{(1)}=\frac{B_+-B_-}{g},\qquad H^{(2)}=\frac{[B_+,B_-]}{g}.
\end{equation}

If $[A^2,R^\perp]=0$, the equation $\operatorname{ad}_A G=R^\perp$ has the solution
$G=(4A^2)^+[A,R^\perp]$ with an operator-valued coefficient, where $(\cdot)^{+}$ denotes the Moore--Penrose pseudoinverse~\cite{SM}.

The input of Theorem~\ref{thm:exactsolution} is the active gap set $\mathcal{G}_A(R^\perp)$ at each perturbative order.
Four practical guidelines help determine this spectral input or a suitable superset (End Matter and SM~\cite{SM}): the block partition, selection rules, commutator closure, which requires no eigenvectors of $A$, and supersets.

For the finest partition, the same input gives the generator and the effective Hamiltonian directly from the source, $S^{(n)}=p(\operatorname{ad}_A)R^{(n)}$ and $H^{(n)}=\prod_i(\mathcal I-\operatorname{ad}_A^2/g_i^2)R^{(n)}$. With Eq.~\eqref{eq:Rgen}, this closes the recursion: every order is a finite-sum of nested commutators~\cite{SM}.
For a coarse-grained partition, the projector in Eq.~\eqref{eq:Hn} must in general be evaluated, and all other steps of the order-by-order recursion require only nested commutators.

\textit{Collective spin model.---}%
We now consider the one-axis-twisting model in a transverse field~\cite{kitagawa1993squeezed,muessel2015twist},
\begin{equation}
	H=\chi J_z^2+\Omega J_x ,
	\label{eq:lmg}
\end{equation}
where $J_\alpha=\tfrac12\sum_{k=1}^{N}\sigma^\alpha_k$ are the collective spin operators of $N$ spin-$1/2$ particles with total spin $j=N/2$, $\chi>0$ is the twisting strength, and $\Omega$ is the transverse field.
We take $N$ even. Here $H_0=\chi J_z^2$, $V=\Omega J_x$, $\Gamma=e^{i\pi(J_z-j)}$, $c_j=j(j+1)$, and $\Lambda\equiv N\chi/|\Omega|$.

When the twisting dominates, the SWT has $A=\chi J_z^2$ and $B=\Omega J_x$. The blocks of the finest partition are the singlet $|0\rangle$ and the doublets $\{|m\rangle,|{-m}\rangle\}$ of $J_z$ eigenstates. The perturbation $B$ is purely nonsecular, with the $j$ active gaps $\mathcal{G}_A(B^{\perp})=\{(2l-1)\chi\}_{l=1}^{j}$, and $g_\Pi=\chi$, $x=N^2/(2\Lambda)$. The nested commutator of Theorem~\ref{thm:exactsolution}, of depth $2j-1$, sums to the closed form
\begin{equation}\label{eq:lmg_S}
	S^{(1)}=p(\operatorname{ad}_A)B=\frac{\Omega}{2\chi}\bigl[J_+(2J_z+1)^{-1}-\mathrm{H.c.}\bigr],
\end{equation}
in which $(2J_z+1)^{-1}$ supplies the energy denominator of each transition $m\to m+1$. The second-order effective Hamiltonian is
\begin{equation}\label{eq:lmg_swt}
	H_{\mathrm{eff}}^{[2]}=\chi J_z^2+\frac{\Omega^2}{2\chi}\,\frac{c_j+J_z^2}{4J_z^2-1}
	+\frac{c_j\Omega^2}{4\chi}\bigl(|1\rangle\langle{-1}|+\mathrm{H.c.}\bigr).
\end{equation}
The second term is the level shift, and the last term, the only intra-block coupling retained by $\mathcal{P}_\Pi$ (the process $-1\to0\to1$), splits the $|m|=1$ doublet by $c_j\Omega^2/(2\chi)$.
Because $|m\rangle$ and $|{-m}\rangle$ have the same $\Gamma$ eigenvalue, the odd orders vanish and the truncation error is $O(\Omega^4/\chi^3)$.

When the transverse field dominates, the dual SWT has $A=\Omega J_x$ and $B=\chi J_z^2$. The blocks are the nondegenerate $J_x$ eigenstates $|m_x\rangle$. Since $B$ changes $m_x$ by $0$ or $\pm2$, its secular part is $B^{\parallel}=\tfrac{\chi}{2}(c_j-J_x^2)$ and its nonsecular part has the single active gap $\mathcal{G}_A(B^{\perp})=\{2|\Omega|\}$, whereas $g_\Pi=|\Omega|$ and $x=N\Lambda/4$.
Equation~\eqref{eq:single} gives
\begin{equation}\label{eq:lmg_Sdual}
	\tilde S^{(1)}=\frac{\chi}{4\Omega}[J_x,J_z^2]=-\frac{i\chi}{4\Omega}\{J_y,J_z\},
\end{equation}
and the second-order effective Hamiltonian
\begin{equation}\label{eq:lmg_dual}
	\tilde H_{\mathrm{eff}}^{[2]}=\Omega J_x+\frac{\chi}{2}\bigl(c_j-J_x^2\bigr)+\frac{\chi^2}{8\Omega}J_x\bigl(2c_j-1-2J_x^2\bigr),
\end{equation}
with truncation error $O(\chi^3/\Omega^2)$. This is the form $V[1+g(V^2)]+f(V^2)$ of Eq.~\eqref{eq:dualform}, with the obstruction $f=B^{\parallel}+O(\chi^3)$. The generator $\tilde S^{(1)}$ is a two-axis countertwisting~\cite{kitagawa1993squeezed}: to first order in $\chi/\Omega$ it squeezes the coherent ground state of $A$ to $2\langle J_z^2\rangle/j=1-(N-1)\chi/(2|\Omega|)$~\cite{SM}.

\begin{figure}[tbp]
	\includegraphics[width=\columnwidth]{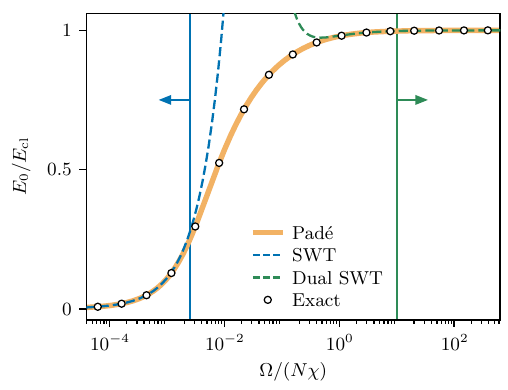}
	\caption{\textbf{Three regimes of the gSWT}. Ground-state energy of the model~\eqref{eq:lmg} for $N=20$, in units of the classical energy $E_{\mathrm{cl}}=-N\Omega/2$~\cite{zibold2010classical}, versus $\Omega/(N\chi)$: exact results (open circles), second-order SWT and dual SWT (blue and green dashed curves), and the two-point $[19/18]$ Pad\'e approximant matching the two series (yellow solid line).
	}
	\label{fig:lmg}
\end{figure}

In Fig.~\ref{fig:lmg}, for $\Omega>0$, the dashed curves are the levels $m=0$ of Eq.~\eqref{eq:lmg_swt} and $m_x=-j$ of Eq.~\eqref{eq:lmg_dual}, and the vertical lines mark $x=\tfrac12$ ($\Lambda=N^2$ and $\Lambda=2/N$). The two-point Pad\'e approximant~\cite{baker1996pade,mccabe1976continued} built from the two series reproduces the exact energy over the whole range~\cite{SM}.

\textit{Transverse-field Ising model.---}%
The Hamiltonian for the transverse-field Ising model (TFIM)~\cite{pfeuty1970one,sachdev2011quantum} on a ring of $L$ sites is
\begin{equation}\label{eq:tfim_H}
	H=-J\sum_{i=1}^{L}\sigma^z_i\sigma^z_{i+1}-h\sum_{i=1}^{L}\sigma^x_i ,
\end{equation}
with $J,h>0$. Here $H_0$ is the Ising term, $V$ is the transverse field, and $\Gamma=\prod_i\sigma^z_i$. Both transformations have a single active gap for every $L$ and collapse to Eq.~\eqref{eq:single}.

Deep in the ferromagnetic phase ($h\ll J$), the SWT has $A=-J\sum_i\sigma_i^z\sigma_{i+1}^z=-JL+2JN_w$ and $B=-h\sum_i\sigma_i^x$. The blocks are the eigenspaces of fixed domain-wall number $N_w$, which is even on a ring, so that $g_\Pi=4J$. A spin flip either moves a wall ($B^{\parallel}$) or creates or annihilates a pair of adjacent walls ($B_\pm$), so $\mathcal{G}_A(B^{\perp})=\{4J\}$. Equation~\eqref{eq:single} then gives, in terms of Jordan--Wigner wall fermions $c_k$,
\begin{equation}\label{eq:tfim}
	\begin{aligned}
		S^{(1)}
		&=\frac{ih}{8J}\sum_i
		\left(
		\sigma_i^y\sigma_{i+1}^z
		+\sigma_i^z\sigma_{i+1}^y
		\right),\\
		H_{\mathrm{eff}}^{[2]}
		&=-L\left(J+\frac{h^2}{4J}\right)\\
		&\quad+\sum_k
		\left(
		2J-2h\cos k+\frac{h^2}{J}\sin^2k
		\right)c_k^\dagger c_k .
	\end{aligned}
\end{equation}
A wall costs $2J$, $B^{\parallel}$ moves it by one bond, giving $-2h\cos k$, and virtual wall pairs give the second-order terms~\cite{SM}. Odd orders survive because a block contains both values of $\Gamma$. Although $x=Lh/(4J)$ grows with $L$, fixed-order terms are local and intensive quantities expand in $h/J$~\cite{SM}.

Deep in the paramagnetic phase ($J\ll h$), the dual SWT has $A=-h\sum_i\sigma_i^x=-hL+2hN_f$, where $N_f$ counts spin flips from the $x$-polarized state. The Ising interaction $B=-J\sum_i\sigma_i^z\sigma_{i+1}^z$ moves a flip ($B^{\parallel}$) or creates or annihilates a neighboring pair ($B_\pm$), so $\mathcal{G}_A(B^{\perp})=\{4h\}$, with $g_\Pi=2h$ and $x=LJ/(2h)$. Equation~\eqref{eq:single} yields
\begin{equation}\label{eq:tfim_dual}
	\begin{aligned}
		\tilde S^{(1)}
		&=-\frac{iJ}{8h}\sum_i
		\left(
		\sigma_i^y\sigma_{i+1}^z
		+\sigma_i^z\sigma_{i+1}^y
		\right),\\
		\tilde H_{\mathrm{eff}}^{[2]}
		&=-L\left(h+\frac{J^2}{4h}\right)\\
		&\quad+\sum_k
		\left(
		2h-2J\cos k+\frac{J^2}{h}\sin^2k
		\right)d_k^\dagger d_k ,
	\end{aligned}
\end{equation}
where $d_k$ are Jordan--Wigner flip fermions~\cite{SM}.

Kramers--Wannier duality~\cite{kramers1941statistics,fradkin1978order}, $\tau_i^x=\sigma_i^z\sigma_{i+1}^z$ and $\tau_{i-1}^z\tau_i^z=\sigma_i^x$, turns walls into flips and exchanges the roles of $H_0$ and $V$. Hence the dual SWT of the TFIM is the SWT of its Kramers--Wannier dual: Eq.~\eqref{eq:tfim} maps onto Eq.~\eqref{eq:tfim_dual} with $J\leftrightarrow h$, up to the ring constraint and boundary conditions.
Both agree with the exact dispersion $\epsilon_k=2\sqrt{J^2+h^2-2Jh\cos k}$~\cite{pfeuty1970one} through second order in $h/J$ and $J/h$, respectively.

\textit{Conclusion and outlook.---}%
We have introduced the dual SWT to construct effective Hamiltonians for strong-coupling problems.
The conventional and dual SWT are unified and generalized to the gSWT for general Hermitian operators, with three ingredients: the dominant operator, the perturbation operator, and the block partition.
Its key physics is the block diagonalization with respect to the block partition, which groups the eigenspaces of the dominant operator, and the canonical gauge choice.
We give the geodesic principle for the canonical gauge condition, which solves an open question about the SWT~\cite{mankodi2024perturbative}.
The exact closed-form solution of the gSWT generator, whose only input is the active gap set, gives the generator a clear physical meaning and provides a practical procedure for gSWT problems, as illustrated by the collective spin model and the transverse-field Ising model.
Our work reveals the geometric and algebraic aspects of the SWT.

The gSWT and its exact generator construction open avenues for applications to concrete physical problems across quantum many-body physics, quantum information, and quantum control, and for extensions to time-dependent, Floquet, and open quantum systems.

\textit{Acknowledgments.---}The author thanks Mircea Trif for helpful and stimulating discussions. This research was supported by the Foundation for Polish Science project MagTop (No. FENG.02.01-IP.05-0028/23) co-financed by the European Union from the funds of Priority 2 of the European Funds for a Smart Economy Program 2021--2027 (FENG).

AI tools (Claude and ChatGPT) were used to assist with the code, proofs, and the preparation of the manuscript. The author takes full responsibility for the scientific content of this paper.

\textit{Data availability.---}The data that support the findings of this article are not publicly available. The data are available from the author upon reasonable request.

\bibliography{ref_main}

\begin{center}\textbf{End Matter}\end{center}

\begin{figure}[htbp]
	\includegraphics[width=\columnwidth]{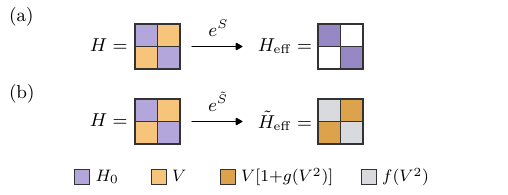}
	\caption{\textbf{Block structure in the eigenbasis of $\Gamma$.} $H_0$ is $\Gamma$-even and $V$ is $\Gamma$-odd. (a)~SWT: $H_{\mathrm{eff}}$ is $\Gamma$-even when each block of the partition lies within a single $\Gamma$ sector. (b)~Dual SWT: the coupling is renormalized to $V[1+g(V^2)]$, and $\tilde H_{\mathrm{eff}}$ is block off-diagonal when $f=0$.}
	\label{fig:blocks}
\end{figure}

\textit{Appendix A: SWT and dual SWT.---}%
In the $\Gamma$ basis (sectors $\Gamma=\pm1$), $H_0=\operatorname{diag}(H_{0+},H_{0-})$ and $V=\bigl(\begin{smallmatrix}0&T\\T^\dagger&0\end{smallmatrix}\bigr)$, so that Eq.~\eqref{eq:H} reads $H=\bigl(\begin{smallmatrix}H_{0+}&T\\T^\dagger&H_{0-}\end{smallmatrix}\bigr)$ (Fig.~\ref{fig:blocks}). The two transformations treat its two parts in exchanged roles (Table~\ref{tab:compare}).

\begin{table}[b]
	\caption{SWT and dual SWT of Hamiltonian~\eqref{eq:H}.}
	\label{tab:compare}
	\begin{ruledtabular}
		\begin{tabular}{lcc}
			& SWT & Dual SWT \\
			\hline
			Reference $A$ & $H_0$ ($\Gamma$-even) & $V$ ($\Gamma$-odd) \\
			Perturbation $B$ & $V$ & $H_0$ \\
			Blocks & eigenspaces of $H_0$ & $|n,\pm\rangle$, $\ker V$ \\
			$H_\mathrm{eff}$ commutes with & $H_0$ & $V$ \\
			Parameter $x$ & $\|V\|/g_{H_0}$ & $\|H_0\|/g_V$ \\
			Canonical gauge & $\mathcal{P}_{H_0}S=0$ & $\mathcal{P}_V\tilde S=0$ \\
			First order $H^{(1)}$ & $V^{\parallel}$ & $H_0^{\parallel}$ \\
			Generator $S^{(1)}$ & $\operatorname{ad}_{H_0}^{-1}V^{\perp}$ & $\operatorname{ad}_V^{-1}H_0^{\perp}$ \\
			Second order $H^{(2)}$ & $\tfrac12\mathcal{P}_{H_0}[S^{(1)},V^{\perp}]$ & $\tfrac12\mathcal{P}_V[\tilde S^{(1)},H_0^{\perp}]$ \\
			$\Gamma$ parity of $H^{(n)}$ & $(-1)^{n}$ & $(-1)^{n+1}$ \\
			Form in $\Gamma$ sectors & block diagonal\footnote{If $H_{0+}$ and $H_{0-}$ share no eigenvalue, so that each block lies within a single $\Gamma$ sector; otherwise the odd orders do not vanish.} & $V[1{+}g(V^2)]{+}f(V^2)$ \\
			Obstruction & none & $f(V^2)$ \\
			Degenerate block & degenerate levels & $\ker V$, Eq.~\eqref{eq:zero} \\
		\end{tabular}
	\end{ruledtabular}
\end{table}

The grading operator constrains the spectrum of $V$: $\Gamma$ maps every eigenstate of $V$ with eigenvalue $\sigma$ to one with eigenvalue $-\sigma$.
The block $T$ has the singular-value decomposition $T=\sum_n\sigma_n|u_n\rangle\langle w_n|$, $\sigma_n>0$. It pairs an even state $|u_n\rangle$ with an odd state $|w_n\rangle$.
Each pair spans a dimer, a two-dimensional invariant subspace of $V$, on which $V=\sigma_n\tau^x$ in the basis $(|u_n\rangle,|w_n\rangle)$.
Its eigenstates $|n,\pm\rangle=(|u_n\rangle\pm|w_n\rangle)/\sqrt2$ have energies $\pm\sigma_n$.
The remaining $k_0=\dim\ker V$ zero modes span the $\Gamma$-invariant subspace $\ker V$, with projector $P_K$, and admit an orthonormal basis of definite parity.

With $H_0$ as the dominant operator, the blocks are the eigenspaces of $H_0$, and the SWT removes the coupling between them order by order. If each block lies in one sector, the odd orders vanish and $H_{\mathrm{eff}}$ is block diagonal in the sectors [Fig.~\ref{fig:blocks}(a)].

With $V$ as the dominant operator, the blocks are the eigenspaces of $V$. The dual SWT removes the part of $H_0$ that connects different eigenvalues of $V$ and leaves $\tilde H_{\mathrm{eff}}=V[1+g(V^2)]+f(V^2)$, Eq.~\eqref{eq:dualform}. The diagonal blocks only contain the obstruction $f$. When $f=0$, $\tilde H_{\mathrm{eff}}$ is purely off-diagonal in the $\Gamma$ basis: the dual SWT then block off-diagonalizes $H$ exactly as the SWT block diagonalizes it [Fig.~\ref{fig:blocks}(b)].

Three structural properties of the dual SWT of Eq.~\eqref{eq:H} follow from the grading.

(i)~\emph{Parity alternation.}
Since $\operatorname{ad}_V$ reverses $\Gamma$ parity
while $\mathcal P_V$ preserves it, the recursion yields alternating $\Gamma$ parity at successive perturbative orders,
\begin{equation}\label{eq:parity}
	\Gamma\tilde S^{(n)}\Gamma=(-1)^n\tilde S^{(n)},\quad
	\Gamma\tilde H^{(n)}\Gamma=(-1)^{n+1}\tilde H^{(n)}.
\end{equation}
Thus odd-order corrections to $\tilde H_{\mathrm{eff}}$ are $\Gamma$-even, even orders $\Gamma$-odd.

(ii)~\emph{Dimer reduction and spectral obstruction.} If $\sigma_n$ are distinct and $k_0\le1$, the exact effective Hamiltonian ($x<\tfrac12$) has the form
\begin{equation}\label{eq:dualform}
	\tilde H_{\mathrm{eff}}=V\bigl[1+g(V^2)\bigr]+f(V^2),
\end{equation}
where $f(V^2)=H_0^{\parallel}+O(H_0^3)$ collects the odd-order corrections in $H_0$ and $Vg(V^2)$, with $g=O(H_0^2)$, the even-order ones. The $f$ term is block diagonal in the $\Gamma$ sectors, the $V[1+g]$ term block off-diagonal. The exact levels continued from $\pm\sigma_n$ are $f(\sigma_n^2)\pm\sigma_n[1+g(\sigma_n^2)]$: odd orders shift the center of each pair, to leading order through $H_0^{\parallel}$, and even orders renormalize its splitting. Thus $\tilde H_{\mathrm{eff}}$ is $\Gamma$-odd if and only if $f(V^2)=0$, i.e., if every such pair is symmetric about zero and any zero mode remains at zero. For a given energy origin, $f$ is fixed by the exact spectrum and cannot be removed by any gauge choice. Equivalently, the dressed grading $\Gamma'=U^\dagger\Gamma U$, with $U=e^{\tilde S}$, satisfies $\Gamma'H\Gamma'=-H+2U^\dagger f(V^2)U$~\cite{SM}.

(iii)~\emph{Zero modes.} With no assumption on the degeneracy of the $\sigma_n$ and $V^{+}$ the pseudoinverse,
\begin{equation}\label{eq:zero}
	P_K\tilde H_{\mathrm{eff}} P_K=P_KH_0P_K-P_KH_0V^{+}H_0P_K+O(H_0^3),
\end{equation}
the dual of second-order downfolding~\cite{lowdin1951note}. Since $P_K$ is $\Gamma$ invariant and $V^{+}$ is $\Gamma$-odd, the second-order term couples only zero modes of opposite parity, so a single zero mode of definite parity moves only through the odd orders. Proofs of (i)--(iii) are given in the SM~\cite{SM}.

\textit{Appendix B: The gSWT order by order.---}%
After substituting $S=\sum_{n\ge1}S^{(n)}$ and $H_{\mathrm{eff}}=A+\sum_{n\ge1}H^{(n)}$, where the superscript denotes the order in $B$, into the Baker--Campbell--Hausdorff expansion
\begin{equation}
	H_{\mathrm{eff}}
	=e^SHe^{-S}
	=\sum_{k\ge0}\frac{1}{k!}\operatorname{ad}_S^k H,
\end{equation}
the only term of order $n$ containing $S^{(n)}$ is $[S^{(n)},A]=-\operatorname{ad}_AS^{(n)}$. The other terms involve $S^{(k)}$ with $k<n$ and are grouped together as $R^{(n)}$, so that $H^{(n)}+\operatorname{ad}_A S^{(n)}=R^{(n)}$, which yields Eqs.~\eqref{eq:Hn} and~\eqref{eq:Sn}.
Recursively replacing the innermost commutators $[S^{(j)},A]$ by $H^{(j)}-R^{(j)}=-\mathcal Q_\Pi R^{(j)}$ ($j<n$) removes explicit occurrences of $A$ from the source,
\begin{align}
	R^{(n)}
	&=\sum_{k=0}^{n-1}\frac{1}{k!}
	\sum_{n_1+\cdots+n_k=n-1}
	\operatorname{ad}_{S^{(n_1)}}\cdots
	\operatorname{ad}_{S^{(n_k)}}B
	\notag\\
	&\;-\sum_{k=2}^{n}\frac{1}{k!}
	\sum_{n_1+\cdots+n_k=n}
	\operatorname{ad}_{S^{(n_1)}}\cdots
	\operatorname{ad}_{S^{(n_{k-1})}}
	\mathcal Q_\Pi R^{(n_k)} ,
	\label{eq:Rgen}
\end{align}
with all $n_i\ge1$ and the $k=0$ contribution equal to $B\delta_{n1}$. Thus $R^{(n)}$ depends only on $B$ and previously determined orders.
Using $\mathcal Q_\Pi R^{(j)}=R^{(j)}-H^{(j)}$ and $H^{(1)}=B^\parallel$, the first three sources are
\begin{align}
	R^{(1)}&=B,
	\notag\\
	R^{(2)}&=[S^{(1)},B^\parallel]
	+\frac12[S^{(1)},B^\perp],
	\notag\\
	R^{(3)}&=[S^{(2)},B^\parallel]
	+\frac12[S^{(2)},B^\perp]
	+\frac12[S^{(1)},H^{(2)}]
	\notag\\
	&\quad+\frac1{12}[S^{(1)},[S^{(1)},B^\perp]].
	\label{eq:R3}
\end{align}
Since each $S^{(j)}$ is block off-diagonal, $\mathcal P_\Pi[S^{(j)},X]=0$ for $X\in\mathcal A_\Pi$. The single commutators with $B^\parallel$ and $H^{(j)}$ therefore vanish under projection, giving
\begin{subequations}\label{eq:Heff-orders}
	\begin{align}
		H^{(1)}
		& =B^\parallel,
		\label{eq:Heff1}\\
		H^{(2)}
		&=\frac12\mathcal P_\Pi[S^{(1)},B^\perp],
		\label{eq:Heff2}\\
		H^{(3)}
		&=\frac12\mathcal P_\Pi[S^{(2)},B^\perp]
		+\frac1{12}\mathcal P_\Pi
		[S^{(1)},[S^{(1)},B^\perp]].
		\label{eq:Heff3}
	\end{align}
\end{subequations}
The fourth order is given in the SM~\cite{SM}.

\textit{Appendix C: Chordal and geodesic distances.---}%
The gauge orbit $\mathcal{O}\cong\prod_\alpha\mathrm{U}(d_\alpha)$ is a compact submanifold of $\mathrm U(N)$, with $d_\alpha$ the dimension of block $\alpha$ and $N=\sum_\alpha d_\alpha$.
The chordal distance between the identity and the block-diagonalizing unitary, $d_c(U)=\|U-\openone\|_{\mathrm{HS}}=[\sum_j4\sin^2(\theta_j/2)]^{1/2}$, can also be used to fix the gauge.
Minimizing $d_c$ yields the least-action transformation of Cederbaum \textit{et al.}~\cite{cederbaum1989block}.
For two blocks, this is the Davis--Kahan direct rotation~\cite{davis1970rotation}, used by Bravyi, DiVincenzo, and Loss to define the exact SWT~\cite{bravyi2011schrieffer}.
Its generator is block off-diagonal and thus satisfies the canonical gauge.
For three or more blocks, however, the least-action generator generally acquires a block-diagonal component at third order in perturbation theory~\cite{mankodi2024perturbative}, separating the chordal and canonical gauge choices.
Theorem~\ref{thm:extremal} establishes a geodesic variational principle for the canonical gauge.

\textit{Appendix D: Determining the active gap set.---}%
The four guidelines for determining the active gap set, or a suitable superset, are as follows~\cite{SM}.
(i)~\emph{Block partition}: only inter-block transitions contribute, with gaps given by eigenvalue differences of $A$.
Grouping near-degenerate eigenspaces into one block retains their mutual couplings within that block and excludes those transitions from $R^\perp$. A good block partition can thus simplify the determination of the active gaps.
(ii)~\emph{Selection rules}: if $A=f(N)$ and $R^\perp=\sum_sR_s$ with $[N,R_s]=sR_s$, then $[A,R_s]=R_s[f(N+s)-f(N)]$.
The active gaps are the nonzero values $|f(n+s)-f(n)|$ for transitions with nonvanishing inter-block matrix elements.
For $f(N)=\omega N$, each allowed shift gives $|s\omega|$, independent of $n$.
(iii)~\emph{Commutator closure}: starting from $C_0=R^\perp$, iterate $C_{k+1}=[A,[A,C_k]]$ until the first linear dependence $C_m+\sum_{k=0}^{m-1}c_kC_k=0$ appears, with scalar coefficients $c_k$.
Since each double commutator multiplies a transition component by its squared gap, the polynomial $\mu(z)=z^m+\sum_{k=0}^{m-1}c_kz^k =\prod_{i=1}^{m}(z-g_i^2)$ has the roots $z_i=g_i^2>0$ and therefore determines $\mathcal{G}_A(R^\perp)=\{\sqrt{z_i}\}_{i=1}^{m}$, without constructing eigenvectors of $A$. Moreover, $G=-\frac{1}{c_0}\sum_{k=1}^{m}c_k[A,C_{k-1}]$, with $c_m=1$.
(iv)~\emph{Supersets}: any finite set of positive gaps known to contain the active set can be used in Eq.~\eqref{eq:ek} and yields the same exact generator.

\textit{Appendix E: Active and passive transformations.---}%
A unitary transformation has two different physical meanings.
In the \emph{active} transformation, the basis $\{|e_i\rangle\}$ is fixed and states and observables are transformed together, $|\psi\rangle \to U|\psi\rangle$ and $O \to U O U^{\dagger}$, which preserves expectation values.
In the \emph{passive} transformation, the physical system does not change and only the representation is changed: the same state and the same observable are re-expressed in the rotated basis $|f_i\rangle=U|e_i\rangle$, with $\langle f_i|\psi\rangle=(U^{\dagger}|\psi\rangle)_i$ and $\langle f_i|O|f_j\rangle=(U^{\dagger}OU)_{ij}$.
The gSWT is active: the basis and the partition $\Pi$ are fixed, $H_{\mathrm{eff}}=e^SHe^{-S}$ is the effective Hamiltonian in the same Hilbert space, and observables obey $\tilde O=e^{S}Oe^{-S}=O+[S,O]+\tfrac12[S,[S,O]]+\cdots$.
Equivalently, in the passive picture with $U=e^{-S}$, one obtains a new Hamiltonian representation $H^\prime = U^\dagger H U=e^SH e^{-S}$ of the system in the new basis $e^{-S}|e_i\rangle$.

\end{document}


\title{Supplemental Material for ``Generalized Schrieffer--Wolff transformation''}
\author{Tie-Cheng Guo\orcidlink{0000-0001-8252-7970}}
\email{gtc16@tsinghua.org.cn}
\affiliation{International Research Centre MagTop, Institute of Physics, Polish Academy of Sciences, Aleja Lotnikow 32/46, PL-02668 Warsaw, Poland}

\begin{abstract}
This Supplemental Material contains the proofs and the supporting material for the main text.
Section~\ref{sec:dual} proves the structural properties (i)--(iii) of the dual Schrieffer--Wolff transformation (SWT), derives its first- and second-order form in the $\Gamma$ basis, and shows when it block off-diagonalizes the Hamiltonian.
Section~\ref{sec:gswt} develops the generalized SWT (gSWT), establishing the sufficient conditions for its existence and convergence, the order-by-order recursion, and the geodesic principle (Theorem~1 of the main text) with the uniqueness of the canonical gauge.
The nested-commutator solution (Theorem~2 of the main text) is proved, with the single-gap collapse, the operator-valued coefficient, and the guidelines for the active gap set. The block-diagonal projector in the effective Hamiltonian is evaluated, including the commutator form of every order for the finest partition.
Section~\ref{sec:models} applies the gSWT to nine models.
Section~\ref{sec:pade} constructs the two-point Pad\'e approximant of Fig.~2 of the main text.
\end{abstract}
\maketitle

\tableofcontents

\section{Dual Schrieffer--Wolff transformation}\label{sec:dual}

The conventional Schrieffer--Wolff transformation (SWT) eliminates a weak coupling $V$ between sectors governed by $H_0$. In the language of Sec.~\ref{sec:gswt}, it is the generalized SWT (gSWT) with $(A,B)=(H_0,V)$. The dual SWT addresses the complementary strong-coupling problem: it takes $V$ as the dominant operator and $H_0$ as the perturbation, i.e., it is the gSWT with $(A,B)=(V,H_0)$.
Throughout this section, perturbative orders are counted in $H_0$ at fixed $V$. In this Supplemental Material, $\|\cdot\|$ denotes the operator norm and $\|\cdot\|_{\mathrm{HS}}$ the Hilbert--Schmidt norm, $\|X\|_{\mathrm{HS}}=(\operatorname{Tr}X^\dagger X)^{1/2}$.

The general construction and bounds of the gSWT are established in Sec.~\ref{sec:gswt}. The results of Sec.~\ref{sec:gswt} used below do not rely on this section.

\subsection{Physical setting and dual transformation}

We consider the Hamiltonian
\begin{equation}\label{eq:dual-setting-sm}
	H=H_0+V,\qquad
	\Gamma H_0\Gamma=H_0,\qquad
	\Gamma V\Gamma=-V,\qquad
	\Gamma=\Gamma^\dagger=\Gamma^{-1}.
\end{equation}
Here $\Gamma$ is a Hermitian unitary grading operator whose eigenvalues $\pm1$ define two sectors.
The Hermitian operators $H_0$ and $V$ describe the dynamics within these sectors and the coupling between them, respectively.
In an eigenbasis of $\Gamma$ ordered by $\Gamma=+1$, $\Gamma=-1$ (the $\Gamma$ basis), we have
\begin{equation}\label{eq:dual-sector-form-sm}
	H_0=
	\begin{pmatrix}
		H_{0+}&0\\
		0&H_{0-}
	\end{pmatrix},
	\qquad
	V=
	\begin{pmatrix}
		0&T\\
		T^\dagger&0
	\end{pmatrix},
\end{equation}
where $T$ maps the $\Gamma=-1$ sector to the $\Gamma=+1$ sector. The general results below are established in a finite-dimensional Hilbert space.

The block $T$ has the singular-value decomposition $T=\sum_n\sigma_n|u_n\rangle\langle w_n|$, with $\sigma_n>0$. It pairs an even ($\Gamma=+1$) state $|u_n\rangle$ with an odd ($\Gamma=-1$) state $|w_n\rangle$. The range of the projector $\tau_n^0=|u_n\rangle\langle u_n|+|w_n\rangle\langle w_n|$ is a two-dimensional invariant subspace of $V$, which we call a dimer. We denote by $\tau_n^{x,y,z}$ the Pauli operators in the ordered basis $(|u_n\rangle,|w_n\rangle)$. Like $\tau_n^0$, they vanish outside the dimer.
On this subspace, $V$ acts as $\sigma_n\tau_n^x$, with eigenstates
\begin{equation}\label{eq:dual-dimer-states-sm}
	|n,s\rangle=\frac{|u_n\rangle+s|w_n\rangle}{\sqrt2},
	\qquad
	V|n,s\rangle=s\sigma_n|n,s\rangle,
	\quad s=\pm1.
\end{equation}
The grading exchanges the two hybridized states: $\Gamma|n,s\rangle=|n,-s\rangle$.

The orthogonal complement of the dimers is $\ker V=\ker T^\dagger\oplus\ker T$, with $\ker T^\dagger$ in the $\Gamma=+1$ sector and $\ker T$ in the $\Gamma=-1$ sector. We choose its basis states $|k\rangle$ within these two subspaces, so that they have definite parity, $\Gamma|k\rangle=\gamma_k|k\rangle$ with $\gamma_k=\pm1$. The dimension and the projector of $\ker V$ are denoted by $k_0$ and $P_K$.

Writing $V=\sum_a v_aQ_a$ with distinct eigenvalues $v_a$, we use its finest partition $\Pi_V$:
\begin{equation}\label{eq:dual-projections-sm}
	\mathcal P_VX=\sum_aQ_aXQ_a,\qquad
	\mathcal Q_VX=X-\mathcal P_VX,\qquad
	X^\parallel=\mathcal P_VX,\qquad
	X^\perp=\mathcal Q_VX.
\end{equation}
Each $Q_a$ projects onto the complete eigenspace of $v_a$, so that degenerate states belong to a single block. Hence $\mathcal P_V$ projects onto the operators commuting with $V$. These blocks are defined by the eigenvalues of $V$, whereas the original sectors are defined by $\Gamma=\pm1$.

The dual SWT eliminates transitions between distinct eigenspaces of $V$ through the unitary transformation
\begin{equation}\label{eq:dual-definition-sm}
	\tilde H_{\mathrm{eff}}
	=e^{\tilde S}He^{-\tilde S}.
\end{equation}
The anti-Hermitian generator $\tilde S^\dagger=-\tilde S$ is determined order by order by the block-diagonal condition, which for the finest partition reads $[V,\tilde H_{\mathrm{eff}}]=0$, and the canonical gauge condition $\mathcal P_V\tilde S=0$.
Expanding about $H_0=0$, where the transformation reduces to the identity, we write
\begin{equation}
	\tilde S=\sum_{n\ge1}\tilde S^{(n)},\qquad
	\tilde H_{\mathrm{eff}}
	=V+\sum_{n\ge1}\tilde H^{(n)},
\end{equation}
where $\tilde S^{(n)}$ and $\tilde H^{(n)}$ are homogeneous of degree $n$ in $H_0$.

At each order, the general gSWT recursion gives
\begin{equation}\label{eq:dual-rec-sm}
	[V,\tilde S^{(n)}]
	=\mathcal Q_V\tilde R^{(n)},\qquad
	\tilde H^{(n)}
	=\mathcal P_V\tilde R^{(n)},
\end{equation}
where $\tilde R^{(n)}$ contains only $V$, $H_0$, and the generators of lower orders, with $\tilde R^{(1)}=H_0$. The first equation removes the transitions between the eigenspaces of $V$, while the second retains the effective dynamics within them. The required inverse is
\begin{equation}\label{eq:dual-inverse-sm}
	\operatorname{ad}_V^{-1}\mathcal Q_VX
	=\sum_{a\ne b}\frac{Q_aXQ_b}{v_a-v_b}.
\end{equation}
The nested-commutator solution of Theorem~\ref{thm:nested} evaluates this inverse without constructing the eigenvectors of $V$.

For $(A,B)=(V,H_0)$, the spectral groups continued from different eigenvalues of $V$ remain separated for $x=\|H_0\|/g_V<\tfrac12$ (Theorem~\ref{thm:exist}), where $g_V=\min_{v_a\ne v_b}|v_a-v_b|$ is the sector gap of $V$, and the canonical expansion converges under the sufficient condition of Theorem~\ref{thm:conv}. For $V=\lambda W$ with $\lambda>0$, this construction expands $H/\lambda=W+\lambda^{-1}H_0$ in powers of the inverse coupling $\lambda^{-1}$, providing the strong-coupling counterpart of the conventional SWT.

\subsection{Parity, dimer structure, and zero modes}

Theorems~\ref{thm:parity}, \ref{thm:dimer}, and~\ref{thm:zero} of this subsection are the properties (i), (ii), and (iii) of the dual SWT stated in the End Matter of the main text.

\begin{theorem}[Parity alternation]\label{thm:parity}
For every $n\ge1$,
\begin{equation}\label{eq:dual-parity-sm}
	\Gamma\tilde S^{(n)}\Gamma
	=(-1)^n\tilde S^{(n)},\qquad
	\Gamma\tilde H^{(n)}\Gamma
	=(-1)^{n+1}\tilde H^{(n)}.
\end{equation}
Odd orders of the effective Hamiltonian act within the $\Gamma$ sectors, while even orders couple the sectors.
\end{theorem}

\begin{proof}
Since $\Gamma^2=\openone$ and $\Gamma(V+H_0)\Gamma=-(V-H_0)$, conjugating $\tilde H_{\mathrm{eff}}=e^{\tilde S}(V+H_0)e^{-\tilde S}$ by $\Gamma$ gives
\begin{equation*}
	e^{\Gamma\tilde S\Gamma}(V-H_0)
	e^{-\Gamma\tilde S\Gamma}
	=-\Gamma\tilde H_{\mathrm{eff}}\Gamma.
\end{equation*}
Because $\Gamma V\Gamma=-V$, conjugation by $\Gamma$ maps each eigenspace of $V$ onto the eigenspace of the opposite eigenvalue. Hence $\Gamma\tilde H_{\mathrm{eff}}\Gamma$ commutes with $V$ and $\mathcal P_V(\Gamma\tilde S\Gamma)=\Gamma(\mathcal P_V\tilde S)\Gamma=0$. The anti-Hermitian $\Gamma\tilde S\Gamma$ thus satisfies, order by order, both conditions of the dual SWT for the perturbation $-H_0$, and by the uniqueness in Theorem~\ref{thm:rec} it is the canonical generator for $-H_0$, with effective Hamiltonian $-\Gamma\tilde H_{\mathrm{eff}}\Gamma$.

Since $\tilde S^{(n)}$ and $\tilde H^{(n)}$ are homogeneous of degree $n$ in $H_0$, the generator and the effective Hamiltonian for $-H_0$ are also $\sum_{n\ge1}(-1)^n\tilde S^{(n)}$ and $V+\sum_{n\ge1}(-1)^n\tilde H^{(n)}$. Comparing the terms of degree $n$ with $\Gamma\tilde S\Gamma$ and $-\Gamma\tilde H_{\mathrm{eff}}\Gamma=V-\sum_{n\ge1}\Gamma\tilde H^{(n)}\Gamma$ gives Eq.~\eqref{eq:dual-parity-sm}.
\end{proof}

The two parities differ by one sign because conjugation by $\Gamma$ reverses both the perturbation and the overall sign of the Hamiltonian. The latter leaves the generator unchanged but reverses the effective Hamiltonian.
The same argument applies to a gapped coarse-grained partition whenever $\Gamma$ permutes its blocks.

\begin{theorem}[Dimer reduction and spectral obstruction]\label{thm:dimer}
Suppose that the positive singular values $\sigma_n$ are pairwise distinct, that $k_0\le1$, and that $x=\|H_0\|/g_V<\tfrac12$, so that the exact levels $E_{n,\pm}$ and $E_K$ continued from $\pm\sigma_n$ and from the zero mode are well defined (Theorem~\ref{thm:exist}). Then, for every block-diagonalizing unitary,
\begin{equation}\label{eq:dimer}
	\tilde H_{\mathrm{eff}}
	=V[1+g(V^2)]+f(V^2),
\end{equation}
where $f$ and $g$ are fixed by the exact levels,
\begin{equation}\label{eq:dual-centers-sm}
	\begin{aligned}
		f(\sigma_n^2)
		&=\frac{E_{n,+}+E_{n,-}}2,\\
		\sigma_n[1+g(\sigma_n^2)]
		&=\frac{E_{n,+}-E_{n,-}}2,
	\end{aligned}
\end{equation}
and $f(0)=E_K$ if a zero mode is present. The operator $f(V^2)$ contains the odd perturbative orders and gives the dimer centers, whereas $Vg(V^2)$ contains the even-order corrections to their splittings. In particular, $\tilde H_{\mathrm{eff}}$ does not depend on the gauge, and it is $\Gamma$-odd if and only if $f(V^2)=0$, i.e., every continued pair is symmetric about zero and any zero mode remains at zero.
\end{theorem}

\begin{proof}
Under the stated assumptions every eigenvalue of $V$ is nondegenerate, so every block is one-dimensional. A block-diagonal $\tilde H_{\mathrm{eff}}$ is then diagonal in the eigenbasis of $V$, and by Theorem~\ref{thm:exist} each block has the exact level continued from it:
\begin{equation*}
	\tilde H_{\mathrm{eff}}
	=\sum_n
	\left(E_{n,+}Q_{n,+}+E_{n,-}Q_{n,-}\right)
	+E_KP_K,
\end{equation*}
where the last term is omitted when $k_0=0$. This expression is the same for every block-diagonalizing unitary. With $Q_{n,\pm}=(\tau_n^0\pm\tau_n^x)/2$, the contribution of the $n$th dimer is
\begin{equation*}
	\frac{E_{n,+}+E_{n,-}}2\,\tau_n^0
	+\frac{E_{n,+}-E_{n,-}}2\,\tau_n^x.
\end{equation*}
Since the points $\sigma_n^2$, together with $0$ when $k_0=1$, are pairwise distinct, Eq.~\eqref{eq:dual-centers-sm} and $f(0)=E_K$ define functions $f$ and $g$ on the spectrum of $V^2$. With $V=\sum_n\sigma_n\tau_n^x$ and $V^2=\sum_n\sigma_n^2\tau_n^0$, the sum over the blocks is $V[1+g(V^2)]+f(V^2)$, which proves Eq.~\eqref{eq:dimer}.

Because $\tau_n^0$ and $P_K$ are $\Gamma$-even and $\tau_n^x$ is $\Gamma$-odd, $f(V^2)$ and $V[1+g(V^2)]$ are the $\Gamma$-even and $\Gamma$-odd parts of $\tilde H_{\mathrm{eff}}$. By Theorem~\ref{thm:parity}, which applies because $\tilde H_{\mathrm{eff}}$ is the same for every block-diagonalizing unitary, they contain the odd and the even perturbative orders, respectively. Hence $\tilde H_{\mathrm{eff}}$ is $\Gamma$-odd if and only if $f(V^2)=0$, i.e., every pair center and, when a zero mode is present, $E_K$ vanish. Being fixed by the exact levels, a nonzero $f$ cannot be removed by any gauge choice.
\end{proof}

At leading order,
\begin{equation}\label{eq:dual-leading-center-sm}
	f(V^2)=H_0^\parallel+O(H_0^3),\qquad
	H_0^\parallel
	=\sum_n c_n\tau_n^0+P_KH_0P_K,
\end{equation}
where
\begin{equation}
	c_n=\frac{
	\langle u_n|H_0|u_n\rangle+
	\langle w_n|H_0|w_n\rangle}{2}.
\end{equation}
The two levels receive the same first-order shift because $H_0$ has no matrix elements between the $\Gamma$ sectors. Only the values of $f$ and $g$ on the spectrum of $V^2$ enter Eq.~\eqref{eq:dimer}, and $g(0)$ does not contribute because $V$ vanishes on its kernel.

The obstruction also has a direct interpretation in the original frame. With $U=e^{\tilde S}$ and $\Gamma'=U^\dagger\Gamma U$, the $\Gamma$ parities of the two terms of Eq.~\eqref{eq:dimer} give
\begin{equation}\label{eq:dual-dressed-grading-sm}
	\Gamma'H\Gamma'
	=-H+2U^\dagger f(V^2)U.
\end{equation}
The obstruction therefore measures the failure of the dressed grading $\Gamma'$ to anticommute with $H$.

\paragraph*{Degenerate reference levels.}
When a positive singular value is degenerate, or when $k_0>1$, the effective Hamiltonian acts as a matrix within the corresponding eigenspaces of $V$, and the scalar form of Eq.~\eqref{eq:dimer} need not hold. For a degenerate singular value, the $\Gamma$-even and $\Gamma$-odd parts of $\tilde H_{\mathrm{eff}}$ on the two eigenspaces at $\pm\sigma_n$ need not commute, so a common basis of independent scalar dimers need not exist. A kernel containing zero modes of both $\Gamma$ parities can support $\Gamma$-odd couplings, which cannot be represented by $Vg(V^2)$ because it vanishes on $\ker V$. The block spectra remain gauge invariant, and parity alternation remains valid in the canonical gauge in all these cases.

\begin{theorem}[Zero-mode dynamics]\label{thm:zero}
Let $V^+=\sum_{v_a\ne0}Q_a/v_a$ be the pseudoinverse of $V$. Without any assumption on the multiplicities of the singular values,
\begin{equation}\label{eq:lowdin}
	P_K\tilde H_{\mathrm{eff}}P_K
	=P_KH_0P_K-P_KH_0V^+H_0P_K+O(H_0^3).
\end{equation}
The second-order term connects only zero modes of opposite $\Gamma$ parity. For a single zero mode, all even-order energy corrections vanish.
\end{theorem}

\begin{proof}
By Eq.~\eqref{eq:dual-rec-sm} with $\tilde R^{(1)}=H_0$, the first-order contribution is $P_K\tilde H^{(1)}P_K=P_KH_0P_K$, and the first-order generator is $\tilde S^{(1)}=\operatorname{ad}_V^{-1}\mathcal Q_VH_0$. For $v_a\ne0$, Eq.~\eqref{eq:dual-inverse-sm} then gives
\begin{equation*}
	P_K\tilde S^{(1)}Q_a
	=-\frac{P_KH_0Q_a}{v_a},\qquad
	Q_a\tilde S^{(1)}P_K
	=\frac{Q_aH_0P_K}{v_a}.
\end{equation*}
Substitution into the second-order term of Eq.~\eqref{eq:H1234} for $(A,B)=(V,H_0)$ yields
\begin{equation}\label{eq:H2zero}
	\begin{aligned}
		P_K\tilde H^{(2)}P_K
		&=\frac12P_K
		[\tilde S^{(1)},H_0^\perp]P_K\\
		&=-\sum_{v_a\ne0}
		\frac{P_KH_0Q_aH_0P_K}{v_a}
		=-P_KH_0V^+H_0P_K.
	\end{aligned}
\end{equation}

Since $\Gamma V^+\Gamma=-V^+$ and $H_0$ is even, the second-order operator is odd. Its matrix elements between zero modes of equal parity therefore vanish. Theorem~\ref{thm:parity} gives the same selection rule at every even order. In particular, the even-order energy corrections of a single zero mode vanish.
\end{proof}

Physically, virtual channels through the paired levels $+\sigma_n$ and $-\sigma_n$ cancel between zero modes of equal parity. A single zero mode can nevertheless shift at first order:
\begin{equation}
	E_K=\langle k|H_0|k\rangle+O(H_0^3).
\end{equation}
The selection rule removes even-order shifts but does not generally fix the level at zero.

\subsection{Second-order dual SWT in the \texorpdfstring{$\Gamma$}{Gamma} basis}\label{sec:dualsecond}

This subsection determines how $H_0$ shifts and mixes the hybridized levels of $V$ and expresses the resulting effective dynamics in the original $\Gamma$ basis. We assume that the positive singular values $\sigma_n$ are pairwise distinct, while allowing an arbitrary number $k_0$ of zero modes.

\paragraph*{Reference states and perturbation matrix elements.}
We use the singular vectors $|u_n\rangle$ and $|w_n\rangle$, the hybridized states $|n,s\rangle$ of Eq.~\eqref{eq:dual-dimer-states-sm}, and an orthonormal basis $|k\rangle$, $k=1,\ldots,k_0$, of the kernel with definite parity $\gamma_k$, so that $P_K=\sum_k|k\rangle\langle k|$. Indices $n,m$ label nonzero dimers, $s,s'=\pm1$ label their hybridized levels, and $k,l$ label zero modes.

The perturbation is specified by its matrix elements within the $\Gamma$ sectors:
\begin{equation}\label{eq:hdef}
	\begin{aligned}
		h^u_{nm}&=\langle u_n|H_0|u_m\rangle,
		&h^w_{nm}&=\langle w_n|H_0|w_m\rangle,\\
		h_{nk}
		&=
		\begin{cases}
			\langle u_n|H_0|k\rangle,&\gamma_k=+1,\\
			\langle w_n|H_0|k\rangle,&\gamma_k=-1.
		\end{cases}
	\end{aligned}
\end{equation}
The center $c_n$ introduced above and the detuning $d_n$ are
\begin{equation}
	c_n=\frac{h^u_{nn}+h^w_{nn}}2,\qquad
	d_n=\frac{h^u_{nn}-h^w_{nn}}2.
\end{equation}
Within a dimer,
\begin{equation}
	\tau_n^0H_0\tau_n^0=c_n\tau_n^0+d_n\tau_n^z.
\end{equation}
The average energy $c_n$ shifts the dimer center, while the detuning $2d_n$ between its two sector states adds a $\tau_n^z$ component to the local Hamiltonian. The off-diagonal elements of $h^u$ and $h^w$ connect different dimers, and $h_{nk}$ connects a dimer to the kernel.

Because $H_0$ has no matrix elements between opposite $\Gamma$ sectors, its matrix elements in the hybridized basis are
\begin{equation}\label{eq:me}
	\begin{aligned}
		\langle n,s|H_0|m,s'\rangle
		&=\frac{h^u_{nm}+ss'h^w_{nm}}2,\\
		\langle n,s|H_0|k\rangle
		&=\frac{s^{(1-\gamma_k)/2}}{\sqrt2}\,h_{nk}.
	\end{aligned}
\end{equation}

\paragraph*{Secular dynamics and the first-order rotation.}
The secular part is $H_0^\parallel=\sum_nc_n\tau_n^0+P_KH_0P_K$, as in Eq.~\eqref{eq:dual-leading-center-sm}. It retains the diagonal matrix elements of the nonzero hybridized levels and the full kernel block, and it gives $\tilde H^{(1)}=H_0^\parallel$. The complementary part, $H_0^\perp=H_0-H_0^\parallel$, contains the intra-dimer detunings $d_n\tau_n^z$, the couplings between different dimers, and the couplings between the dimers and the kernel, all of which connect distinct eigenspaces of $V$.

The first-order generator eliminates these transitions through $[V,\tilde S^{(1)}]=H_0^\perp$. Its $\Gamma$-odd parity and anti-Hermiticity give
\begin{equation}\label{eq:constraint}
	\begin{aligned}
		\tilde S^{(1)}
		&=\begin{pmatrix}0&Y\\-Y^\dagger&0\end{pmatrix},\\
		TY^\dagger+YT^\dagger&=-H_{0+}^\perp,\\
		T^\dagger Y+Y^\dagger T&=H_{0-}^\perp,
	\end{aligned}
\end{equation}
where $Y$ maps the negative $\Gamma$ sector to the positive one, and $H_{0\pm}^\perp$ are the corresponding diagonal blocks of $H_0^\perp$.

For $Y_{nm}=\langle u_n|Y|w_m\rangle$, these equations become
\begin{equation}\label{eq:pairs}
	\begin{aligned}
		\sigma_mY_{nm}+\sigma_nY_{mn}^*
		&=-h^u_{nm}+c_n\delta_{nm},\\
		\sigma_nY_{nm}+\sigma_mY_{mn}^*
		&=h^w_{nm}-c_n\delta_{nm}.
	\end{aligned}
\end{equation}
Solving them and imposing $\mathcal P_V\tilde S^{(1)}=0$ gives
\begin{equation}\label{eq:Ysol}
	\begin{aligned}
		Y_{nn}&=-\frac{d_n}{2\sigma_n},\\
		Y_{nm}
		&=\frac{\sigma_mh^u_{nm}+\sigma_nh^w_{nm}}
		{\sigma_n^2-\sigma_m^2},
		&&n\ne m,\\
		Y_{km}&=-\frac{h_{mk}^*}{\sigma_m},
		&&\gamma_k=+1,\\
		Y_{nk}&=\frac{h_{nk}}{\sigma_n},
		&&\gamma_k=-1.
	\end{aligned}
\end{equation}
Here $Y_{km}=\langle k|Y|w_m\rangle$ and $Y_{nk}=\langle u_n|Y|k\rangle$ in the last two lines. The kernel-to-kernel block vanishes in the canonical gauge.

The denominators identify the physical transition energies. The detuning $d_n$ couples the two levels of one dimer, separated by $2\sigma_n$. The combinations $h^u_{nm}+h^w_{nm}$ and $h^u_{nm}-h^w_{nm}$ connect equal-sign and opposite-sign levels, respectively, with denominators $\sigma_n-\sigma_m$ and $\sigma_n+\sigma_m$. Their combination produces $\sigma_n^2-\sigma_m^2$ in Eq.~\eqref{eq:Ysol}, a product that is not itself an active gap. Transitions involving a zero mode have gap $\sigma_n$. For an isolated dimer, the generator reduces to
\begin{equation}
	\tilde S^{(1)}
	=-i\frac{d_n}{2\sigma_n}\tau_n^y,
\end{equation}
the infinitesimal rotation that removes its detuning in the hybridized basis.

\paragraph*{Virtual processes and the second-order Hamiltonian.}
At second order,
\begin{equation}
	\tilde H^{(2)}
	=\frac12\mathcal P_V[\tilde S^{(1)},H_0^\perp].
\end{equation}
For a spectral projector $Q_a$ of $V$ with eigenvalue $v_a$, substitution of the first-order inverse gives
\begin{equation}\label{eq:dual-H2-block-sm}
	Q_a\tilde H^{(2)}Q_a
	=\sum_{b\ne a}
	\frac{Q_aH_0Q_bH_0Q_a}{v_a-v_b}.
\end{equation}
This expression also holds for degenerate blocks. It retains the virtual processes that leave a reference eigenspace and return to it.

For the simple nonzero levels considered here,
\begin{equation}\label{eq:H2diag}
	\begin{aligned}
		\langle n,s|\tilde H^{(2)}|n,s\rangle
		&=\sum_{(m,s')\ne(n,s)}
		\frac{|\langle n,s|H_0|m,s'\rangle|^2}
		{s\sigma_n-s'\sigma_m}\\
		&\quad+\sum_k
		\frac{|\langle n,s|H_0|k\rangle|^2}
		{s\sigma_n}
		=s\Delta_n,
	\end{aligned}
\end{equation}
where
\begin{equation}\label{eq:Delta}
	\begin{aligned}
		\Delta_n
		&=\frac{d_n^2}{2\sigma_n}
		+\sum_{m\ne n}\sum_{s'=\pm1}
		\frac{|h^u_{nm}+s'h^w_{nm}|^2}
		{4(\sigma_n-s'\sigma_m)}\\
		&\quad+\sum_k\frac{|h_{nk}|^2}{2\sigma_n}.
	\end{aligned}
\end{equation}
The three terms describe virtual transitions to the opposite level of the same dimer, to other dimers, and to zero modes, respectively. Reversing the signs of both hybridized levels preserves the transition probability and reverses the energy denominator, and the zero-mode contribution changes sign in the same way. Thus
\begin{equation}
	\tau_n^0\tilde H^{(2)}\tau_n^0=\Delta_n\tau_n^x.
\end{equation}
The second order therefore changes the splitting without shifting the pair center. The intra-dimer and zero-mode contributions to $\Delta_n$ are nonnegative, whereas the inter-dimer contribution can have either sign. The effective dynamics within the kernel is given by Eq.~\eqref{eq:H2zero}.

In general, the projector $\mathcal P_V$ in $\tilde H^{(2)}$ cannot be omitted. Although $[\tilde S^{(1)},H_0^\perp]$ is $\Gamma$-odd, it need not commute with $V$: it can connect different dimers or the opposite levels of one dimer. Together with $[\tilde S^{(1)},H_0^\parallel]$, its block-off-diagonal components determine
\begin{equation}\label{eq:dual-S2-source-sm}
	[V,\tilde S^{(2)}]
	=\mathcal Q_V\left(
	[\tilde S^{(1)},H_0^\parallel]
	+\frac12[\tilde S^{(1)},H_0^\perp]
	\right).
\end{equation}
They are canceled at second order by $[\tilde S^{(2)},V]$ and re-enter $\tilde H_{\mathrm{eff}}$ at third order through $\tfrac12\mathcal P_V[\tilde S^{(2)},H_0^\perp]$. By Theorem~\ref{thm:parity} this term is $\Gamma$-even, so on the kernel it acts only between zero modes of equal parity, in contrast to the second-order term of Eq.~\eqref{eq:H2zero}.

\paragraph*{Effective dimer and zero-mode dynamics.}
Combining the retained blocks gives
\begin{equation}\label{eq:dualresult}
	\begin{aligned}
		\tilde H_{\mathrm{eff}}
		&=\sum_n\left[
		c_n\tau_n^0+(\sigma_n+\Delta_n)\tau_n^x
		\right]\\
		&\quad+P_K(H_0-H_0V^+H_0)P_K
		+O(H_0^3).
	\end{aligned}
\end{equation}
Each nonzero dimer therefore has levels $c_n\pm(\sigma_n+\Delta_n)+O(H_0^3)$. Under the simple-spectrum assumptions of Theorem~\ref{thm:dimer}, parity further implies
\begin{equation}
	f(\sigma_n^2)=c_n+O(H_0^3),\qquad
	\sigma_ng(\sigma_n^2)=\Delta_n+O(H_0^4).
\end{equation}
The leading spectral obstruction is the average intra-sector energy $c_n$, while $\Delta_n$ renormalizes the hybridization through the allowed virtual channels.

In the $\Gamma$ basis, Eq.~\eqref{eq:dualresult} becomes
\begin{equation}
	\tilde H_{\mathrm{eff}}
	=
	\begin{pmatrix}
		F_+&\tilde T\\
		\tilde T^\dagger&F_-
	\end{pmatrix}
	+O(H_0^3).
\end{equation}
Here $F_\pm$ are the restrictions of $H_0^\parallel$ to the $\Gamma=\pm1$ sectors, and
\begin{equation}
	\begin{aligned}
		\tilde T
		&=\sum_n(\sigma_n+\Delta_n)
		|u_n\rangle\langle w_n|\\
		&\quad-
		\sum_{\substack{k,l\\\gamma_k=+1,\ \gamma_l=-1}}
		\left(\sum_n\frac{h_{nk}^*h_{nl}}{\sigma_n}\right)
		|k\rangle\langle l|.
	\end{aligned}
\end{equation}
The second term is the coupling between opposite-parity zero modes induced by virtual transitions through the nonzero dimers. Thus, through second order, the nonzero dimers retain the singular-vector directions of $T$, acquire gauge-invariant center shifts, and have renormalized coupling strengths. The kernel retains its full effective matrix dynamics, and every retained term commutes with $V$.

\subsection{Block off-diagonalization and the minimal obstruction}

An operator that is block off-diagonal in the $\Gamma$ basis is $\Gamma$-odd, and its spectrum is therefore symmetric about zero. The SWT block diagonalizes $H$ with respect to the eigenspaces of $H_0$. The following proposition states when the dual SWT block off-diagonalizes $H$ in the $\Gamma$ basis.

\begin{proposition}[Block off-diagonalization]\label{prop:offdiag}
Under the assumptions of Theorem~\ref{thm:dimer}, the following statements are equivalent:
\begin{enumerate}[label=(\alph*),nosep,leftmargin=2.2em]
\item The obstruction vanishes, $f(V^2)=0$.
\item The dual SWT block off-diagonalizes $H$ in the $\Gamma$ basis,
\begin{equation}\label{eq:dual-offdiag-sm}
	\tilde H_{\mathrm{eff}}=V[1+g(V^2)]
	=\begin{pmatrix}0&\tilde T\\\tilde T^\dagger&0\end{pmatrix},
	\qquad
	\tilde T=\sum_n\tilde\sigma_n|u_n\rangle\langle w_n|,
	\qquad
	\tilde\sigma_n=\sigma_n[1+g(\sigma_n^2)]=\frac{E_{n,+}-E_{n,-}}2.
\end{equation}
\item Some unitary $W$ makes $WHW^\dagger$ block off-diagonal in the $\Gamma$ basis.
\item The spectrum of $H$ is symmetric about zero.
\end{enumerate}
\end{proposition}

\begin{proof}
(a)$\Rightarrow$(b): With $f(V^2)=0$, Eq.~\eqref{eq:dimer} reduces to $\tilde H_{\mathrm{eff}}=V[1+g(V^2)]$. This operator acts as $\tilde\sigma_n\tau_n^x$ on the $n$th dimer and vanishes on $\ker V$, and Eq.~\eqref{eq:dual-centers-sm} gives $\tilde\sigma_n=(E_{n,+}-E_{n,-})/2$.

(b)$\Rightarrow$(c): A block-diagonalizing unitary $W$ exists by Theorem~\ref{thm:exist}, and $WHW^\dagger=\tilde H_{\mathrm{eff}}$ is block off-diagonal by (b).

(c)$\Rightarrow$(d): The operator $M=WHW^\dagger$ has the spectrum of $H$ and obeys $\Gamma M\Gamma=-M$. Hence $\Gamma$ maps every eigenvector of $M$ with eigenvalue $E$ to one with eigenvalue $-E$.

(d)$\Rightarrow$(a): By Weyl's inequality, as in the proof of Theorem~\ref{thm:exist}, the exact level continued from an eigenvalue $v$ of $V$ lies within $\|H_0\|<g_V/2$ of $v$, while distinct eigenvalues of $V$ are separated by at least $g_V$. If the spectrum is symmetric, $-E_{n,+}$ is an exact level within $\|H_0\|$ of $-\sigma_n$, so it equals $E_{n,-}$. Likewise, when a zero mode is present, $-E_K$ is an exact level within $\|H_0\|$ of zero, so it equals $E_K$. Thus $E_{n,+}+E_{n,-}=0$ for every $n$ and $E_K=0$, and Eq.~\eqref{eq:dual-centers-sm} gives $f(V^2)=0$.
\end{proof}

Thus the dual SWT block off-diagonalizes $H$ whenever any unitary transformation can do so, in analogy with the block diagonalization of $H$ by the SWT. When $f(V^2)\ne0$, no unitary transformation block off-diagonalizes $H$, and the $\Gamma$-even part retained by the dual SWT is the smallest one, in Hilbert--Schmidt norm, compatible with the level pairing inherited from $V$. Indeed, a Hermitian operator on the $n$th dimer with eigenvalues $E_{n,\pm}$ has diagonal elements $a_n$ and $b_n$ in the basis $(|u_n\rangle,|w_n\rangle)$. They constitute its $\Gamma$-even part and satisfy $a_n+b_n=E_{n,+}+E_{n,-}$, so that
\begin{equation}\label{eq:dual-minimum-sm}
	a_n^2+b_n^2\ge\frac{(E_{n,+}+E_{n,-})^2}{2},
\end{equation}
with equality for the scalar pair center $a_n=b_n=f(\sigma_n^2)$ of Eq.~\eqref{eq:dimer}.

\paragraph*{Leading obstruction.}
By Eq.~\eqref{eq:dual-leading-center-sm}, $f(V^2)=H_0^\parallel+O(H_0^3)$. The obstruction therefore vanishes at leading order if and only if $H_0^\parallel=\mathcal P_VH_0=0$, i.e., $c_n=0$ for every dimer and $P_KH_0P_K=0$. In that case $\tilde H^{(1)}=0$ and the first-order generator solves $[V,\tilde S^{(1)}]=H_0$ with the full $H_0$ on the right-hand side. The effective Hamiltonian truncated at second order (superscript $[2]$, as in the main text),
\begin{equation}\label{eq:dual-sector-second}
	\tilde H_{\mathrm{eff}}^{[2]}=V+\tfrac12\mathcal P_V[\tilde S^{(1)},H_0],
\end{equation}
is then $\Gamma$-odd by Theorem~\ref{thm:parity}: through second order the dual SWT block off-diagonalizes $H$ in the $\Gamma$ basis, and $f(V^2)=O(H_0^3)$.

\section{Generalized Schrieffer--Wolff transformation}\label{sec:gswt}

The gSWT applies to a general Hermitian operator $H=A+B$, with $A$ the dominant operator and $B$ the perturbation. It block diagonalizes $H$ by a unitary transformation with respect to a chosen gapped spectral partition of $A$ and retains the dynamics within each block. The following four subsections establish the existence, perturbative construction, and convergence of the gSWT, characterize the canonical gauge by a geodesic principle, derive an exact finite nested-commutator solution for the generator at each order, and evaluate the projection that determines the effective operator within the blocks.

\subsection{Definition and perturbative construction}\label{sec:gswtdef}

Let $H=A+B$, with $A$ and $B$ Hermitian on $\mathcal H=\mathbb C^N$. We choose a partition $\Pi=\{P_\alpha\}$ into at least two nonzero orthogonal projectors satisfying $\sum_\alpha P_\alpha=\openone$ and $[A,P_\alpha]=0$, and let $d_\alpha=\operatorname{rank}P_\alpha$. If $A_\alpha$ is the restriction of $A$ to $\operatorname{ran}P_\alpha$, define the sector gap and the expansion parameter by
\begin{equation}
	g_\Pi=\min_{\alpha\ne\beta}\operatorname{dist}
	\bigl(\operatorname{spec}A_\alpha,\operatorname{spec}A_\beta\bigr)>0,
	\qquad
	x=\frac{\|B\|}{g_\Pi}.
\end{equation}

The blocks of $\Pi$ are the spectral sectors of $A$. In the spectral decomposition $A=\sum_aE_aQ_a$, each $Q_a$ projects onto the complete eigenspace of a distinct eigenvalue $E_a$. Since $g_\Pi>0$, every block consists of complete eigenspaces, $P_\alpha=\sum_{a\in I_\alpha}Q_a$ with disjoint index sets $I_\alpha$, and $g_\Pi=\min\{|E_a-E_b|:a\in I_\alpha,\ b\in I_\beta,\ \alpha\neq\beta\}$, as in the main text. A block may contain one eigenspace or several nearby levels, whose mutual couplings are retained in the effective operator rather than eliminated at the cost of small energy denominators.

The block-diagonal projector $\mathcal P_\Pi$ and the complementary block-off-diagonal projector $\mathcal Q_\Pi$ give the block-diagonal and block-off-diagonal parts of an operator,
\begin{equation}
	\mathcal P_\Pi X=\sum_\alpha P_\alpha XP_\alpha,\qquad
	\mathcal Q_\Pi X=X-\mathcal P_\Pi X,\qquad
	X^\parallel=\mathcal P_\Pi X,\qquad X^\perp=\mathcal Q_\Pi X.
\end{equation}
In particular, the secular part $B^\parallel$ acts within the blocks and the nonsecular part $B^\perp$ couples them. We write $\mathcal A_\Pi=\{X:[P_\alpha,X]=0\ \text{for every }\alpha\}$ for the set of block-diagonal operators.

In the \emph{finest partition} every block is a single eigenspace of $A$, so that $\mathcal P_\Pi=\mathcal P_A$ retains exactly the operators commuting with $A$, and we write $\mathcal Q_AX=X-\mathcal P_AX$. In a \emph{coarse-grained partition} some blocks unite several eigenspaces.

Whenever the spectral groups remain separated along $A+tB$, $0\le t\le1$, let $\Pi_\alpha$ be the exact spectral projector obtained by continuing $P_\alpha$ to $t=1$. The sufficient condition $x<\tfrac12$ for this continuation is established in Theorem~\ref{thm:exist}.

\begin{definition}[gSWT]\label{def:gswt}
A unitary $U$ is \emph{block diagonalizing} for $(A,B,\Pi)$ if it satisfies the block-diagonal condition
\begin{equation}\label{eq:gswtdef}
	U\Pi_\alpha U^\dagger=P_\alpha
	\qquad\text{for every }\alpha.
\end{equation}
It is \emph{canonical} if, in addition, $U=e^S$ with a principal logarithm $S=\log U$ satisfying the canonical gauge condition
\begin{equation}\label{eq:gswtgauge}
	S^\dagger=-S,\qquad \|S\|<\pi,\qquad
	\mathcal P_\Pi S=0.
\end{equation}
A gSWT of $(A,B,\Pi)$ is a transformation $H_{\mathrm{eff}}=UHU^\dagger$ by a canonical block-diagonalizing unitary.
\end{definition}

These are the two conditions of the main text. The block-diagonal condition assigns the exact spectral groups to the reference blocks, and the canonical gauge condition restricts the generator to inter-block rotations. By Corollary~\ref{cor:unique}, at most one canonical block-diagonalizing unitary lies within the geodesic distance $\pi/(2\sqrt2)$ of the identity, and we call the transformation by this unitary the gSWT. The geodesic distance is defined in Sec.~\ref{sec:geodesic}. We construct the gSWT perturbatively from $S=0$ at $B=0$ and denote the resulting unitary by $U_g$. Its geodesic characterization is given in Theorem~\ref{thm:geodesic}.

\begin{theorem}[Spectral separation and existence]\label{thm:exist}
If $x<\tfrac12$, the spectral groups of $A+tB$, $0\le t\le1$, continued from different reference blocks remain separated by at least
\begin{equation}
	g_\Pi-2t\|B\|>0.
\end{equation}
Their spectral projectors $\Pi_\alpha(t)$ retain ranks $d_\alpha$, and block-diagonalizing unitaries exist. For each of them, every block of $UHU^\dagger$ has the exact spectrum continued from the corresponding reference block.
\end{theorem}

\begin{proof}
Weyl's inequality bounds the shift of every ordered eigenvalue of $A+tB$ by $t\|B\|$~\cite{bhatia1997matrix}. Two eigenvalues continued from different blocks, which are at least $g_\Pi$ apart at $t=0$, therefore remain separated by at least $g_\Pi-2t\|B\|>0$. The spectral projectors $\Pi_\alpha(t)$ of the continued groups are thus continuous in $t$ and retain the ranks $d_\alpha$.

Mapping an orthonormal basis of each $\operatorname{ran}\Pi_\alpha$ onto one of $\operatorname{ran}P_\alpha$ defines a unitary $U$ satisfying Eq.~\eqref{eq:gswtdef}. For every such $U$, $[H,\Pi_\alpha]=0$ gives $[UHU^\dagger,P_\alpha]=0$, so $UHU^\dagger$ is block diagonal, and its block $\alpha$ is unitarily equivalent to the restriction of $H$ to $\operatorname{ran}\Pi_\alpha$.
\end{proof}

For any block-diagonalizing unitary $U_0$, every other one is $KU_0$, where $K$ is block diagonal and unitary: indeed, $K=UU_0^\dagger$ commutes with every $P_\alpha$. These unitaries form the \emph{gauge orbit} $\mathcal O=\{KU_0:K\in\mathcal A_\Pi\ \text{unitary}\}\cong\prod_\alpha\mathrm U(d_\alpha)$, a compact submanifold of $\mathrm U(N)$. Along this orbit, $H_{\mathrm{eff}}\mapsto KH_{\mathrm{eff}}K^\dagger$, so the block spectra are fixed, while the basis within each multidimensional block remains arbitrary.

We write
\begin{equation}
	S=\sum_{n\ge1}S^{(n)},\qquad
	H_{\mathrm{eff}}=A+\sum_{n\ge1}H^{(n)},
\end{equation}
where order $n$ means degree $n$ in $B$ at fixed $A$ and $\Pi$. With $\operatorname{ad}_XY=[X,Y]$, the construction follows directly from the Baker--Campbell--Hausdorff (BCH) expansion,
\begin{equation}\label{eq:exact}
	\begin{aligned}
		e^SHe^{-S}
		&=H+[S,H]+\frac12[S,[S,H]]+\cdots\\
		&=\sum_{k\ge0}\frac1{k!}\operatorname{ad}_S^k(A+B).
	\end{aligned}
\end{equation}
The series are initially formal, and their convergence is established in Theorem~\ref{thm:conv}.

\begin{theorem}[Order-by-order construction]\label{thm:rec}
At each order $n$, the block-diagonal and canonical gauge conditions uniquely determine
\begin{equation}\label{eq:rec}
	H^{(n)}=\mathcal P_\Pi R^{(n)},\qquad
	[A,S^{(n)}]=\mathcal Q_\Pi R^{(n)},\qquad
	S^{(n)}=\operatorname{ad}_A^{-1}\mathcal Q_\Pi R^{(n)},
\end{equation}
where the source term $R^{(n)}$ contains only $A$, $B$, and the generators of lower orders:
\begin{equation}\label{eq:Rn}
	\begin{aligned}
		R^{(n)}
		&=\sum_{k=0}^{n-1}\frac1{k!}
		\sum_{n_1+\cdots+n_k=n-1}
		\operatorname{ad}_{S^{(n_1)}}\cdots
		\operatorname{ad}_{S^{(n_k)}}B\\
		&\quad+\sum_{k=2}^{n}\frac1{k!}
		\sum_{n_1+\cdots+n_k=n}
		\operatorname{ad}_{S^{(n_1)}}\cdots
		\operatorname{ad}_{S^{(n_k)}}A.
	\end{aligned}
\end{equation}
All $n_i\ge1$, the $k=0$ term is $B\delta_{n1}$, and empty sums vanish. The coefficients are homogeneous of degree $n$ in $B$, with $H^{(n)}$ Hermitian and $S^{(n)}$ anti-Hermitian.
\end{theorem}

\begin{proof}
\emph{Order-$n$ equation.} At order $n$ of the BCH expansion~\eqref{eq:exact}, the only term containing the unknown $S^{(n)}$ is $[S^{(n)},A]$. All other terms form the source in Eq.~\eqref{eq:Rn}, giving
\begin{equation}\label{eq:ordern}
	H^{(n)}=R^{(n)}-[A,S^{(n)}].
\end{equation}
The block-diagonal condition requires $H^{(n)}\in\mathcal A_\Pi$. In the canonical gauge $S^{(n)}$ is block off-diagonal, and so is $[A,S^{(n)}]$, because $A$ preserves the blocks. The block-diagonal and block-off-diagonal parts of Eq.~\eqref{eq:ordern} are therefore the first two relations of Eq.~\eqref{eq:rec}.

\emph{Uniqueness.} For $Q_a$ and $Q_b$ belonging to different blocks, the second relation reads
\begin{equation*}
	Q_aS^{(n)}Q_b=\frac{Q_aR^{(n)}Q_b}{E_a-E_b},
\end{equation*}
with $|E_a-E_b|\ge g_\Pi>0$, while the part of $S^{(n)}$ within the blocks vanishes. This fixes $S^{(n)}$ uniquely and defines $\operatorname{ad}_A^{-1}$ on the block-off-diagonal operators.

\emph{Hermiticity and homogeneity.} We use induction on $n$. The source $R^{(1)}=B$ is Hermitian and of degree one in $B$. If $S^{(j)}$ is anti-Hermitian and of degree $j$ for every $j<n$, every term of Eq.~\eqref{eq:Rn} is Hermitian and of degree $n$, because the commutator of an anti-Hermitian operator with a Hermitian one is Hermitian. Hence $H^{(n)}=\mathcal P_\Pi R^{(n)}$ is Hermitian, and $S^{(n)}$ is anti-Hermitian, because $(Q_aR^{(n)}Q_b)^\dagger=Q_bR^{(n)}Q_a$ and the denominator changes sign under $a\leftrightarrow b$. Both are of degree $n$.
\end{proof}

\paragraph*{Explicit low orders.}
At the lower orders, Eqs.~\eqref{eq:rec} and~\eqref{eq:ordern} give the already determined identities
\begin{equation}
	[S^{(j)},A]=-\mathcal Q_\Pi R^{(j)}=H^{(j)}-R^{(j)},
	\qquad j<n.
\end{equation}
Substituting them into the innermost commutator of the second sum in Eq.~\eqref{eq:Rn} gives the equivalent form free of $A$,
\begin{equation}\label{eq:RnA}
	R^{(n)}=\sum_{k=0}^{n-1}\frac1{k!}\sum_{n_1+\cdots+n_k=n-1}\operatorname{ad}_{S^{(n_1)}}\cdots\operatorname{ad}_{S^{(n_k)}}B
	-\sum_{k=2}^{n}\frac1{k!}\sum_{n_1+\cdots+n_k=n}\operatorname{ad}_{S^{(n_1)}}\cdots\operatorname{ad}_{S^{(n_{k-1})}}\mathcal Q_\Pi R^{(n_k)}.
\end{equation}
This is the form given in the End Matter of the main text. The recursion is built from $B$, the block projectors, and nested commutators, and the spectrum of $A$ enters only through the inversion of $\operatorname{ad}_A$.

Direct expansion through fourth order gives
\begin{subequations}\label{eq:R1234}
\begin{align}
	R^{(1)}&=B,\\
	R^{(2)}&=[S^{(1)},B^\parallel]
	+\tfrac12[S^{(1)},B^\perp],\\
	R^{(3)}&=[S^{(2)},B^\parallel]
	+\tfrac12[S^{(2)},B^\perp]
	+\tfrac12[S^{(1)},H^{(2)}]
	+\tfrac1{12}[S^{(1)},[S^{(1)},B^\perp]],\\
	R^{(4)}&=[S^{(3)},B^\parallel]
	+\tfrac12[S^{(3)},B^\perp]
	+\tfrac12[S^{(2)},H^{(2)}]
	+\tfrac12[S^{(1)},H^{(3)}]\notag\\
	&\quad+\tfrac1{12}\bigl(
	[S^{(1)},[S^{(2)},B^\perp]]
	+[S^{(2)},[S^{(1)},B^\perp]]
	-[S^{(1)},[S^{(1)},H^{(2)}]]\bigr).
\end{align}
\end{subequations}
For a block-off-diagonal $X$ and a block-diagonal $Y$, $\mathcal P_\Pi[X,Y]=0$, because $[X,Y]$ is again block off-diagonal. Thus
\begin{subequations}\label{eq:H1234}
\begin{align}
	H^{(1)}&=B^\parallel,\\
	H^{(2)}&=\tfrac12\mathcal P_\Pi[S^{(1)},B^\perp],\\
	H^{(3)}&=\tfrac12\mathcal P_\Pi[S^{(2)},B^\perp]
	+\tfrac1{12}\mathcal P_\Pi[S^{(1)},[S^{(1)},B^\perp]],\\
	H^{(4)}&=\tfrac12\mathcal P_\Pi[S^{(3)},B^\perp]
	+\tfrac1{12}\mathcal P_\Pi\bigl(
	[S^{(1)},[S^{(2)},B^\perp]]
	+[S^{(2)},[S^{(1)},B^\perp]]
	-[S^{(1)},[S^{(1)},H^{(2)}]]\bigr).
\end{align}
\end{subequations}
The second-order effective term depends only on $B^\perp$. However, $B^\parallel$ enters $S^{(2)}$ through $[S^{(1)},B^\parallel]$ and can affect the effective terms from third order onward. The nested term containing $H^{(2)}$ in $H^{(4)}$ generally survives projection. The inversion of $\operatorname{ad}_A$ and the projection $\mathcal P_\Pi$ are treated in Secs.~\ref{sec:nested} and~\ref{sec:projector}, respectively.

Spectral separation guarantees the existence of a block diagonalization but does not by itself establish convergence of the canonical series. The following sufficient condition ensures that the order-by-order construction sums to the exact gSWT.

\begin{theorem}[Convergence]\label{thm:conv}
If
\begin{equation}
	\|B\|_{\mathrm{HS}}<c\,g_\Pi,\qquad c=e^{-1/2}-\tfrac12,
\end{equation}
the canonical series $S=\sum_{n\ge1}S^{(n)}$ and $A+\sum_{n\ge1}H^{(n)}$ generated by the recursion~\eqref{eq:rec} converge absolutely and satisfy
\begin{equation}
	\|S\|_{\mathrm{HS}}<\tfrac14,\qquad
	\mathcal P_\Pi S=0,\qquad
	e^SHe^{-S}=A+\sum_{n\ge1}H^{(n)}\in\mathcal A_\Pi.
\end{equation}
The resulting unitary maps the continued spectral subspaces onto their reference blocks and satisfies the canonical gauge condition. It is the gSWT unitary $U_g$.
\end{theorem}

\begin{proof}
\emph{Bound on each order.} The transition-energy denominators and orthogonality of the components $Q_aR^{(n)}Q_b$ give
\begin{equation}\label{eq:sylv}
	\|S^{(n)}\|_{\mathrm{HS}}
	\le\frac{\|\mathcal Q_\Pi R^{(n)}\|_{\mathrm{HS}}}{g_\Pi}
	\le\frac{\|R^{(n)}\|_{\mathrm{HS}}}{g_\Pi}.
\end{equation}

\emph{Recursive inequality.} Write $r_n=\|R^{(n)}\|_{\mathrm{HS}}$ and define the partial sums
\begin{equation*}
	t_M=\frac1{g_\Pi}\sum_{n=1}^{M}r_n,\qquad t_0=0.
\end{equation*}
We use the source in the form of Eq.~\eqref{eq:RnA}, in which each innermost $[S^{(j)},A]$ of Eq.~\eqref{eq:Rn}, with $j<n$, is replaced by $-\mathcal Q_\Pi R^{(j)}$. Bounding every term by $\|[X,Y]\|_{\mathrm{HS}}\le2\|X\|_{\mathrm{HS}}\|Y\|_{\mathrm{HS}}$ and Eq.~\eqref{eq:sylv}, summing through order $M$, and dropping the order constraints on the nonnegative products gives
\begin{equation*}
	\begin{aligned}
		t_M
		&\le\sum_{k\ge2}\frac{2^{k-1}}{k!}t_{M-1}^k
		+\frac{\|B\|_{\mathrm{HS}}}{g_\Pi}\sum_{k\ge0}\frac{2^k}{k!}t_{M-1}^k\\
		&=\frac{e^{2t_{M-1}}-1-2t_{M-1}}2
		+\frac{\|B\|_{\mathrm{HS}}}{g_\Pi}e^{2t_{M-1}}.
	\end{aligned}
\end{equation*}

\emph{Induction.} The right-hand side increases with $t_{M-1}\ge0$. At $t_{M-1}=\tfrac14$, it equals
\begin{equation*}
	\left(\tfrac12+\frac{\|B\|_{\mathrm{HS}}}{g_\Pi}\right)e^{1/2}-\tfrac34<\tfrac14
\end{equation*}
under the stated condition. Starting from $t_0=0$, induction therefore bounds all partial sums by a common number smaller than $\tfrac14$. Consequently,
\begin{equation*}
	\sum_{n\ge1}\|S^{(n)}\|_{\mathrm{HS}}<\tfrac14,
	\qquad
	\sum_{n\ge1}\|H^{(n)}\|_{\mathrm{HS}}
	\le\sum_{n\ge1}r_n<\tfrac{g_\Pi}{4}.
\end{equation*}

\emph{Summation.} The series $S=\sum_{n\ge1}S^{(n)}$ therefore converges absolutely to an anti-Hermitian, block-off-diagonal operator with $\|S\|_{\mathrm{HS}}<\tfrac14$. Substituting it into the BCH expansion~\eqref{eq:exact} gives a multiple series of nested commutators, and the estimates above bound the sum of the norms of all its terms. The series may thus be regrouped by order. By Eq.~\eqref{eq:ordern}, the terms of order $n$ sum to $H^{(n)}$, so that $e^SHe^{-S}=A+\sum_{n\ge1}H^{(n)}$ is block diagonal.

\emph{Identification of the blocks.} It remains to show that $e^S$ maps each $\Pi_\alpha$ onto $P_\alpha$. By homogeneity, the same construction for $A+tB$, $0\le t\le1$, gives $S(t)=\sum_{n\ge1}t^nS^{(n)}$, which is continuous in $t$, with $S(0)=0$ and $S(1)=S$. The projectors $\Pi'_\alpha(t)=e^{-S(t)}P_\alpha e^{S(t)}$ commute with $A+tB$, because $e^{S(t)}(A+tB)e^{-S(t)}$ is block diagonal. They therefore commute with the spectral projectors $\Pi_\alpha(t)$, which exist by Theorem~\ref{thm:exist} because $\|B\|\le\|B\|_{\mathrm{HS}}<g_\Pi/2$ and are polynomials in $A+tB$. For two commuting orthogonal projectors, the square of their difference is again an orthogonal projector, so that $\|\Pi'_\alpha(t)-\Pi_\alpha(t)\|$ is either $0$ or $1$. Both families are continuous in $t$ and coincide at $t=0$. Hence $\Pi'_\alpha(t)=\Pi_\alpha(t)$ for $0\le t\le1$, and at $t=1$ the unitary $e^S$ satisfies Eq.~\eqref{eq:gswtdef}.
\end{proof}

This sufficient condition involves the full perturbation because $B^\parallel$ also enters the higher-order generator. The constant $e^{-1/2}-\tfrac12$ need not be optimal. For two blocks, a convergence radius in the operator norm is given in Ref.~\cite{bravyi2011schrieffer}.

\paragraph*{Truncation error.}
Under the condition of Theorem~\ref{thm:conv}, the truncation error is bounded explicitly:
\begin{equation}\label{eq:trunc}
	\begin{aligned}
		\left\|H_{\mathrm{eff}}-A-\sum_{k=1}^{n}H^{(k)}\right\|_{\mathrm{HS}}
		&\le\frac{g_\Pi}{4}
		\left(\frac{\|B\|_{\mathrm{HS}}}{c g_\Pi}\right)^{n+1},\\
		\left\|S-\sum_{k=1}^{n}S^{(k)}\right\|_{\mathrm{HS}}
		&\le\frac14
		\left(\frac{\|B\|_{\mathrm{HS}}}{c g_\Pi}\right)^{n+1}.
	\end{aligned}
\end{equation}
These bounds follow by applying the absolute-sum bounds of Theorem~\ref{thm:conv} to $\lambda B$, with $1<\lambda<c g_\Pi/\|B\|_{\mathrm{HS}}$, and using homogeneity.

\paragraph*{Order bounds.}
Inserting Eq.~\eqref{eq:sylv} into the recursion~\eqref{eq:RnA} and using $\|[X,Y]\|_{\mathrm{HS}}\le2\|X\|_{\mathrm{HS}}\|Y\|_{\mathrm{HS}}$ gives, by induction, $\|S^{(n)}\|_{\mathrm{HS}}\le a_n(\|B\|_{\mathrm{HS}}/g_\Pi)^n$ and $\|H^{(n)}\|_{\mathrm{HS}}\le a_ng_\Pi(\|B\|_{\mathrm{HS}}/g_\Pi)^n$, with combinatorial constants $a_n$ independent of $A$, $B$, and $\Pi$.

In the operator norm, the Sylvester equation $A_\alpha G_{\alpha\beta}-G_{\alpha\beta}A_\beta=R_{\alpha\beta}$ for each pair of blocks obeys the Bhatia--Davis--McIntosh bound $\|G_{\alpha\beta}\|\le\pi\|R_{\alpha\beta}\|/(2g_\Pi)$~\cite{bhatia1983perturbation}, so that $\|\operatorname{ad}_A^{-1}R^\perp\|\le(\pi/2)|\Pi|(|\Pi|-1)\|R^\perp\|/g_\Pi$ for a partition into $|\Pi|$ blocks. Since $\|\mathcal P_\Pi X\|\le\|X\|$ and hence $\|\mathcal Q_\Pi X\|\le2\|X\|$, the same induction then gives
\begin{equation}\label{eq:orderbounds}
	\|S^{(n)}\|=O(x^n),\qquad \|H^{(n)}\|=O(g_\Pi x^n),\qquad x=\frac{\|B\|}{g_\Pi},
\end{equation}
the form quoted in the main text, with constants depending only on $n$ and the number of blocks $|\Pi|$.

\begin{theorem}[Symmetry and cancellation of odd orders]\label{thm:chiral}
Suppose a unitary $\Theta$ commutes with $A$ and every $P_\alpha$, and satisfies $\Theta B\Theta^\dagger=-B$. Then
\begin{equation}
	\Theta S^{(n)}\Theta^\dagger=(-1)^nS^{(n)},\qquad
	\Theta H^{(n)}\Theta^\dagger=(-1)^nH^{(n)}.
\end{equation}
On every block where $\Theta$ is scalar, $P_\alpha H^{(n)}P_\alpha=0$ for odd $n$. If $\Theta=\sum_\alpha e^{i\phi_\alpha}P_\alpha$ with real $\phi_\alpha$, then $B^\parallel=0$ and all odd effective orders vanish.
\end{theorem}

\begin{proof}
Conjugation by $\Theta$ leaves $A$ and every $P_\alpha$ invariant. It therefore commutes with $\mathcal P_\Pi$, with $\mathcal Q_\Pi$, and with the inverse of $\operatorname{ad}_A$ on the block-off-diagonal operators, and it maps the canonical recursion for $B$ onto that for $\Theta B\Theta^\dagger=-B$. By the uniqueness in Theorem~\ref{thm:rec}, $\Theta S^{(n)}\Theta^\dagger$ and $\Theta H^{(n)}\Theta^\dagger$ are the coefficients of order $n$ for $-B$, which by homogeneity are $(-1)^nS^{(n)}$ and $(-1)^nH^{(n)}$.

On a block where $\Theta$ is scalar, $P_\alpha\Theta H^{(n)}\Theta^\dagger P_\alpha=P_\alpha H^{(n)}P_\alpha$, so that $P_\alpha H^{(n)}P_\alpha=0$ for odd $n$. If $\Theta$ is scalar on every block, all odd orders therefore vanish, and $n=1$ gives $B^\parallel=H^{(1)}=0$.
\end{proof}

For the SWT, $(A,B)=(H_0,V)$, the grading operator $\Gamma$ of Sec.~\ref{sec:dual} is such a $\Theta$. For the dual SWT, $(A,B)=(V,H_0)$, it is not, because $\Gamma$ reverses the dominant operator $V$. The parity of the orders is then given by Theorem~\ref{thm:parity}.

Within the convergence domain, blocks where the symmetry is scalar have a fourth-order remainder after second-order truncation. On a block where $\Theta$ is not scalar, the odd orders need not vanish. The theorem does not require $\Theta^2=\openone$.

\subsection{Geodesic principle}\label{sec:geodesic}

The block-diagonal condition leaves the basis within each block arbitrary: $U\mapsto KU$, with a block-diagonal unitary $K$, preserves the block spectra and spans the gauge orbit $\mathcal O$ of Sec.~\ref{sec:gswtdef}. The canonical gauge condition $\mathcal P_\Pi S=0$ fixes this freedom perturbatively by selecting a representative on $\mathcal O$. Here we establish the variational meaning of this choice: under the conditions below, the gSWT is the unitary on $\mathcal O$ closest to the identity in geodesic distance.

We measure this distance on the unitary group $\mathrm U(N)$ with the bi-invariant Hilbert--Schmidt metric $\langle X,Y\rangle=\operatorname{Re}\operatorname{Tr}X^\dagger Y$ on anti-Hermitian generators:
\begin{equation}
	d_g(U)=\left(\sum_{j=1}^N\theta_j^2\right)^{1/2},
	\qquad
	U|j\rangle=e^{i\theta_j}|j\rangle,
	\quad \theta_j\in(-\pi,\pi].
\end{equation}
Here $\theta_j$ are the eigenphases of $U$. This distance is continuous on $\mathrm U(N)$. If $-1$ is absent from the spectrum of $U$, its principal logarithm $S=\log U$ is unique, $d_g(U)=\|S\|_{\mathrm{HS}}$, and the shortest geodesic is $U(t)=e^{tS}$, $0\le t\le1$, because for a bi-invariant metric the geodesics through the identity are the one-parameter subgroups~\cite{milnor1976curvatures}. As the Riemannian distance from the identity for a bi-invariant metric, $d_g$ satisfies $d_g(U^\dagger)=d_g(U)$ and the triangle inequality $d_g(UW)\le d_g(U)+d_g(W)$.

The theorem below is Theorem~1 of the main text. Its last statement refers to the principal angles $\vartheta_{\alpha,k}\in[0,\pi/2]$, $k=1,\ldots,d_\alpha$, between the bare block subspaces $\operatorname{ran}P_\alpha$ and the exact subspaces $\operatorname{ran}\Pi_\alpha$. Their cosines are the singular values of $P_\alpha\Pi_\alpha$ on $\operatorname{ran}\Pi_\alpha$. The principal-angle condition is
\begin{equation}\label{eq:angle}
	\sum_{\alpha,k}\sin^2(\vartheta_{\alpha,k}/2)<\tfrac18.
\end{equation}
The proof shows that it is implied by $\|B\|_{\mathrm{HS}}<(e^{-1/2}-\tfrac12)g_\Pi$.

\begin{theorem}[Geodesic principle]\label{thm:geodesic}
Let $x<\tfrac12$ and $U=e^{S}\in\mathcal{O}$ with $S=\log U$, $\|S\|<\pi$. (i)~$U$ is a stationary point of $d_g^2$ on $\mathcal{O}$ if and only if $\mathcal{P}_\Pi S=0$. (ii)~The stationary point is a strict local minimum of $d_g$ if $\max_j\theta_j-\min_j\theta_j<\pi$. (iii)~A stationary point with $d_g(U)<\pi/(2\sqrt2)$ is the unique global minimizer of $d_g$ on $\mathcal{O}$.
Such a $U$ exists whenever $\|B\|_{\mathrm{HS}}<(e^{-1/2}-\tfrac12)g_\Pi$, and more generally under a principal-angle condition.
\end{theorem}

\begin{proof}
Throughout, $X$ denotes an anti-Hermitian element of $\mathcal A_\Pi$. The curves $e^{tX}U$ remain on $\mathcal O$, and their velocities $XU$ span its tangent space at $U$. Let $F=\tfrac12d_g^2$ and $Z(t)=\log(e^{tX}U)$. The principal logarithm is smooth as long as $-1$ is not an eigenvalue of its argument, which holds near $t=0$ because $\|S\|<\pi$ and at every point of $\mathcal O$ with $d_g<\pi$.

\emph{Variations of $F$.} Differentiating $e^{Z(t)}=e^{tX}U$ gives
\begin{equation}\label{eq:dexp}
	X=\int_0^1e^{sZ}\dot Ze^{-sZ}\,ds.
\end{equation}
Taking the Hilbert--Schmidt inner product with $Z$, which commutes with $e^{sZ}$, yields $\langle Z,X\rangle=\langle Z,\dot Z\rangle$, so that
\begin{equation}\label{eq:firstvar}
	\frac{d}{dt}F(e^{tX}U)=\langle Z,X\rangle,
	\qquad
	\frac{d^2}{dt^2}F(e^{tX}U)=\langle\dot Z,X\rangle.
\end{equation}
In an eigenbasis of $Z$, with $Z|j\rangle=i\theta_j|j\rangle$ and $\mu_{jk}=\theta_j-\theta_k$, Eq.~\eqref{eq:dexp} reads $X_{jk}=\dot Z_{jk}(e^{i\mu_{jk}}-1)/(i\mu_{jk})$, and the second variation becomes
\begin{equation}\label{eq:hess}
	\frac{d^2}{dt^2}F(e^{tX}U)
	=\sum_{j,k}
	\frac{\mu_{jk}}2\cot\frac{\mu_{jk}}2\,|X_{jk}|^2.
\end{equation}
The coefficient equals one at $\mu_{jk}=0$ and is positive for $|\mu_{jk}|<\pi$.

\emph{Proof of (i).} At $t=0$, $Z=S$ and the first variation is $\langle S,X\rangle=\langle\mathcal P_\Pi S,X\rangle$, because $X$ is block diagonal. It vanishes for every $X$ if and only if $\mathcal P_\Pi S=0$, since $\mathcal P_\Pi S$ is itself an admissible $X$.

\emph{Proof of (ii).} If $\max_j\theta_j-\min_j\theta_j<\pi$, every coefficient in Eq.~\eqref{eq:hess} is positive at $t=0$. At a stationary point the second variation of $F$ is then positive for every $X\ne0$, so the point is a strict local minimum.

\emph{Proof of (iii).} Let $U$ be stationary with $d_g(U)<\rho$, where $\rho=\pi/(2\sqrt2)$. Since $\mathcal O$ is compact and $d_g$ is continuous, a global minimizer $U_*$ exists, and $d_g(U_*)\le d_g(U)<\rho$. Because $F$ is smooth at $U_*$, the minimizer is stationary. Write $U_*=e^XU$ with $X=\log(U_*U^\dagger)\in\mathcal A_\Pi$. The triangle inequality gives $\|X\|_{\mathrm{HS}}=d_g(U_*U^\dagger)\le d_g(U_*)+d_g(U)<2\rho$. The curve $\gamma(t)=e^{tX}U$, $0\le t\le1$, joins $U$ to $U_*$ on $\mathcal O$, and the triangle inequalities from its two endpoints give
\begin{equation*}
	d_g(\gamma(t))
	\le\frac{d_g(U)+d_g(U_*)+\|X\|_{\mathrm{HS}}}{2}
	<2\rho=\frac{\pi}{\sqrt2}.
\end{equation*}
The eigenphases of $\gamma(t)$ therefore obey $|\theta_j-\theta_k|\le\sqrt2\,d_g(\gamma(t))<\pi$, and by Eq.~\eqref{eq:hess} the function $F(\gamma(t))$ is strictly convex unless $X=0$. Its derivative vanishes at $t=0$ and at $t=1$, because both endpoints are stationary. A strictly convex function cannot have two stationary points, so $X=0$ and $U_*=U$.

\emph{Existence under the angle condition.} Let $d_c(U)=\|U-\openone\|_{\mathrm{HS}}$ be the chordal distance. The inequalities $2|\sin(\theta/2)|\le|\theta|\le\pi|\sin(\theta/2)|$ for $|\theta|\le\pi$ give
\begin{equation}\label{eq:dcdg}
	d_c(U)\le d_g(U)\le\frac\pi2d_c(U).
\end{equation}
For $U\in\mathcal O$, $P_\alpha UP_\alpha=U\Pi_\alpha P_\alpha$, so that $d_c^2(U)=2N-2\operatorname{Re}\operatorname{Tr}(Y^\dagger U)$ with $Y=\sum_\alpha P_\alpha\Pi_\alpha$. The real trace is at most the sum of the singular values of $Y$~\cite{fan1955some}, which are $\cos\vartheta_{\alpha,k}$, with equality for a unitary polar factor of $Y$. Since $Y$ maps each $\operatorname{ran}\Pi_\alpha$ into $\operatorname{ran}P_\alpha$, this factor can be chosen to map $\operatorname{ran}\Pi_\alpha$ onto $\operatorname{ran}P_\alpha$ for every $\alpha$, i.e., in $\mathcal O$. Consequently,
\begin{equation}\label{eq:mindc}
	\min_{U\in\mathcal O}d_c^2(U)
	=2\sum_{\alpha,k}(1-\cos\vartheta_{\alpha,k})
	=4\sum_{\alpha,k}\sin^2(\vartheta_{\alpha,k}/2).
\end{equation}
Under the angle condition, Eqs.~\eqref{eq:dcdg} and~\eqref{eq:mindc} give a point of $\mathcal O$ with $d_g<\rho$. The global minimizer $U_*$ then also satisfies $d_g(U_*)<\rho$. It is stationary, hence canonical by (i), and unique by (iii).

\emph{Perturbative condition.} If $\|B\|_{\mathrm{HS}}<(e^{-1/2}-\tfrac12)g_\Pi$, Theorem~\ref{thm:conv} gives a canonical $U_g=e^S\in\mathcal O$ with $\|S\|_{\mathrm{HS}}<\tfrac14$. Equations~\eqref{eq:mindc} and~\eqref{eq:dcdg} then yield
\begin{equation*}
	\sum_{\alpha,k}\sin^2(\vartheta_{\alpha,k}/2)
	\le\frac14d_c^2(U_g)
	\le\frac14\|S\|_{\mathrm{HS}}^2
	<\frac1{64}<\frac18,
\end{equation*}
which is the angle condition.
\end{proof}

Under the canonical gauge condition, the geodesic $e^{tS}$, $0\le t\le1$, from the identity to $U$ therefore meets the gauge orbit orthogonally: its endpoint tangent $SU$ is orthogonal to every intra-block rotation $XU$. Under the global-minimum conditions, this selects the dressed basis with the smallest root-sum-square rotation angle from the bare basis. The geodesic principle is thus a variational criterion for fixing the gSWT gauge while retaining the full effective dynamics within each block.

\begin{corollary}[Uniqueness of the gSWT]\label{cor:unique}
Let $x<\tfrac12$. (i) The canonical series $S=\sum_{n\ge1}S^{(n)}$ of Theorem~\ref{thm:rec} is unique order by order. (ii) At most one $U\in\mathcal O$ with $d_g(U)<\pi/(2\sqrt2)$ is canonical. Under the condition of Theorem~\ref{thm:conv} this point exists, equals the sum $U_g=e^S$ of the canonical series, and satisfies $\|S\|_{\mathrm{HS}}<\tfrac14$.
\end{corollary}

\begin{proof}
Part (i) is the uniqueness statement of Theorem~\ref{thm:rec}. For (ii), a canonical $U$ with $d_g(U)<\pi/(2\sqrt2)$ is stationary by Theorem~\ref{thm:geodesic}(i) and is the unique global minimizer of $d_g$ on $\mathcal O$ by Theorem~\ref{thm:geodesic}(iii). Two such points would both be this minimizer and therefore coincide. Theorem~\ref{thm:conv} provides $U_g$ with $d_g(U_g)=\|S\|_{\mathrm{HS}}<\tfrac14<\pi/(2\sqrt2)$.
\end{proof}

The chordal criterion measures the total squared displacement of an orthonormal basis, $d_c^2(U)=\sum_j4\sin^2(\theta_j/2)$, whereas the geodesic criterion measures $\sum_j\theta_j^2$. For invertible $Y=\sum_\alpha P_\alpha\Pi_\alpha$, the unique chordal minimizer is $U_c=Y(Y^\dagger Y)^{-1/2}$, the least-action transformation of Cederbaum \emph{et al.}~\cite{cederbaum1989block}. It obeys $\mathcal P_\Pi U_c>0$ and, near the identity, the stationarity condition $\mathcal P_\Pi\sinh(\log U_c)=0$.

For two blocks, $U_c$ coincides with the canonical transformation when $\|S\|<\pi/2$: odd powers of $S$ remain block off-diagonal and $\mathcal P_\Pi e^S=\cosh S>0$. This is the Davis--Kahan direct rotation used to define the exact two-block SWT~\cite{davis1970rotation,bravyi2011schrieffer}.

For three or more blocks, closed paths through three distinct blocks can give $\mathcal P_\Pi(S^{(1)})^3\ne0$, so the chordal generator generically acquires an intra-block component beginning at third order~\cite{mankodi2024perturbative}. The two effective Hamiltonians remain related by a block-diagonal unitary and have identical block spectra. Their matrix elements may differ within multidimensional blocks, while one-dimensional effective blocks are unchanged. Thus the canonical gauge minimizes the geodesic rotation under the conditions of Theorem~\ref{thm:geodesic}, even when it differs from the chordal least-action prescription.

\subsection{Exact solution of the generator}\label{sec:nested}

At each perturbative order, the gSWT removes inter-block transitions by solving $[A,G]=R^{\perp}$, with $R^{\perp}=\mathcal Q_\Pi R^{(n)}$ and $G=S^{(n)}$. Physically, this requires dividing each transition amplitude by its reference energy difference. The sector gap $g_\Pi$ is a lower bound on the magnitude of every inter-block denominator, but a given source generally involves only a subset of the possible transition energies.

\begin{definition}[Active gap set]\label{def:active}
For a block-off-diagonal source $R^\perp=\mathcal Q_\Pi R$, its active gap set is
\begin{equation}
	\mathcal G_A(R^\perp)
	=\{|E_a-E_b|:Q_aR^\perp Q_b\ne0\}
	=\{g_1<\cdots<g_m\}.
\end{equation}
\end{definition}

This finite set contains only positive gaps, each at least $g_\Pi$, and is empty precisely when $R^\perp=0$.

To classify the transitions induced by the perturbation, we group its matrix elements by their transition energy $\mu$, which we also call their frequency:
\begin{equation}
	B^\perp=\sum_\mu B_\mu,\qquad
	B_\mu=\sum_{E_a-E_b=\mu}Q_aB^\perp Q_b,
	\qquad [A,B_\mu]=\mu B_\mu.
\end{equation}
Here $\mu$ runs over the nonzero signed active gaps. Hermiticity gives $B_{-\mu}=B_\mu^\dagger$, pairing each energy-raising process with its reverse. Since $\operatorname{ad}_A^k(Q_aXQ_b)=(E_a-E_b)^kQ_aXQ_b$, repeated commutators only multiply these components by powers of their transition energies:
\begin{equation}
	\operatorname{ad}_A^k B^\perp=\sum_\mu\mu^kB_\mu.
\end{equation}

The same statement holds for each block-off-diagonal source $\mathcal Q_\Pi R^{(n)}$, where higher-order virtual processes can activate gaps absent from $B^\perp$ itself. The inverse of $\operatorname{ad}_A$ acting on such a source is therefore given by a polynomial in $\operatorname{ad}_A$ that reproduces $1/\omega$ at the transition energies $\omega$ present in the source. This gives an exact finite expression in nested commutators, without requiring eigenvectors of $A$ once the active gaps are known. The precise statement is Theorem~2 of the main text.

\begin{theorem}[Nested-commutator solution]\label{thm:nested}
For the gSWT problem with $(A,B,\Pi)$, let $R^{\perp}$ be block off-diagonal, with active gap set $\mathcal{G}_A(R^{\perp})=\{g_1,\dots,g_m\}$. The equation $\operatorname{ad}_AG=R^{\perp}$ has the unique block-off-diagonal solution $G=p(\operatorname{ad}_A)\,R^{\perp}$ with $p(\omega)=\omega^{-1}\bigl[1-\prod_{i=1}^{m}(1-\omega^2/g_i^2)\bigr]$. Explicitly,
\begin{equation}\label{eq:ek}
	G=\sum_{k=1}^{m}(-1)^{k+1}e_k\bigl(g_1^{-2},\dots,g_m^{-2}\bigr)\,\operatorname{ad}_A^{2k-1}R^{\perp},
\end{equation}
with $e_k$ the elementary symmetric polynomials, and $p(\omega)$ is the unique odd polynomial of degree $\le2m-1$ with $p(\pm g_i)=\pm1/g_i$.
\end{theorem}

\begin{proof}
Decompose the source into components of definite transition energy,
\begin{equation*}
	R^{\perp}=\sum_{\mu}R_\mu,\qquad
	R_\mu=\sum_{E_a-E_b=\mu}Q_aR^{\perp}Q_b,
	\qquad [A,R_\mu]=\mu R_\mu.
\end{equation*}
The nonzero components have $\mu\in\{\pm g_i\}$ and remain block off-diagonal. The polynomial $p$ is odd: $\omega p(\omega)$ is an even polynomial that vanishes at $\omega=0$. At every active frequency it satisfies $\mu p(\mu)=1$, and therefore
\begin{equation*}
	[A,p(\operatorname{ad}_A)R^{\perp}]
	=\sum_\mu\mu p(\mu)R_\mu=R^{\perp}.
\end{equation*}
The sector gap $g_\Pi>0$ makes $\operatorname{ad}_A$ invertible on the block-off-diagonal operators, proving uniqueness there.

Expanding the product gives Eq.~\eqref{eq:ek}. Finally, two interpolants of degree at most $2m-1$ would have a difference vanishing at all $2m$ distinct nodes, and hence must coincide. This is Lagrange interpolation applied to the transition-frequency components, equivalently the diagonalizable case of Sylvester's formula~\cite{higham2008functions}, and degeneracies of $A$ do not affect the argument.
\end{proof}

In Eq.~\eqref{eq:ek}, $e_k$ is the $k$th elementary symmetric polynomial, the sum of all products of $k$ distinct arguments. For $m=3$, for example, $e_1=g_1^{-2}+g_2^{-2}+g_3^{-2}$, $e_2=(g_1g_2)^{-2}+(g_1g_3)^{-2}+(g_2g_3)^{-2}$, and $e_3=(g_1g_2g_3)^{-2}$.

For the Hermitian sources of the gSWT recursion, $R_{-g_i}=R_{+g_i}^\dagger$, so $G$ is anti-Hermitian. Moreover, the depth $2m-1$ is minimal among all scalar-coefficient polynomials in $\operatorname{ad}_A$ acting on this source. Indeed, every such polynomial $q$ must satisfy $q(\pm g_i)=\pm1/g_i$. If its degree were smaller, the nonzero polynomial $\omega q(\omega)-1$ would have $2m$ roots but degree less than $2m$.

When only one or two gaps are active, or when $[A^2,R^{\perp}]=0$, the solution collapses to short closed forms.

\begin{corollary}[Collapse formulas]\label{cor:collapse}
If $\mathcal G_A(B^{\perp})=\{g\}$, its raising and lowering components are
\begin{equation}
	B_\pm=\frac12
	\left(B^{\perp}\pm\frac{[A,B^{\perp}]}{g}\right),
	\qquad
	[A,B_\pm]=\pm gB_\pm,
	\qquad B_-=B_+^\dagger.
\end{equation}
The first-order generator and second-order effective Hamiltonian reduce to
\begin{equation}\label{eq:single}
	\begin{aligned}
		S^{(1)}
		&=\frac{[A,B^{\perp}]}{g^2}
		=\frac{B_+-B_-}{g},\\
		H^{(2)}
		&=\frac{[B_+,B_-]}{g},\\
		H_{\mathrm{eff}}
		&=A+B^{\parallel}
		+\frac{[B_+,B_-]}{g}+O(B^3).
	\end{aligned}
\end{equation}
No explicit projector is needed in $H^{(2)}$ for any allowed partition. For two active gaps,
\begin{equation}
	S^{(1)}
	=(g_1^{-2}+g_2^{-2})[A,B^{\perp}]
	-g_1^{-2}g_2^{-2}\operatorname{ad}_A^3B^{\perp}.
\end{equation}
More generally, if $[A^2,R^{\perp}]=0$, the inverse admits an operator-valued coefficient:
\begin{equation}\label{eq:opval}
	G=(4A^2)^+[A,R^{\perp}]
	=[A,R^{\perp}](4A^2)^+,
\end{equation}
where $(\cdot)^+$ is the Moore--Penrose pseudoinverse.
\end{corollary}

\begin{proof}
For one gap, the frequency components of $B^{\perp}$ satisfy $B^{\perp}=B_{+g}+B_{-g}$ and $[A,B^{\perp}]=g(B_{+g}-B_{-g})$, so that $B_\pm=B_{\pm g}$, with $B_-=B_+^\dagger$ by Hermiticity. Hence $\operatorname{ad}_A^2B^{\perp}=g^2B^{\perp}$, and $p(\omega)=\omega/g^2$ gives $S^{(1)}$. Moreover, $[S^{(1)},B^{\perp}]=2[B_+,B_-]/g$ and $[A,[B_+,B_-]]=0$. Since $g_\Pi>0$, an operator commuting with $A$ is block diagonal, proving Eq.~\eqref{eq:single}. The two-gap expression follows by expanding $p$.

If $[A^2,R^{\perp}]=0$, an active transition obeys $E_a^2=E_b^2$ and $E_a\ne E_b$, hence $E_b=-E_a\ne0$. Therefore $\operatorname{ad}_A^2R^{\perp}=4A^2R^{\perp}$, and Eq.~\eqref{eq:opval} divides each component by $E_a-E_b$. The coefficient commutes with $[A,R^{\perp}]$, so it may be placed on either side.
\end{proof}

Equation~\eqref{eq:opval} uses an operator-valued coefficient and is therefore consistent with the minimum depth established for scalar coefficients.

An all-order resummation is possible when the operator algebra closes. If $H_0$ is, up to a conserved operator, proportional to the grading operator, the order-by-order solution sums to an arctangent for both the SWT and the dual SWT.

\begin{proposition}[Exact arctangent rotation]\label{prop:arctan}
Let $H=H_0+V$ be graded by $\Gamma$ as in Eq.~\eqref{eq:dual-setting-sm}, with
\begin{equation}\label{eq:arctan-H}
	H_0=\Xi+\frac\Delta2\Gamma,
\end{equation}
where $\Delta\ne0$ is real and the Hermitian operator $\Xi$ commutes with $\Gamma$ and $V$. We write $|V|=(V^2)^{1/2}$.

(i) \emph{SWT.} For the dominant operator $H_0$, with the two sectors $\Gamma=\pm1$ as blocks,
\begin{equation}\label{eq:arctan-S}
	S=\frac{\arctan(2|V|/\Delta)}{2|V|}\,\Gamma V,
	\qquad
	e^SHe^{-S}=\Xi+\frac{\operatorname{sgn}(\Delta)}2\,\Gamma\sqrt{\Delta^2+4V^2}.
\end{equation}

(ii) \emph{Dual SWT.} If $V$ is invertible, for the dominant operator $V$, with its positive and negative spectral subspaces as blocks,
\begin{equation}\label{eq:arctan-Sd}
	\tilde S=-\frac{\arctan[\Delta/(2|V|)]}{2|V|}\,\Gamma V,
	\qquad
	e^{\tilde S}He^{-\tilde S}=\Xi+V\sqrt{1+\frac{\Delta^2}{4V^2}}.
\end{equation}

Both generators are anti-Hermitian and block off-diagonal, with $\|S\|<\pi/4$ and $\|\tilde S\|<\pi/4$, and both transformed Hamiltonians are block diagonal. The rotation angles $\theta=\arctan(2|V|/\Delta)$ and $\tilde\theta=\arctan[\Delta/(2|V|)]$ are complementary, $\theta+\tilde\theta=\operatorname{sgn}(\Delta)\pi/2$.
\end{proposition}

\begin{proof}
The operator $J=\Gamma V$ is anti-Hermitian, anticommutes with $\Gamma$ and with $V$, commutes with $\Xi$ and $|V|$, and satisfies $J^2=-V^2$. Let $S_\varphi=\varphi J/|V|$, where $\varphi$ is a real function of $|V|$. On $\ker V$, $J=0$ and $S_\varphi=0$. The operator $S_\varphi$ is anti-Hermitian and commutes with $\Xi$, and $(J/|V|)^2=-\openone$ on the range of $V$ gives $e^{2S_\varphi}=\cos2\varphi+(J/|V|)\sin2\varphi$.

Only $M=\Delta\Gamma/2+V$ is therefore rotated. Since $J$ anticommutes with $M$, $e^{S_\varphi}Me^{-S_\varphi}=e^{2S_\varphi}M$. Using $J\Gamma=-V$ and $JV=\Gamma V^2$,
\begin{equation}
	e^{S_\varphi}Me^{-S_\varphi}
	=\Gamma\left(\frac\Delta2\cos2\varphi+|V|\sin2\varphi\right)
	+\frac{V}{|V|}\left(|V|\cos2\varphi-\frac\Delta2\sin2\varphi\right).
\end{equation}

The choice $2\varphi=\arctan(2|V|/\Delta)$ removes the second term and gives Eq.~\eqref{eq:arctan-S}. The choice $2\varphi=-\arctan[\Delta/(2|V|)]$ removes the first term and gives Eq.~\eqref{eq:arctan-Sd}. The two results commute with $\Gamma$ and with $V/|V|$, respectively, and are therefore block diagonal. Finally, $S_\varphi$ anticommutes with $\Gamma$ and with $V/|V|$, so it is block off-diagonal for either partition, and $\|S_\varphi\|=\max|\varphi|<\pi/4$.
\end{proof}

The operators $\Xi$ and $V^2$ commute with $H$. On each of their joint eigenspaces with $V\ne0$, the dominant operator has two eigenvalues, the stated blocks are its eigenspaces, and the single active gap is $|\Delta|$ in (i) and $2|V|$ in (ii). There $S$ and $\tilde S$ are the canonical generators of the SWT and of the dual SWT to all orders. Their first-order terms, $S^{(1)}=\Gamma V/\Delta$ and $\tilde S^{(1)}=-\Delta\Gamma V/(4V^2)$, are those of Eqs.~\eqref{eq:single} and~\eqref{eq:opval}. The expansions of $S$ in powers of $V$ and of $\tilde S$ in powers of $\Delta$ converge for $2|V|<|\Delta|$ and for $|\Delta|<2|V|$, respectively, whereas the closed forms hold for all real parameters. The dual result has the form of Eq.~\eqref{eq:dimer}, with the obstruction $\Xi$.

The simplest instance is the two-level system $H=(\Delta/2)\sigma_z+b\sigma_x$, with $\Gamma=\sigma_z$, $\Xi=0$, and $V=b\sigma_x$, for which $S=(i/2)\arctan(2b/\Delta)\,\sigma_y$ and $e^SHe^{-S}=\tfrac12\operatorname{sgn}(\Delta)\sqrt{\Delta^2+4b^2}\,\sigma_z$. Further instances are every excitation sector of the Jaynes--Cummings model (Sec.~\ref{sec:m-jc}), where $\Gamma=\sigma_z$ and $V^2=g^2N$, and the Dirac Hamiltonian (Sec.~\ref{sec:m-dirac}), where $\Gamma=\beta$, $\Delta=2m$, and $V=\bm\alpha\cdot\bm p$, so that $S$ and $\tilde S$ are the Foldy--Wouthuysen and Cini--Touschek rotations.

For the gSWT problem $(A,B,\Pi)$, the input of Theorem~\ref{thm:nested} is the active gap set $\mathcal G_A(R^{\perp})$ at each perturbative order. The following four remarks show how to determine this spectral input or a suitable superset.

\begin{remark}[Block partition]\label{rem:partition}
Theorem~\ref{thm:nested} applies to a source that is block off-diagonal with respect to the chosen partition. For the finest partition, $p(0)=0$ implies $p(\operatorname{ad}_A)R^{(n)}=p(\operatorname{ad}_A)\mathcal Q_A R^{(n)}$, so that no separate projection of the source is needed (Corollary~\ref{cor:finest}). For a coarse-grained partition, intra-block terms can carry nonzero transition energies and must first be removed by $\mathcal Q_\Pi$. Merging eigenspaces removes their mutual transitions from the block-off-diagonal source, but a gap value disappears only if no remaining inter-block transition carries it. All denominators involve actual eigenvalue differences of $A$.
\end{remark}

\begin{remark}[Selection rules]\label{rem:selection}
Selection rules often determine the active gaps directly. If $A=f(N)$ and a component $R_s$ changes the quantum number by $s$, $[N,R_s]=sR_s$, then
\begin{equation}
	[A,R_s]=R_s[f(N+s)-f(N)].
\end{equation}
The gaps are the nonzero absolute values of these energy changes on states connected across blocks. For $A=\omega N$, they are $|s\omega|$, independently of Hilbert-space dimension.
\end{remark}

\begin{remark}[Commutator closure]\label{rem:krylov}
For a nonzero Hermitian source, form $C_0=R^{\perp}$ and $C_{k+1}=[A,[A,C_k]]$ until the first linear relation. Each double commutator multiplies a transition by its squared energy, so $C_k=\sum_i g_i^{2k}(R_{+g_i}+R_{-g_i})$. Independence of these components gives
\begin{equation}\label{eq:krylovrel}
	\begin{aligned}
		C_m+\sum_{k=0}^{m-1}c_kC_k&=0,\\
		\mu(z)=z^m+\sum_{k=0}^{m-1}c_kz^k
		&=\prod_{i=1}^{m}(z-g_i^2).
	\end{aligned}
\end{equation}
Thus closure occurs after exactly $m$ steps, and the positive square roots of the roots of $\mu$ are the active gaps. No eigenvectors are required: this is the Krylov-subspace construction for the superoperator $\operatorname{ad}_A^2$~\cite{saad2003iterative}.

The same relation gives the generator without extracting these roots. With $c_m=1$ and $c_0=(-1)^m\prod_i g_i^2\ne0$,
\begin{equation}\label{eq:krylovG}
	G=-\frac1{c_0}\sum_{k=1}^{m}c_k[A,C_{k-1}].
\end{equation}
Indeed, the commutator of $A$ with this expression is $-c_0^{-1}\sum_{k=1}^m c_kC_k=C_0$. This is Eq.~\eqref{eq:ek}, since $p(\omega)=[1-\mu(\omega^2)/\mu(0)]/\omega$. More generally, any polynomial satisfying $\nu(\operatorname{ad}_A^2)R^{\perp}=0$ and $\nu(0)\ne0$ can replace $\mu$.

For $A=\chi J_z^2$, $R^{\perp}=\Omega J_x$, integer $j\ge1$, and nonzero $\chi,\Omega$, the finest partition gives $\mu(z)=\prod_{l=1}^{j}[z-(2l-1)^2\chi^2]$. For the transverse-field Ising model of Sec.~\ref{sec:m-tfim} on a ring, instead $C_1=16J^2C_0$, giving the single active gap $4J$.

The relation~\eqref{eq:krylovrel} also characterizes a variational truncation. For $G_\ell=\sum_{k=1}^{\ell}\alpha_k\operatorname{ad}_A^{2k-1}R^{\perp}$, the residual is minimized by
\begin{equation}
	\min_{\alpha_k}\|[A,G_\ell]-R^{\perp}\|_{\mathrm{HS}}
	=
	\min_{\alpha_k}
	\left\|\sum_{k=1}^{\ell}\alpha_kC_k-C_0\right\|_{\mathrm{HS}}.
\end{equation}
It is nonzero for $\ell<m$ and vanishes at $\ell=m$, where $\alpha_k=-c_k/c_0$. Exactness is therefore controlled by the number of active gaps. A series of nested commutators of the same form is used to build approximate adiabatic gauge potentials~\cite{claeys2019floquet}, on which variational constructions of the SWT generator build~\cite{wurtz2020variational}. Here the active gap set fixes the coefficients $\alpha_k$ in closed form, and the series terminates at $\ell=m$.
\end{remark}

\begin{remark}[Supersets]\label{rem:supersets}
Supersets preserve exactness. Any finite positive gap set containing $\mathcal G_A(R^{\perp})$ can be inserted into Eq.~\eqref{eq:ek}. Extra nodes constrain only absent components: they increase the formal commutator depth, but simplification gives the same operator,
\begin{equation}
	G=\sum_{\substack{a,b\\
	Q_a,Q_b\text{ in different blocks}}}
	\frac{Q_aR^{\perp}Q_b}{E_a-E_b}.
\end{equation}
Thus the absolute values of all inter-block eigenvalue differences of $A$ provide a common admissible superset at every order. Once the source is known, these gaps are the only spectral input needed for the coefficients.

At higher orders, virtual processes can activate gaps absent from $B^{\perp}$. An order-$n$ source connects reference eigenspaces only through paths containing at most $n$ perturbation matrix elements: projection and inverse commutation preserve matrix-element support, while products concatenate paths. Cancellations may remove such processes. In particular, for the finest partition and $\mathcal G_A(B^{\perp})=\{g\}$, every step carries frequency $0$ or $\pm g$, giving
\begin{equation}
	\mathcal G_A(\mathcal Q_A R^{(n)})
	\subseteq\{g,2g,\ldots,ng\}.
\end{equation}
This set is a sufficient input, and not every listed gap need occur. The low-order recursion gives the sharper supersets $\{g\}$, $\{g,2g\}$, and $\{g,2g,3g\}$ at orders two, three, and four, respectively.
\end{remark}

The four guidelines of the main text (End Matter) correspond to the statements of this subsection: (i)~the block partition fixes which transitions are inter-block and hence which eigenvalue differences of $A$ can be active (Definition~\ref{def:active} and Remark~\ref{rem:partition}), (ii)~selection rules for $A=f(N)$ give the gaps as $|f(n+s)-f(n)|$ (Remark~\ref{rem:selection}), (iii)~commutator closure determines the active gaps and the generator without eigenvectors (Remark~\ref{rem:krylov}), and (iv)~any finite superset of the active gap set preserves exactness (Remark~\ref{rem:supersets}).

\subsection{The projector in \texorpdfstring{$H_{\mathrm{eff}}$}{H\_eff}}\label{sec:projector}

This subsection evaluates the block-diagonal projector $\mathcal P_\Pi$ in $H^{(n)}=\mathcal P_\Pi R^{(n)}$, which retains the processes whose initial and final states belong to the same reference block. We identify when the projection is automatic or reduces to selecting whole commutator terms, and otherwise evaluate it using transition frequencies or block labels.

Since $\mathcal P_\Pi$ retains exactly the components $P_\alpha YP_\alpha$ of an operator $Y=\sum_{\alpha,\beta}P_\alpha YP_\beta$, $\mathcal P_\Pi Y=Y$ if and only if $[P_\alpha,Y]=0$ for every $\alpha$, and $\mathcal P_\Pi Y=0$ if and only if $P_\alpha YP_\alpha=0$ for every $\alpha$. An expression decomposed into block-diagonal and block-off-diagonal terms can thus be projected by retaining the former terms in full.

This termwise criterion is sufficient but not necessary: cancellations can make a sum block diagonal even when its individual terms contain both types of components. Cases~(i)--(iii) below give structural simplifications, whereas case~(iv) is the generic situation, in which the projector is needed.

\paragraph*{(i) Two blocks and disjoint pairs.}
For two blocks, define $\Gamma_\Pi=P_1-P_2$. Then
\begin{equation}
	\mathcal P_\Pi Y
	=\frac12(Y+\Gamma_\Pi Y\Gamma_\Pi).
\end{equation}
The generators $S^{(k)}$ and $B^{\perp}$ are odd under conjugation by $\Gamma_\Pi$, whereas $A$, $B^{\parallel}$, and $H^{(k)}$ are even. After resolving $B=B^{\parallel}+B^{\perp}$, every nested commutator in the recursion has definite parity: conjugation preserves commutators, so the factor parities multiply. Projection therefore retains terms containing an even number of block-off-diagonal factors. After this selection, no explicit projector is needed at any order. In particular,
\begin{equation}\label{eq:Stwoblock}
	\begin{aligned}
		H^{(2)}
		&=\frac12[S^{(1)},B^{\perp}],\\
		H^{(3)}
		&=\frac12[S^{(2)},B^{\perp}],\\
		H^{(4)}
		&=\frac12[S^{(3)},B^{\perp}]
		-\frac1{12}[S^{(1)},[S^{(1)},H^{(2)}]],
	\end{aligned}
\end{equation}
in agreement with the two-block SWT expansion~\cite{bravyi2011schrieffer}. The same argument applies separately to disjoint pairs of coupled blocks, with isolated blocks requiring no rotation: the recursion cannot generate couplings between different pairs.

\paragraph*{(ii) A single active gap.}
If $\mathcal G_A(B^{\perp})=\{g\}$, write $B^{\perp}=B_++B_-$ with $[A,B_\pm]=\pm gB_\pm$. Corollary~\ref{cor:collapse} gives
\begin{equation}
	\frac12[S^{(1)},B^{\perp}]
	=\frac{[B_+,B_-]}{g}=H^{(2)}.
\end{equation}
The commutator has zero transition energy and is therefore block diagonal for any allowed partition. A single active gap thus removes the second-order projector, but does not guarantee the same simplification at higher orders.

For the finest partition, $B^{\parallel}$ has zero frequency and
\begin{equation}
	S^{(2)}
	=\operatorname{ad}_A^{-1}[S^{(1)},B^{\parallel}].
\end{equation}
Consequently, $[S^{(2)},B^{\perp}]$ can contain frequencies $0$ and $\pm2g$ and may require projection. In contrast, $[S^{(1)},[S^{(1)},B^{\perp}]]$ has only frequencies $\pm g$ and is removed in full. For a coarse-grained partition, $B^{\parallel}$ can itself carry nonzero frequencies, so this higher-order classification need not apply.

\paragraph*{(iii) Closure at all orders.}
For the finest partition, a further sufficient condition is that the complex Lie algebra generated by $B^{\parallel}$ and $B_{\pm g}$ contains only transition frequencies $0,\pm g$. Its zero-frequency and nonzero-frequency parts form an even--odd grading: commutators of two odd operators have frequency zero because their $\pm2g$ components vanish by closure. The algebra is invariant under $\operatorname{ad}_A$, since its generators have definite frequencies and commutation with $A$ is a derivation. Frequency projection and inversion on nonzero frequencies also remain within it. The recursion therefore preserves this grading, and the projector selects the even terms at every order, as in case~(i).

For quadratic fermions with a flat reference quasiparticle spectrum, bilinears conserve quasiparticle number or create or annihilate a pair, and their commutators remain bilinear. This gives the required closure in both constructions for the transverse-field Ising model of Sec.~\ref{sec:m-tfim}, within each fixed fermion-parity sector with its corresponding boundary conditions.

The dual SWT of the collective spin model provides a contrasting example (Sec.~\ref{sec:m-oat}): there $\mathcal Q_\Pi\tfrac12[\tilde S^{(2)},B^{\perp}]$ contains the frequencies $\pm4\Omega$ for $j\ge2$, because $[[B_+,B^{\parallel}],B_+]=\chi^3(J_y+iJ_z)^4/4\ne0$, and closure fails. For $j=1$ or $\tfrac32$, the allowed transitions instead form disjoint pairs.

\paragraph*{(iv) Multiblock virtual processes.}
In a generic multiblock problem, the projector is already necessary at second order: two successive inter-block transitions can end in a third block. In an eigenbasis of $A$, let $\alpha_a$ label the block containing $|a\rangle$. Since $S^{(1)}_{ab}=B_{ab}/(E_a-E_b)$ across blocks,
\begin{equation}\label{eq:Spath}
	[S^{(1)},B^{\perp}]_{ac}
	=
	\sum_{\substack{b\\
	\alpha_b\notin\{\alpha_a,\alpha_c\}}}
	B_{ab}B_{bc}
	\left(
	\frac1{E_a-E_b}+\frac1{E_c-E_b}
	\right).
\end{equation}
If $a,c$ belong to the same block, half this expression is retained in $H^{(2)}$. Otherwise it is removed. For a nonzero individual path amplitude, the coefficient vanishes precisely when $2E_b=E_a+E_c$. Different paths can also cancel after summation.

For example, consider the finest partition of $A=\operatorname{diag}(0,1,3)$ and $B=\varepsilon(|1\rangle\langle2|+|2\rangle\langle3|+\mathrm{H.c.})$, with real $\varepsilon$. Direct evaluation gives
\begin{equation}\label{eq:threelevel}
	\begin{aligned}
		S^{(1)}
		&=\varepsilon
		\begin{pmatrix}
			0&-1&0\\
			1&0&-1/2\\
			0&1/2&0
		\end{pmatrix},\\
		\frac12[S^{(1)},B]
		&=\varepsilon^2
		\begin{pmatrix}
			-1&0&-1/4\\
			0&1/2&0\\
			-1/4&0&1/2
		\end{pmatrix}.
	\end{aligned}
\end{equation}
The $(1,3)$ entry is removed from $H^{(2)}$ and gives $S^{(2)}_{13}=\varepsilon^2/12$. Diagonalizing the unprojected second-order expression instead produces additional shifts $\mp\varepsilon^4/48$ in the levels continued from $0$ and $3$, relative to the projected second-order Hamiltonian. These shifts are artifacts of the retained inter-block term and are not the correct fourth-order corrections.


\paragraph*{Evaluation for the finest partition.}
Here $\mathcal P_\Pi=\mathcal P_A$ retains the part commuting with $A$, equivalently its infinite-time average,
\begin{equation}
	\mathcal P_AY
	=\lim_{T\to\infty}\frac1T
	\int_0^T e^{iAt}Ye^{-iAt}\,dt.
\end{equation}
For $[A,X_\mu]=\mu X_\mu$ and $[A,Y_\nu]=\nu Y_\nu$, the Jacobi identity gives
\begin{equation}
	[A,[X_\mu,Y_\nu]]
	=(\mu+\nu)[X_\mu,Y_\nu].
\end{equation}
Thus a nonzero nested commutator carries the sum of its factor frequencies, and projection retains total frequency zero. With $B^{\perp}=\sum_\mu B_\mu$ and $S^{(1)}=\sum_\mu B_\mu/\mu$,
\begin{equation}\label{eq:SH2freq}
	H^{(2)}
	=\frac12\sum_{\mu+\nu=0}
	\frac{[B_\mu,B_\nu]}{\mu}
	=\sum_{i=1}^{m}
	\frac{[B_{+g_i},B_{-g_i}]}{g_i}.
\end{equation}
The nonzero-frequency cross terms, together with $[S^{(1)},B^{\parallel}]$, determine the second-order generator:
\begin{equation}\label{eq:SS2freq}
	S^{(2)}
	=\sum_\mu\frac{[B_\mu,B^{\parallel}]}{\mu^2}
	+\frac12\sum_{\mu+\nu\ne0}
	\frac{[B_\mu,B_\nu]}{\mu(\mu+\nu)}.
\end{equation}
At third order, frequency selection gives
\begin{equation}\label{eq:SH3freq}
	\begin{aligned}
		H^{(3)}
		&=\frac12\sum_\mu
		\frac{[[B_\mu,B^{\parallel}],B_{-\mu}]}{\mu^2}\\
		&\quad+\frac13
		\sum_{\mu+\nu+\rho=0}
		\frac{[B_\mu,[B_\nu,B_\rho]]}{\mu\nu}.
	\end{aligned}
\end{equation}
All frequency indices run over the nonzero signed active gaps, and $B_\mu=0$ outside that set. The last sum describes closed three-transition processes. Its coefficient combines the zero-frequency parts of $\tfrac12[S^{(2)},B^{\perp}]$ and $\tfrac1{12}[S^{(1)},[S^{(1)},B^{\perp}]]$: relabeling the frequencies in the former gives $1/4+1/12=1/3$.

For a single active gap, three nonzero frequencies cannot sum to zero, so
\begin{equation}\label{eq:SH3single}
	H^{(3)}
	=
	\frac{
	[[B_+,B^{\parallel}],B_-]
	+[[B_-,B^{\parallel}],B_+]
	}{2g^2},
\end{equation}
where the two terms are Hermitian conjugates. If also $B^{\parallel}=0$, the recursion gives
\begin{equation}
	S^{(2)}=H^{(3)}=0,\qquad
	R^{(3)}=\frac23[S^{(1)},H^{(2)}],\qquad
	S^{(3)}=\frac{2}{3g^2}[B^{\perp},H^{(2)}].
\end{equation}
Substitution into Eq.~\eqref{eq:H1234} yields
\begin{equation}\label{eq:SH4single}
	H^{(4)}
	=
	-\frac{
	[B_+,[B_-,H^{(2)}]]
	+[B_-,[B_+,H^{(2)}]]
	}{4g^2}.
\end{equation}
For $A=(g/2)\sigma_z$ and $B=\lambda\sigma_x$, these formulas give $H^{(2)}=\lambda^2\sigma_z/g$ and $H^{(4)}=-\lambda^4\sigma_z/g^3$, in agreement with the expansion of the exact levels $\pm\tfrac12\sqrt{g^2+4\lambda^2}$.

Instead of resolving $Y$ into frequency components and retaining $Y_0$, one may remove its nonzero-frequency components by a polynomial in $\operatorname{ad}_A$. Combined with Theorem~\ref{thm:nested}, this gives every order of the gSWT in nested-commutator form.

\begin{corollary}[Nested-commutator form for the finest partition]\label{cor:finest}
Let $\Pi$ be the finest partition, so that $\mathcal P_\Pi=\mathcal P_A$. Let $\{g_1,\dots,g_m\}$ be a finite set of positive numbers and $p(\omega)=\omega^{-1}\bigl[1-\prod_{i=1}^{m}(1-\omega^2/g_i^2)\bigr]$, as in Theorem~\ref{thm:nested}.

(i) For every operator $Y$ with $\mathcal G_A(\mathcal Q_\Pi Y)\subseteq\{g_1,\dots,g_m\}$,
\begin{equation}\label{eq:Sfilter}
	\mathcal P_AY
	=\prod_{i=1}^{m}
	\left(
	\mathcal I-\frac{\operatorname{ad}_A^2}{g_i^2}
	\right)Y
	=\bigl[\mathcal I-\operatorname{ad}_A\,p(\operatorname{ad}_A)\bigr]Y,
\end{equation}
where $\mathcal I$ is the identity superoperator.

(ii) If $\mathcal G_A(\mathcal Q_\Pi R^{(n)})\subseteq\{g_1,\dots,g_m\}$, the generator and the effective Hamiltonian of order $n$ are
\begin{equation}\label{eq:finest}
	S^{(n)}=p(\operatorname{ad}_A)R^{(n)},\qquad
	H^{(n)}=\prod_{i=1}^{m}
	\left(
	\mathcal I-\frac{\operatorname{ad}_A^2}{g_i^2}
	\right)R^{(n)}.
\end{equation}
\end{corollary}

\begin{proof}
(i) Write $Y=\sum_\mu Y_\mu$ with $[A,Y_\mu]=\mu Y_\mu$, so that a polynomial $q(\operatorname{ad}_A)$ multiplies each $Y_\mu$ by $q(\mu)$. For the finest partition, $\mathcal P_AY=Y_0$, and by assumption every nonzero $Y_\mu$ with $\mu\ne0$ has $\mu\in\{\pm g_i\}$. The first equality follows because $\prod_i(1-\omega^2/g_i^2)$ equals $1$ at $\omega=0$ and $0$ at $\omega=\pm g_i$. The second is the definition of $p$, $1-\omega p(\omega)=\prod_i(1-\omega^2/g_i^2)$.

(ii) By Theorem~\ref{thm:rec}, $H^{(n)}=\mathcal P_\Pi R^{(n)}$ and $S^{(n)}=\operatorname{ad}_A^{-1}\mathcal Q_\Pi R^{(n)}$. The first is part~(i) with $Y=R^{(n)}$. The second follows from $p(0)=0$ and $p(\pm g_i)=\pm1/g_i$, which give $p(\operatorname{ad}_A)R^{(n)}=\sum_{\mu\ne0}R^{(n)}_\mu/\mu=\operatorname{ad}_A^{-1}\mathcal Q_\Pi R^{(n)}$.
\end{proof}

For the finest partition, the projection and the inversion therefore require neither eigenvectors nor spectral projectors of $A$: both are polynomials in $\operatorname{ad}_A$ acting on the source, and their only input is the active gap set of the source or a finite superset of it. The two polynomials are related by the order-$n$ equation~\eqref{eq:ordern}, which is the second equality of Eq.~\eqref{eq:Sfilter} for $Y=R^{(n)}$.

The source in turn requires only $B$ and the generators and effective Hamiltonians of lower orders. For $n\ge2$,
\begin{equation}\label{eq:Rall}
	R^{(n)}
	=\tfrac12[S^{(n-1)},B]
	+\tfrac12\sum_{j+k=n}[S^{(j)},H^{(k)}]
	+\sum_{l\ge1}\frac{b_{2l}}{(2l)!}
	\sum_{j_1+\cdots+j_{2l}+m=n}
	\operatorname{ad}_{S^{(j_1)}}\cdots\operatorname{ad}_{S^{(j_{2l})}}X^{(m)},
\end{equation}
where $X^{(1)}=B-H^{(1)}$ and $X^{(m)}=-H^{(m)}$ for $m\ge2$, all indices are at least $1$, and $b_{2l}$ are the Bernoulli numbers ($b_2=1/6$, $b_4=-1/30$). Only the terms with $2l\le n-1$ contribute. Like Eq.~\eqref{eq:RnA}, Eq.~\eqref{eq:Rall} holds for every partition.

Equation~\eqref{eq:Rall} is Eq.~\eqref{eq:RnA} solved for the source: the sources of lower orders, which enter Eq.~\eqref{eq:RnA} through $\mathcal Q_\Pi R^{(j)}=R^{(j)}-H^{(j)}$, are eliminated. Indeed, with $R=\sum_{n\ge1}R^{(n)}$, Eq.~\eqref{eq:RnA} summed over $n$ reads $R=e^{\operatorname{ad}_S}B-\psi(\operatorname{ad}_S)(R-H_{\mathrm{eff}}+A)$, where $\psi(x)=(e^x-1-x)/x$. Since $1+\psi(x)=(e^x-1)/x$ is invertible as a power series, with inverse $x/(e^x-1)=1-x/2+\sum_{l\ge1}b_{2l}x^{2l}/(2l)!$, solving for $R$ gives
\begin{equation*}
	\sum_{n\ge1}R^{(n)}
	=B+\tfrac12[S,B+H_{\mathrm{eff}}-A]
	+\sum_{l\ge1}\frac{b_{2l}}{(2l)!}\operatorname{ad}_S^{2l}(H-H_{\mathrm{eff}}),
\end{equation*}
whose term of order $n$ is Eq.~\eqref{eq:Rall}.

The lowest orders read
\begin{subequations}\label{eq:Rlow}
\begin{align}
	R^{(1)}&=B,\\
	R^{(2)}&=\tfrac12[S^{(1)},B+H^{(1)}],\\
	R^{(3)}&=\tfrac12[S^{(2)},B+H^{(1)}]
	+\tfrac12[S^{(1)},H^{(2)}]
	+\tfrac1{12}[S^{(1)},[S^{(1)},B-H^{(1)}]],\\
	R^{(4)}&=\tfrac12[S^{(3)},B+H^{(1)}]
	+\tfrac12[S^{(2)},H^{(2)}]
	+\tfrac12[S^{(1)},H^{(3)}]\notag\\
	&\quad+\tfrac1{12}\bigl(
	[S^{(1)},[S^{(2)},B-H^{(1)}]]
	+[S^{(2)},[S^{(1)},B-H^{(1)}]]
	-[S^{(1)},[S^{(1)},H^{(2)}]]\bigr).
\end{align}
\end{subequations}
Through fourth order, these are Eqs.~\eqref{eq:R1234} with $B^\parallel=H^{(1)}$ and $B^\perp=B-H^{(1)}$: the partition enters the source only through the generators and effective Hamiltonians of lower orders.

For the finest partition, Eqs.~\eqref{eq:finest} and~\eqref{eq:Rall} thus form a closed recursion whose only operations are commutators: $R^{(n)}$ gives $S^{(n)}$ and $H^{(n)}$, which enter $R^{(n+1)}$. By induction on $n$, every $S^{(n)}$ and $H^{(n)}$ is a finite linear combination of $B$ and of nested commutators of $A$ and $B$, with scalar coefficients fixed by the active gap sets $\mathcal G_A(\mathcal Q_\Pi R^{(k)})$, $k\le n$.

Nested commutators of sums of local operators are again sums of local operators. Hence, when $A$ and $B$ are sums of local operators and the number of active gaps does not grow with the system size, every order of the gSWT is a sum of local operators, as for the transverse-field Ising model of Sec.~\ref{sec:m-tfim}.

For a coarse-grained partition, intra-block and inter-block transitions can carry the same frequency, which a polynomial in $\operatorname{ad}_A$ cannot separate. The projection is then evaluated from the block labels, as described next.

\paragraph*{Evaluation for a coarse-grained partition.}
A coarse-grained partition also retains transitions between different reference energies within one block. For $a,c$ in block $\alpha$, Eq.~\eqref{eq:Spath} gives
\begin{equation}\label{eq:SH2coarse}
	H^{(2)}_{ac}
	=
	\frac12\sum_{b:\alpha_b\ne\alpha}
	B_{ab}B_{bc}
	\left(
	\frac1{E_a-E_b}+\frac1{E_c-E_b}
	\right),
\end{equation}
the Hermitian van Vleck matrix element of quasidegenerate perturbation theory~\cite{klein1974degenerate,shavitt1980quasidegenerate}. The denominators contain the actual reference energies. Internal couplings are retained as $B^{\parallel}$ rather than divided by their own small energy differences. Refining the blocks amounts to an additional diagonalization within them. When both constructions follow the same continued coarse-grained spectral subspaces, their effective Hamiltonians are related by a unitary within the coarse-grained blocks, although their canonical generators need not coincide.

For an operator expression, projection can be evaluated directly from block labels. We number the $|\Pi|$ blocks by $\alpha=0,\ldots,|\Pi|-1$ and let $\ell_\Pi=\sum_\alpha\alpha P_\alpha$. Then
\begin{equation}\label{eq:Slabel}
	\mathcal P_\Pi X
	=
	\frac1{|\Pi|}\sum_{t=0}^{|\Pi|-1}
	e^{2\pi i\ell_\Pi t/|\Pi|}Xe^{-2\pi i\ell_\Pi t/|\Pi|}.
\end{equation}
The finite average removes components between unequal labels. If blocks are labeled by a function $\ell(N)$ of a quantum number, a sequence of shifts $s_1,\ldots,s_r$ is retained on states satisfying
\begin{equation}
	\ell\left(N+\sum_i s_i\right)=\ell(N).
\end{equation}
This condition covers the return to the same quantum number as well as transitions within a degenerate manifold. For the finest partition of $A=f(N)$, one may take $\ell=f$, and for $A=\chi J_z^2$ equivalently $\ell(m)=|m|$.

When $A=f(N)$ and $[N,B_s]=sB_s$, the same shift decomposition also gives the generator directly:
\begin{equation}\label{eq:Sfn}
	S^{(1)}
	=
	\sum_s B_s^{\perp}[f(N+s)-f(N)]^{-1},
	\qquad
	B_s^{\perp}=\mathcal Q_\Pi B_s.
\end{equation}
Here the blocks are labeled by $\ell(N)$, and the inverse acts only on initial states of nonzero inter-block transitions. It may be set to zero elsewhere. Projection of subsequent products imposes return of $\ell$, rather than only $\sum_i s_i=0$. For the finest partition this condition is $f(N+\sum_i s_i)=f(N)$.

\section{Applications}\label{sec:models}

We apply the gSWT to nine models. Each subsection first states the Hamiltonian and its parameters. For each construction we then specify the dominant operator $A$, the perturbation $B$, and the block partition (the finest partition or a coarse-grained partition), followed by the active gap set $\mathcal G_A(B^\perp)$, the sector gap $g_\Pi$, and the expansion parameter $x=\|B\|/g_\Pi$. We then give the generator and the effective Hamiltonian of the gSWT.

\subsection{One-axis-twisting model in a transverse field}\label{sec:m-oat}

\paragraph*{Physical setting.}
The collective spin model of the main text is the one-axis-twisting model in a transverse field~\cite{kitagawa1993squeezed,muessel2015twist},
\begin{equation}\label{eq:H-oat}
	H=\chi J_z^2+\Omega J_x.
\end{equation}
Here $J_\alpha=\tfrac12\sum_{k=1}^N\sigma_k^\alpha$ are the collective spin operators of $N$ spin-$1/2$ particles with total spin $j=N/2$, $\chi>0$ is the twisting strength, and $\Omega$ is the transverse field. We write $c_j=j(j+1)$ and $\Lambda=N\chi/|\Omega|$, and denote the eigenstates of $J_z$ and $J_x$ by $|m\rangle$ and $|m_x\rangle$. In the graded form of Sec.~\ref{sec:dual}, $H_0=\chi J_z^2$, $V=\Omega J_x$, and $\Gamma=e^{i\pi(J_z-j)}$. Unless stated otherwise, $j\ge1$ is an integer ($N$ even, as in the main text).

This Hamiltonian is also known as the twist-and-turn Hamiltonian. Up to scalar terms, the same operator form describes the Bose--Hubbard dimer with on-site interaction $U$ and hopping amplitude $J$, for $(\chi,\Omega)=(U,-2J)$, and, after relabeling spin axes, the Lipkin--Meshkov--Glick model $-(\lambda/N)J_x^2-hJ_z$, for $(\chi,\Omega)=(-\lambda/N,-h)$.

\paragraph*{SWT: interaction-dominated regime.}
The dominant operator is $A=\chi J_z^2$ and the perturbation is $B=\Omega J_x$. We use the finest partition: its blocks are the eigenspaces $\operatorname{span}\{|m\rangle,|-m\rangle\}$, $m=1,\ldots,j$, and $\operatorname{span}\{|0\rangle\}$. For $\Omega\ne0$ the field changes $m$ by one, so $B^{\parallel}=0$. The active gap set is $\mathcal G_A(B^{\perp})=\{\chi,3\chi,\ldots,(2j-1)\chi\}$, the sector gap is $g_\Pi=\chi$, and $x=|\Omega|j/\chi=N^2/(2\Lambda)$, so the sufficient separation condition $x<\tfrac12$ reads $\Lambda>N^2$.

Theorem~\ref{thm:nested} gives the first-order generator with commutator depth $2j-1$, and the second-order effective Hamiltonian follows:
\begin{subequations}\label{eq:m-oat}
\begin{align}
	S^{(1)}&=\sum_{k=1}^{j}
	\frac{(-1)^{k+1}e_k\bigl(1,3^{-2},\ldots,(2j-1)^{-2}\bigr)}{\chi^{2k}}
	\operatorname{ad}_A^{2k-1}(\Omega J_x)
	=\frac{\Omega}{2\chi}\bigl[J_+(2J_z+1)^{-1}-\mathrm{H.c.}\bigr],
	\label{eq:m-oat-S}\\
	H_{\mathrm{eff}}^{[2]}&=\chi J_z^2
	+\frac{\Omega^2}{2\chi}\,\frac{c_j+J_z^2}{4J_z^2-1}
	+\frac{c_j\Omega^2}{4\chi}
	\bigl(|1\rangle\langle-1|+\mathrm{H.c.}\bigr),
	\label{eq:m-oat-H}
\end{align}
\end{subequations}
with truncation error $O(\Omega^4/\chi^3)$. These are the generator and the second-order effective Hamiltonian given in the main text.

For $j=2$, $(e_1,e_2)=(10/9,1/9)$, and for $j=3$, $(e_1,e_2,e_3)=(259/225,7/45,1/225)$. The closed form of the generator follows from $[A,J_\pm]=\pm\chi J_\pm(2J_z\pm1)$: the transition $m\to m+1$ changes the energy by $(2m+1)\chi$, and $(2J_z+1)^{-1}$ provides the corresponding energy denominator. Equivalently, $S^{(1)}_{m+1,m}=\Omega\langle m+1|J_x|m\rangle/[\chi(2m+1)]$.

The diagonal part of Eq.~\eqref{eq:m-oat-H} assigns to $|m\rangle$ the energy
\begin{equation}\label{eq:m-oat-E}
	E_m=\chi m^2+
	\frac{(c_j+m^2)\Omega^2}{2(4m^2-1)\chi}.
\end{equation}
Its second-order shift is the sum of the two virtual paths $m\to m\pm1\to m$,
\begin{equation}
	\frac{\Omega^2}{4\chi}
	\left[\frac{c_j-m^2+m}{2m-1}
	-\frac{c_j-m^2-m}{2m+1}\right].
\end{equation}
The projector additionally retains the path $-1\to0\to1$, because its endpoints belong to the same degenerate block, which gives the last term of Eq.~\eqref{eq:m-oat-H}. To this order, the level $m=0$ is $E_0=-c_j\Omega^2/(2\chi)$, and the $\pm1$ doublet has levels $E_1\pm c_j\Omega^2/(4\chi)$. The doublet $\pm m$ first splits at order $2|m|$. In particular, the leading splitting of $\pm j$ is
\begin{equation}
	\frac{2\chi(2j)!}{[(2j-1)!]^2}
	\left(\frac{|\Omega|}{2\chi}\right)^{2j},
\end{equation}
the tunnel splitting of the large-spin doublet. Since $\Gamma$ is constant on each integer-spin block, odd orders of $H_{\mathrm{eff}}$ vanish.

For half-integer $j\ge\tfrac32$, the dominant operator, the perturbation, and the finest partition into the $\pm m$ doublets are the same, but the central doublet $m=\pm\tfrac12$ contains a field matrix element,
\begin{equation}
	B^{\parallel}=\frac{\Omega}{2}\left(j+\frac12\right)
	\bigl(|\tfrac12\rangle\langle-\tfrac12|+\mathrm{H.c.}\bigr),
\end{equation}
which splits it by $|\Omega|(j+\tfrac12)$ at first order. The active gap set is $\mathcal G_A(B^{\perp})=\{2\chi,4\chi,\ldots,(2j-1)\chi\}$, the sector gap is $g_\Pi=2\chi$, and $x=|\Omega|j/(2\chi)$. Then
\begin{subequations}\label{eq:m-oat-half}
\begin{align}
	S^{(1)}&=\sum_{\substack{m=-j,\ldots,j-1\\m\ne-1/2}}
	\frac{\Omega\sqrt{c_j-m(m+1)}}{2\chi(2m+1)}
	\bigl(|m+1\rangle\langle m|-\mathrm{H.c.}\bigr),\\
	H_{\mathrm{eff}}^{[2]}&=\sum_{m=-j}^{j}E_m|m\rangle\langle m|+B^{\parallel},
\end{align}
\end{subequations}
with truncation error $O(\Omega^3/\chi^2)$. Here $E_m$ is given by Eq.~\eqref{eq:m-oat-E} for $|m|\ge\tfrac32$, while
\begin{equation}
	E_{\pm1/2}=\frac\chi4-\frac{(c_j-3/4)\Omega^2}{8\chi}.
\end{equation}
The grading is not constant on these doublets, so third-order terms are allowed.

\paragraph*{Dual SWT: field-dominated regime.}
For $\Omega\ne0$, the dominant operator is $A=\Omega J_x$ and the perturbation is $B=\chi J_z^2$. We use the finest partition, which consists of the one-dimensional eigenspaces spanned by $|m_x\rangle$. The operators $J'_\pm=J_y\pm iJ_z$ obey $[J_x,J'_\pm]=\pm J'_\pm$, and
\begin{equation}
	B^{\parallel}=\frac\chi2(c_j-J_x^2),\qquad
	B^{\perp}=-\frac\chi4(J_+^{\prime2}+J_-^{\prime2}).
\end{equation}
Thus the active gap set is $\mathcal G_A(B^{\perp})=\{2|\Omega|\}$, independently of $j\ge1$, whereas the sector gap is $g_\Pi=|\Omega|$ and $x=\chi j^2/|\Omega|=N\Lambda/4$, so the sufficient separation condition $x<\tfrac12$ reads $\Lambda<2/N$.

The single-gap formula, Eq.~\eqref{eq:single}, gives
\begin{subequations}\label{eq:m-oat-d}
\begin{align}
	\tilde S^{(1)}&=\frac\chi{4\Omega}[J_x,J_z^2]
	=-\frac{i\chi}{4\Omega}\{J_y,J_z\},\\
	\tilde H_{\mathrm{eff}}^{[2]}&=\Omega J_x+
	\frac\chi2(c_j-J_x^2)
	+\frac{\chi^2}{8\Omega}J_x(2c_j-1-2J_x^2),
	\label{eq:m-oat-dH}
\end{align}
\end{subequations}
with truncation error $O(\chi^3/\Omega^2)$, as given in the main text. Indeed, for $\Omega>0$, $B_\pm=-\chi J_\pm^{\prime2}/4$ and $[J_+^{\prime2},J_-^{\prime2}]=4J_x(2c_j-1-2J_x^2)$. The matrix element of this commutator at $m_x$ is $C(m_x)^2-C(m_x+2)^2$, where $C(m)^2=(j+m)(j-m+1)(j+m-1)(j-m+2)$. Equation~\eqref{eq:m-oat-d} remains valid for either sign of $\Omega$.

The levels continued from $\pm m_x$ have pair center $\chi(c_j-m_x^2)/2+O(\chi^3/\Omega^2)$ and signed half-splitting $\Omega m_x+\chi^2m_x(2c_j-1-2m_x^2)/(8\Omega)+O(\chi^4/\Omega^3)$. The nonzero center is the secular obstruction described by Theorem~\ref{thm:dimer}. The second-order projector is redundant because there is one active gap, but at third order it is needed for $j\ge2$ (Sec.~\ref{sec:projector}).

\paragraph*{Squeezing.}
A rotation of the $y,z$ axes by $\pi/4$ gives $\{J_y,J_z\}=J_{y'}^2-J_{z'}^2$, so $\tilde S^{(1)}$ generates two-axis countertwisting~\cite{kitagawa1993squeezed}. For $\Omega>0$, evaluating the transformed observable $e^{\tilde S}J_z^2e^{-\tilde S}$ in the reference ground state $|m_x=-j\rangle$ gives
\begin{equation}
	\frac{2\langle J_z^2\rangle}{j}
	=1-\frac{(N-1)\chi}{2|\Omega|}
	+O(\chi^2/\Omega^2).
\end{equation}
Its linear term differs from that of the Gaussian result $(1+\Lambda)^{-1/2}$ only by the finite-$N$ replacement $N\to N-1$.

\subsection{Transverse-field Ising model}\label{sec:m-tfim}

\paragraph*{Physical setting.}
The transverse-field Ising model of the main text describes the competition between ferromagnetic order and field polarization~\cite{pfeuty1970one},
\begin{equation}\label{eq:H-tfim}
	H=-J\sum_{i=1}^{L}\sigma_i^z\sigma_{i+1}^z
	-h\sum_{i=1}^{L}\sigma_i^x.
\end{equation}
Here $\sigma_i^\alpha$ are Pauli operators on a ring of $L\ge3$ sites, $\sigma^\alpha_{L+1}\equiv\sigma^\alpha_1$, $J>0$ is the Ising coupling, and $h>0$ is the transverse field. In the graded form of Sec.~\ref{sec:dual}, $H_0$ is the Ising term, $V$ is the transverse field, and $\Gamma=\prod_i\sigma_i^z$. The conserved Ising parity $\prod_i\sigma_i^x$ is a different operator and fixes the fermionic boundary condition. Exact free-fermion results provide a benchmark~\cite{pfeuty1970one}.

\paragraph*{SWT: domain walls in the ferromagnetic regime.}
For $h\ll J$, the dominant operator is $A=-J\sum_i\sigma_i^z\sigma_{i+1}^z=-JL+2JN_w$ and the perturbation is $B=-h\sum_i\sigma_i^x$. We use the finest partition: each block contains all configurations with the same domain-wall number $N_w$, which is even on a ring, and their degeneracy is retained in full. The sector gap is $g_\Pi=4J$, and $\|B\|=Lh$ gives $x=Lh/(4J)$.

On bond $i$, define $\tau_i^x=\sigma_i^z\sigma_{i+1}^z=1-2n_i$, $n_i=s_i^+s_i^-$, where $s_i^\pm$ create and annihilate a hard-core wall. The wall constraint and the Ising-sector boundary twist are understood in the cyclic sums below. A spin flip either moves one wall by one bond or creates or annihilates a pair of neighboring walls:
\begin{equation}
	B^{\parallel}=-h\sum_i(s_i^+s_{i+1}^-+\mathrm{H.c.}),\qquad
	B_\pm=-h\sum_i s_i^\pm s_{i+1}^\pm,\qquad
	\mathcal G_A(B^{\perp})=\{4J\}.
\end{equation}
Equivalently, $B_\pm=-h\sum_i\sigma_i^xP_i^\pm$, where $P_i^\pm=(1\pm\sigma_{i-1}^z\sigma_i^z)(1\pm\sigma_i^z\sigma_{i+1}^z)/4$ select no walls or two walls adjacent to site $i$.

There is a single active gap, $4J$, for every $L$, and Eq.~\eqref{eq:single} gives
\begin{subequations}\label{eq:m-tfim}
\begin{align}
	S^{(1)}&=\frac{ih}{8J}\sum_i
	(\sigma_i^y\sigma_{i+1}^z+\sigma_i^z\sigma_{i+1}^y),\\
	H_{\mathrm{eff}}^{[2]}&=A+B^{\parallel}
	+\frac{h^2}{4J}\sum_i\left[
	n_i+n_{i+1}-1
	-(s_i^+\tau_{i+1}^xs_{i+2}^-+\mathrm{H.c.})\right],
	\label{eq:m-tfim-H}
\end{align}
\end{subequations}
with truncation error $O(h^3/J^2)$. The commutator on the same bond pair gives $[s_i^+s_{i+1}^+,s_i^-s_{i+1}^-]=n_i+n_{i+1}-1$, overlapping pairs give $-s_i^+\tau_{i+1}^xs_{i+2}^-$, and disjoint pairs commute.

The correction therefore has three parts. Virtual wall pairs lower the vacuum energy by $h^2/(4J)$ per site. A wall blocks them on its two adjacent sites, which raises its energy by $h^2/(2J)$, and it hops by two bonds through them with amplitude $-h^2/(4J)$. The string $\tau_{i+1}^x$ in this hop is the Jordan--Wigner sign, which is required for the free-fermion form given below. The wall-energy shift and the hop together give the term $(h^2/J)\sin^2k$ of the wall dispersion.

Disconnected contributions cancel in these commutators: $H^{(2)}$ has support on at most three consecutive bonds, irrespective of $L$. Although the global expansion parameter $x$ grows with $L$, fixed-order local contributions to intensive quantities are organized in powers of $h/J$.

After the Jordan--Wigner transformation, Eq.~\eqref{eq:m-tfim-H} takes the free-fermion form given in the main text,
\begin{equation}\label{eq:m-tfim-ff}
	H_{\mathrm{eff}}^{[2]}=-L\left(J+\frac{h^2}{4J}\right)
	+\sum_k\varepsilon_kc_k^\dagger c_k,
\end{equation}
where $c_k$ are the wall fermions and
\begin{equation}
	\varepsilon_k=2J-2h\cos k+\frac{h^2}{J}\sin^2k.
\end{equation}
The second-order ground-state energy per site is $E_0/L=-J-h^2/(4J)$. Two-wall excitation energies are $\varepsilon_{k_1}+\varepsilon_{k_2}$, subject to the physical wall constraint. The allowed momenta are antiperiodic, $k=2\pi(n+\tfrac12)/L$, in the $\prod_i\sigma_i^x=+1$ sector and periodic, $k=2\pi n/L$, in the opposite sector, with the corresponding zero-mode and parity constraints.

The exact quasiparticle dispersion is
\begin{equation}
	\epsilon_k=2\sqrt{J^2+h^2-2Jh\cos k}.
\end{equation}
Its next term is $(h^3/J^2)\cos k\sin^2k$. Thus the dispersion error is $O(h^3/J^2)$: $\Gamma$ commutes with $N_w$ but is not constant within its eigenspaces, and wall hopping allows odd perturbative orders.

In each fixed fermion-parity sector the generated algebra remains quadratic, with transition frequencies $0,\pm4J$. Projection therefore reduces, at every order, to retaining the terms that are even under the grading described in Sec.~\ref{sec:projector} and discarding the odd ones.

\paragraph*{Dual SWT: spin flips in the paramagnetic regime.}
For $J\ll h$, the dominant operator is $A=-h\sum_i\sigma_i^x=-hL+2hN_f$ and the perturbation is $B=-J\sum_i\sigma_i^z\sigma_{i+1}^z$. We use the finest partition: each block contains all configurations with the same number $N_f$ of spins flipped against the field. On the full spin Hilbert space, the sector gap is $g_\Pi=2h$, and $\|B\|=LJ$ gives $x=LJ/(2h)$. If the calculation is restricted to one conserved Ising-parity sector, the spacing is instead $4h$ and $x=LJ/(4h)$. With $s_i^+=(\sigma_i^z+i\sigma_i^y)/2$, $s_i^-=(s_i^+)^\dagger$, $n_i=s_i^+s_i^-$, and $\sigma_i^x=1-2n_i$,
\begin{equation}
	B^{\parallel}=-J\sum_i(s_i^+s_{i+1}^-+\mathrm{H.c.}),\qquad
	B_\pm=-J\sum_i s_i^\pm s_{i+1}^\pm,\qquad
	\mathcal G_A(B^{\perp})=\{4h\}.
\end{equation}

There is again a single active gap, $4h$, and Eq.~\eqref{eq:single} gives
\begin{subequations}\label{eq:m-tfim-d}
\begin{align}
	\tilde S^{(1)}&=-\frac{iJ}{8h}\sum_i
	(\sigma_i^y\sigma_{i+1}^z+\sigma_i^z\sigma_{i+1}^y),\\
	\tilde H_{\mathrm{eff}}^{[2]}&=A+B^{\parallel}
	+\frac{J^2}{4h}\sum_i\left[
	n_i+n_{i+1}-1
	-(s_i^+\sigma_{i+1}^xs_{i+2}^-+\mathrm{H.c.})\right],
	\label{eq:m-tfim-dH}
\end{align}
\end{subequations}
with truncation error $O(J^3/h^2)$.

After the Jordan--Wigner transformation, Eq.~\eqref{eq:m-tfim-dH} takes the free-fermion form given in the main text,
\begin{equation}\label{eq:m-tfim-dff}
	\tilde H_{\mathrm{eff}}^{[2]}=-L\left(h+\frac{J^2}{4h}\right)
	+\sum_k\left(2h-2J\cos k+\frac{J^2}{h}\sin^2k\right)d_k^\dagger d_k,
\end{equation}
where $d_k$ are the flip fermions. The allowed momenta are antiperiodic, $k=2\pi(n+\tfrac12)/L$, in the sector $\prod_i\sigma_i^x=+1$, which contains an even number of flips, and periodic, $k=2\pi n/L$, in the opposite sector.

The dispersion error is $O(J^3/h^2)$, whereas the next correction to the ground-state energy in the thermodynamic limit is of fourth order. The dual parity rule is explicit: $B^{\parallel}$ is $\Gamma$-even, and $\tilde S^{(1)}$ and $\tilde H^{(2)}$ are $\Gamma$-odd. As in the wall representation, quadratic closure reduces all-order projection to parity selection.

\paragraph*{Duality and the exact energy benchmark.}
The Kramers--Wannier relations $\tau_i^x=\sigma_i^z\sigma_{i+1}^z$ and $\tau_{i-1}^z\tau_i^z=\sigma_i^x$ exchange $J$ and $h$~\cite{kramers1941statistics}. Let $W$ be the normalized duality map between corresponding parity sectors, with the wall constraint and the boundary twist of the ring retained. Then $WA=A'W$ and $WB=B'W$, with $A'=-J\sum_i\tau_i^x$ and $B'=-h\sum_i\tau_{i-1}^z\tau_i^z$, and the block projectors and the inverse of $\operatorname{ad}_A$ on the block-off-diagonal operators transform in the same way. Induction in the recursion consequently gives $WS^{(n)}=\tilde S'^{(n)}W$ and $WH^{(n)}=\tilde H'^{(n)}W$: the ferromagnetic SWT and the paramagnetic dual SWT coincide order by order after exchanging variables and boundary data.

In the thermodynamic limit the exact energy density is~\cite{pfeuty1970one}
\begin{equation}
	\frac{E_0}{L}=-\frac1{2\pi}\int_0^\pi\epsilon_k\,dk
	=-J\sum_{n\ge0}\binom{1/2}{n}^{\!2}(h/J)^{2n}
	=-J-\frac{h^2}{4J}-\frac{h^4}{64J^3}-\cdots
\end{equation}
for $h\le J$, and follows by $J\leftrightarrow h$ for $h\ge J$. Both energy series converge at the critical point, where $E_0/L=-4J/\pi$. This does not imply convergence of a global canonical generator there.

Finite rings retain sector-dependent tunneling corrections, first possible at order $h^L$. For even $L$, the two states in the $N_w=0$ block both have $\Gamma=+1$, so its odd-order corrections vanish and the second-order ground-state energy has a fourth-order remainder.

\subsection{Jaynes--Cummings model}\label{sec:m-jc}

\paragraph*{Physical setting.}
The Jaynes--Cummings Hamiltonian describes excitation exchange between a qubit and one cavity mode~\cite{jaynes1963comparison},
\begin{equation}\label{eq:H-jc}
	H=\omega_ca^\dagger a+\frac{\omega_q}{2}\sigma_z
	+g(a\sigma_++a^\dagger\sigma_-).
\end{equation}
Here $a$ annihilates a cavity photon, $\sigma_\pm$ raise and lower the qubit, $\omega_c$ and $\omega_q$ are the cavity and qubit frequencies, and $g>0$ is the coupling, taken positive by a choice of phase. We write $\Delta=\omega_q-\omega_c$ for the detuning and $N=a^\dagger a+\sigma_+\sigma_-$ for the conserved excitation number. In the graded form of Sec.~\ref{sec:dual}, $H_0=\omega_ca^\dagger a+\omega_q\sigma_z/2$, $V=g(a\sigma_++a^\dagger\sigma_-)$, and $\Gamma=\sigma_z$.

For $N=n\ge1$, the invariant space is $\{|n-1,e\rangle,|n,g\rangle\}$ and $H=\omega_c(n-\tfrac12)+\Delta\tau_z/2+g\sqrt n\,\tau_x$. The vacuum $|0,g\rangle$ is an uncoupled one-dimensional sector. This reduction makes the following rotations exact sector by sector.

\paragraph*{SWT: dispersive regime.}
The dominant operator is $A=H_0=\omega_ca^\dagger a+\omega_q\sigma_z/2$ and the perturbation is $B=V=g(a\sigma_++a^\dagger\sigma_-)$. Within a fixed sector $N=n\ge1$ we use the finest partition, which consists of the two one-dimensional subspaces spanned by the bare states and coincides with the qubit blocks when $\Delta\ne0$. Since $a\sigma_+$ changes the reference energy by $\Delta$, $B^{\parallel}=0$. For $\Delta\ne0$ the active gap set is $\mathcal G_A(B^{\perp})=\{|\Delta|\}$, the sector gap is $g_\Pi=|\Delta|$, and $x=g\sqrt n/|\Delta|$.

The resulting dispersive Hamiltonian is
\begin{subequations}\label{eq:m-jc}
\begin{align}
	S^{(1)}&=\frac g\Delta(a\sigma_+-a^\dagger\sigma_-),\\
	H_{\mathrm{eff}}^{[2]}&=\omega_ca^\dagger a+
	\frac12\left(\omega_q+\frac{g^2}{\Delta}\right)\sigma_z
	+\frac{g^2}{\Delta}a^\dagger a\,\sigma_z
	+\frac{g^2}{2\Delta},
\end{align}
\end{subequations}
with sectorwise truncation error $O(g^4n^2/|\Delta|^3)$. The absence of odd orders follows from the qubit grading. The same operator generator applies to the direct sum of sectors, but the expansion parameter grows as $\sqrt n$ and gives no uniform dispersive bound on the entire ladder.

The exact all-order canonical generator and effective Hamiltonian are~\cite{boissonneault2009dispersive}
\begin{subequations}\label{eq:m-jc-exact}
\begin{align}
	S&=\frac{\arctan(2g\sqrt N/\Delta)}{2\sqrt N}\,(a\sigma_+-a^\dagger\sigma_-),\\
	H_{\mathrm{eff}}&=\omega_c(N-\tfrac12)
	+\frac{\operatorname{sgn}(\Delta)}2\,
	\sigma_z\sqrt{\Delta^2+4g^2N},
\end{align}
\end{subequations}
where the coefficient of $S$ has the regular limit $g/\Delta$ at $N=0$ and $S|0,g\rangle=0$. The factor $\operatorname{sgn}(\Delta)$ retains the bare ordering for either sign of detuning, as in Proposition~\ref{prop:arctan}. At $\Delta=0$, the uncoupled doublet is a single degenerate block and this dispersive two-block expansion is not defined. The second-order dispersive shift $g^2/\Delta$ is therefore the leading term in the expansion of the exact, photon-number-dependent splitting and is not an exact constant shift.

\paragraph*{Dual SWT: interaction-dominated doublets.}
The dominant operator is $A=V=g(a\sigma_++a^\dagger\sigma_-)$ and the perturbation is $B=H_0=\omega_ca^\dagger a+\omega_q\sigma_z/2$. In the sector $N=n\ge1$ we use the finest partition, whose blocks are spanned by $|n,\pm\rangle=(|n-1,e\rangle\pm|n,g\rangle)/\sqrt2$, with eigenvalues $\pm g\sqrt n$. The sector scalar $\omega_c(n-\tfrac12)$ commutes with the full Hamiltonian. After its subtraction, the reduced problem has $A=g\sqrt n\,\tau_x$ and $B=\Delta\tau_z/2$, so for $\Delta\ne0$ the active gap set is $\mathcal G_A(B^{\perp})=\{2g\sqrt n\}$, the sector gap is $g_\Pi=2g\sqrt n$, and $x=|\Delta|/(4g\sqrt n)$. On the full ladder the active gap set is $\{2g\sqrt n:n\ge1\}$, and the spacing between neighboring positive eigenvalues of $V$ tends to zero, so no uniform finite-gap theorem applies across all sectors.

The secular part is $B^{\parallel}=\omega_c(N-\tfrac12)-\Delta|0,g\rangle\langle0,g|/2$. Since $V^2=g^2N$ commutes with $H_0$, the operator-valued inverse of Eq.~\eqref{eq:opval} gives, on the doublets $N\ge1$,
\begin{subequations}\label{eq:m-jc-d}
\begin{align}
	\tilde S^{(1)}&=-\frac\Delta{4g}N^+
	(a\sigma_+-a^\dagger\sigma_-),\\
	\tilde H_{\mathrm{eff}}^{[2]}&=\omega_c(N-\tfrac12)
	+V\left(1+\frac{\Delta^2}{8V^2}\right),
\end{align}
\end{subequations}
with truncation error $O(\Delta^4/(g^3n^{3/2}))$ in the sector $N=n$. The vacuum $|0,g\rangle$ is a one-dimensional block with the exact energy $-\omega_q/2$.

The exact all-order generator and effective Hamiltonian are, on the doublets $N\ge1$,
\begin{subequations}\label{eq:m-jc-d-exact}
\begin{align}
	\tilde S&=-\frac{\arctan\bigl(\Delta/(2g\sqrt N)\bigr)}{2\sqrt N}\,(a\sigma_+-a^\dagger\sigma_-),\\
	\tilde H_{\mathrm{eff}}&=\omega_c(N-\tfrac12)
	+V\sqrt{1+\frac{\Delta^2}{4V^2}},
\end{align}
\end{subequations}
and $\tilde S|0,g\rangle=0$. Together with the unchanged vacuum energy, this gives exact levels $\omega_c(n-\tfrac12)\pm\sqrt{g^2n+\Delta^2/4}$. The two positive rotation angles are complementary for $\Delta>0$, as in Proposition~\ref{prop:arctan}.

The pair centers and the vacuum energy remain entirely secular: $f(V^2)=\omega_c(N-\tfrac12)-\Delta|0,g\rangle\langle0,g|/2$. Thus the obstruction is given exactly by the first-order secular term and commutes with $H$, whereas the splitting is renormalized to all orders.

\subsection{Quantum Rabi model}\label{sec:m-rabi}

\paragraph*{Physical setting.}
The quantum Rabi model retains both excitation exchange and counter-rotating processes~\cite{braak2011integrability},
\begin{equation}\label{eq:H-rabi}
	H=\omega a^\dagger a+\frac{\omega_q}{2}\sigma_z
	+g\sigma_x(a+a^\dagger),
\end{equation}
where $a$ annihilates a photon of the cavity mode of frequency $\omega>0$, $\omega_q>0$ is the qubit frequency, and $g\in\mathbb R$ is the coupling. We write $\Delta=\omega_q-\omega$ for the detuning and $\Sigma=\omega_q+\omega$ for the sum frequency. In the graded form of Sec.~\ref{sec:dual}, $H_0=\omega a^\dagger a+\omega_q\sigma_z/2$, $V=g\sigma_x(a+a^\dagger)$, and $\Gamma=\sigma_z$. Unlike the Jaynes--Cummings model, it has no conserved excitation number, so finite-dimensional norm estimates require a photon cutoff.

\paragraph*{SWT: dispersive qubit blocks.}
The dominant operator is $A=H_0=\omega a^\dagger a+\omega_q\sigma_z/2$ and the perturbation is $B=V=g\sigma_x(a+a^\dagger)$. We use a coarse-grained partition whose two blocks are the qubit sectors $\sigma_z=\pm1$. Here $B^{\parallel}=0$. For $g\ne0$ and a nonresonant dominant operator, the rotating and counter-rotating transitions give the active gap set $\mathcal G_A(B^{\perp})=\{|\Delta|,\Sigma\}$, with repeated values counted once.

For the cutoff $0\le n\le n_{\max}$, with $n_{\max}\ge1$, the sector gap is $g_\Pi=\min_{|k|\le n_{\max}}|\omega_q-k\omega|$ and $x=|g|\,\|a+a^\dagger\|/g_\Pi$, where $a$ denotes the truncated oscillator matrix when evaluating the norm. A simple upper bound is $\|B\|\le2|g|\sqrt{n_{\max}+1}$. On the full oscillator ladders, the corresponding gap is $\min_{k\in\mathbb Z}|\omega_q-k\omega|$, which is positive away from multiphoton degeneracies and equals $|\Delta|$ only when the one-photon resonance is the nearest one. The perturbation, however, is unbounded there.

The formulas below use the full oscillator commutation relation and give the low-lying matrix elements whose virtual paths do not reach beyond the cutoff. At the cutoff boundary, the consistent second-order expression is $A+\tfrac12[S^{(1)},B]$ evaluated with the truncated operators.

The two active gaps give
\begin{subequations}\label{eq:m-rabi}
\begin{align}
	S^{(1)}&=\frac g\Delta(a\sigma_+-a^\dagger\sigma_-)
	+\frac g\Sigma(a^\dagger\sigma_+-a\sigma_-),\\
	H_{\mathrm{eff}}^{[2]}&=\omega a^\dagger a+
	\frac12\left(\omega_q+\frac{g^2}{\Delta}+\frac{g^2}{\Sigma}\right)\sigma_z
	\notag\\
	&\quad+g^2\left(\frac1\Delta+\frac1\Sigma\right)
	\left[a^\dagger a+\frac12(a^2+a^{\dagger2})\right]\sigma_z
	+\frac{g^2}{2}\left(\frac1\Delta-\frac1\Sigma\right).
	\label{eq:m-rabi-H}
\end{align}
\end{subequations}
In $\tfrac12[S^{(1)},B]$, $[a\sigma_+,a^\dagger\sigma_-]=\sigma_+\sigma_-+a^\dagger a\sigma_z$ and $[a^\dagger\sigma_+,a\sigma_-]=a^\dagger a\sigma_z-\sigma_-\sigma_+$, and the mixed commutators produce the two-photon term. The counter-rotating contribution $g^2/\Sigma$ is the vacuum Bloch--Siegert correction to the qubit frequency. The term $(a^2+a^{\dagger2})\sigma_z$ is retained within each qubit block even though it changes the oscillator energy by $2\omega$.

For the finest partition of a nondegenerate $A$, projection removes this term and leaves the remaining terms in Eq.~\eqref{eq:m-rabi}. The sector gap $g_\Pi$ is then the minimum of the inter-qubit separations and the intra-qubit spacing $\omega$. Both effective Hamiltonians are second-order results, with fourth-order remainders at fixed nonresonant frequencies and cutoff, because $\Gamma$ is constant on the chosen blocks.

\paragraph*{Strong coupling: displaced-oscillator blocks.}
The strict dual choice $A=g\sigma_x(a+a^\dagger)$, $B=\omega a^\dagger a+\omega_q\sigma_z/2$ admits no gapped spectral partition, because $V=g\sigma_x(a+a^\dagger)$ has a continuous spectrum on the untruncated oscillator. Instead, we retain the interaction in
\begin{equation}
	A=\omega a^\dagger a+g\sigma_x(a+a^\dagger)
	=\omega(a+\alpha\sigma_x)^\dagger(a+\alpha\sigma_x)-g^2/\omega,
	\qquad \alpha=g/\omega,
\end{equation}
as the dominant operator and take the perturbation $B=\omega_q\sigma_z/2$.

The exact eigenstates of $A$, $|n,\pm\rangle=|\pm x\rangle\otimes D(\mp\alpha)|n\rangle$ with $\sigma_x|\pm x\rangle=\pm|\pm x\rangle$ and $D(\beta)=e^{\beta(a^\dagger-a)}$, form degenerate doublets at $\omega n-g^2/\omega$. We use the finest partition, whose blocks are these doublets. The sector gap is $g_\Pi=\omega$ and, for the full bounded perturbation $B$, $x=\omega_q/(2\omega)$. A finite truncation in these exact displaced doublets has $x=\|B\|/\omega\le\omega_q/(2\omega)$, with $B$ understood as the compressed operator.

For $g\ne0$ and $\omega_q\ne0$, the nonsecular perturbation $B^{\perp}$ has the infinite active gap set $\mathcal G_A(B^{\perp})=\{k\omega:k\ge1\}$, since
\begin{equation}
	\langle n,s|B|m,-s\rangle
	=\frac{\omega_q}{2}\langle n|D(2s\alpha)|m\rangle.
\end{equation}
Keeping the displaced doublets $0\le n\le n_{\max}$ gives the finite active gap set $\{\omega,2\omega,\ldots,n_{\max}\omega\}$, to which Theorem~\ref{thm:nested} applies. On the untruncated oscillator the generator below is the spectral inverse $\operatorname{ad}_A^{-1}B^{\perp}$ rather than a finite polynomial in $\operatorname{ad}_A$.

This displacement diagonalizes the dominant operator exactly,
\begin{equation}
	e^{\alpha\sigma_x(a^\dagger-a)}A
	e^{-\alpha\sigma_x(a^\dagger-a)}
	=\omega a^\dagger a-g^2/\omega.
\end{equation}

Writing $\tau_n^x=|n,+\rangle\langle n,-|+\mathrm{H.c.}$ and $\tau_n^0$ for the doublet projector, the first-order generator and the second-order effective Hamiltonian are
\begin{subequations}\label{eq:m-rabi-d}
\begin{align}
	S^{(1)}&=\sum_{n\ne m}\sum_{s=\pm}
	\frac{\langle n,s|B|m,-s\rangle}{\omega(n-m)}
	|n,s\rangle\langle m,-s|,\\
	H_{\mathrm{eff}}^{[2]}&=\sum_n\left[
	\left(\omega n-\frac{g^2}{\omega}+\delta_n\right)\tau_n^0
	+\frac{\omega_q}{2}e^{-2\alpha^2}L_n(4\alpha^2)\tau_n^x
	\right],
\end{align}
\end{subequations}
with truncation error $O(\omega_q^3/\omega^2)$, where
\begin{equation}\label{eq:m-rabi-delta}
	\delta_n=\frac{\omega_q^2}{4\omega}
	\sum_{m\ne n}\frac{|\langle n|D(2\alpha)|m\rangle|^2}{n-m}
\end{equation}
and $L_n$ is the Laguerre polynomial. The first-order term mixes the two states of a doublet. Two virtual qubit flips return to the original qubit branch, giving the same second-order shift $\delta_n$ for both states, so there is no second-order correction to their splitting. The sum in Eq.~\eqref{eq:m-rabi-delta} is for the full oscillator and is restricted correspondingly in a truncated calculation.

The first-order Hamiltonian $A+B^{\parallel}$ is the adiabatic approximation of Ref.~\cite{irish2005dynamics}, with doublet splittings $\omega_qe^{-2\alpha^2}|L_n(4\alpha^2)|$. For $\alpha=0.6$ and $\omega_q=0.05\omega$, the ground-doublet splitting through second order is $0.024338\omega$, compared with $0.024328\omega$ from numerical diagonalization. Equation~\eqref{eq:m-rabi-d} is exact in $g/\omega$ at each order in $\omega_q$.

The generalized rotating-wave approximation cited in the main text retains selected couplings between adjacent dressed doublets when their levels approach resonance. It can be organized at leading order using $A+B^{\parallel}$ as the dominant operator and retaining those near-resonant states together in a coarse-grained partition. The remaining inter-block couplings then generate further gSWT corrections. These are strong-coupling gSWT frames, not the strict dual SWT.

After a rotation of the spin axes, the Hamiltonian reads $-t\sigma_x+\omega a^\dagger a+g\sigma_z(a+a^\dagger)$ with $t=\omega_q/2$, which is the two-site Holstein model with one electron: $\sigma_z$ is the difference of the electron occupations of the two sites, $a$ annihilates a phonon of their relative vibration, and $t$ is the hopping amplitude. The displacement that diagonalizes $A$ is then the Lang--Firsov transformation~\cite{lang1963kinetic}, and $B^{\parallel}$ is the hopping reduced by the Franck--Condon factor, $te^{-2\alpha^2}$ in the phonon vacuum.

\subsection{Three-level \texorpdfstring{$\Lambda$}{Lambda} system}\label{sec:m-lambda}
\paragraph*{Physical setting.}
A three-level $\Lambda$ system connects two ground states to a common excited state~\cite{brion2007adiabatic}:
\begin{equation}\label{eq:H-lambda}
	H=\Delta|e\rangle\langle e|
	+\frac12\sum_{i=1,2}\Omega_i
	(|e\rangle\langle i|+\mathrm{H.c.}).
\end{equation}
Here $|1\rangle$ and $|2\rangle$ are the ground states, $|e\rangle$ is the excited state, $\Delta$ is the detuning, and $\Omega_i$ are the Rabi frequencies, chosen real by fixing the phases of the ground states. We write $\Omega=(\Omega_1^2+\Omega_2^2)^{1/2}>0$. The bright state $|b\rangle=(\Omega_1|1\rangle+\Omega_2|2\rangle)/\Omega$ couples to $|e\rangle$, whereas $|d\rangle=(\Omega_2|1\rangle-\Omega_1|2\rangle)/\Omega$ is an exact dark state. In the graded form of Sec.~\ref{sec:dual}, $H_0=\Delta|e\rangle\langle e|$, $V$ is the optical coupling, and $\Gamma=\openone-2|e\rangle\langle e|$.

\paragraph*{SWT: off-resonant Raman coupling.}
The dominant operator is $A=\Delta|e\rangle\langle e|$ and the perturbation is $B=(\Omega/2)(|e\rangle\langle b|+\mathrm{H.c.})$. For $\Delta\ne0$ we use the finest partition, whose blocks are the two-dimensional ground eigenspace and the excited state. Here $B^{\parallel}=0$, the active gap set is $\mathcal G_A(B^{\perp})=\{|\Delta|\}$, the sector gap is $g_\Pi=|\Delta|$, and $x=\Omega/(2|\Delta|)$.

The first-order generator and the second-order effective Hamiltonian are
\begin{subequations}\label{eq:m-lambda}
\begin{align}
	S^{(1)}&=\frac1{2\Delta}\sum_i\Omega_i
	(|e\rangle\langle i|-|i\rangle\langle e|),\\
	H_{\mathrm{eff}}^{[2]}&=\left(\Delta+\frac{\Omega^2}{4\Delta}\right)
	|e\rangle\langle e|
	-\frac1{4\Delta}\sum_{i,j}\Omega_i\Omega_j
	|i\rangle\langle j|,
\end{align}
\end{subequations}
with truncation error $O(\Omega^4/|\Delta|^3)$. The ground-block term contains the light shifts and the Raman coupling obtained by adiabatic elimination~\cite{brion2007adiabatic}.

Only the bright state participates, so a two-level rotation gives the exact all-order result:
\begin{subequations}\label{eq:m-lambda-exact}
\begin{align}
	S&=\frac12\arctan\frac\Omega\Delta
	(|e\rangle\langle b|-|b\rangle\langle e|),\\
	H_{\mathrm{eff}}&=
	\frac{\Delta+\operatorname{sgn}(\Delta)\sqrt{\Delta^2+\Omega^2}}2
	|e\rangle\langle e|
	+\frac{\Delta-\operatorname{sgn}(\Delta)\sqrt{\Delta^2+\Omega^2}}2
	|b\rangle\langle b|.
\end{align}
\end{subequations}
This result leaves $|d\rangle$ at zero energy and follows the levels of $A$ continuously for either sign of the detuning.

\paragraph*{Dual SWT: the driven doublet.}
The dominant operator is $A=(\Omega/2)(|e\rangle\langle b|+\mathrm{H.c.})$ and the perturbation is $B=\Delta|e\rangle\langle e|$. We use the finest partition, whose blocks are the three one-dimensional eigenspaces: $|\pm\rangle=(|b\rangle\pm|e\rangle)/\sqrt2$ at $\pm\Omega/2$, together with $|d\rangle$ at zero. The perturbation does not couple to the dark state:
\begin{equation}
	B^{\parallel}=\frac\Delta2(|+\rangle\langle+|+|-\rangle\langle-|),\qquad
	B^{\perp}=-\frac\Delta2(|+\rangle\langle-|+\mathrm{H.c.}).
\end{equation}
Thus the sector gap is $g_\Pi=\Omega/2$ and $x=2|\Delta|/\Omega$, but for $\Delta\ne0$ the active gap set is $\mathcal G_A(B^{\perp})=\{\Omega\}$: the smaller gaps to the dark state are inactive.

The first-order generator and the second-order effective Hamiltonian show how detuning shifts the center and increases the splitting of the Autler--Townes doublet~\cite{autler1955stark}:
\begin{subequations}\label{eq:m-lambda-d}
\begin{align}
	\tilde S^{(1)}&=-\frac\Delta{2\Omega}
	(|+\rangle\langle-|-|-\rangle\langle+|),\\
	\tilde H_{\mathrm{eff}}^{[2]}&=
	\left(\frac\Omega2+\frac\Delta2+\frac{\Delta^2}{4\Omega}\right)
	|+\rangle\langle+|
	+\left(-\frac\Omega2+\frac\Delta2-\frac{\Delta^2}{4\Omega}\right)
	|-\rangle\langle-|,
\end{align}
\end{subequations}
with truncation error $O(\Delta^4/\Omega^3)$.

The exact all-order generator and effective Hamiltonian are
\begin{subequations}\label{eq:m-lambda-d-exact}
\begin{align}
	\tilde S&=-\frac12\arctan\frac\Delta\Omega
	(|+\rangle\langle-|-|-\rangle\langle+|),\\
	\tilde H_{\mathrm{eff}}&=
	\frac{\Delta+\sqrt{\Omega^2+\Delta^2}}2|+\rangle\langle+|
	+\frac{\Delta-\sqrt{\Omega^2+\Delta^2}}2|-\rangle\langle-|.
\end{align}
\end{subequations}
The spectral obstruction is $f(\Omega^2/4)=\Delta/2$, while $f(0)=0$: the doublet has a nonzero center but the dark state remains at zero. The scalar secular term in the bright--excited subspace explains the absence of a third-order correction in Eq.~\eqref{eq:m-lambda-d}.

\subsection{Dirac Hamiltonian}\label{sec:m-dirac}
\paragraph*{Physical setting.}
At fixed momentum, the free Dirac Hamiltonian is the sum of a mass term and a kinetic term:
\begin{equation}\label{eq:H-dirac}
	H=\beta m+\bm\alpha\cdot\bm p.
\end{equation}
Here $\bm\alpha$ and $\beta$ are the Dirac matrices, $m>0$ is the mass, and $\bm p$ is the momentum, in units $\hbar=c=1$. We write $p=|\bm p|$ and $E=\sqrt{p^2+m^2}$. In the graded form of Sec.~\ref{sec:dual}, $H_0=\beta m$, $V=\bm\alpha\cdot\bm p$, and $\Gamma=\beta$. The Clifford relations give $H^2=E^2\openone$, so that both generators are obtained in closed form.

\paragraph*{SWT: the nonrelativistic expansion.}
The dominant operator is $A=\beta m$ and the perturbation is $B=\bm\alpha\cdot\bm p$. We use the finest partition, whose blocks are the two spin-degenerate eigenspaces of $\beta$, each retained in full. Here $B^{\parallel}=0$. For $p>0$ the active gap set is $\mathcal G_A(B^{\perp})=\{2m\}$, while at $p=0$ it is empty. The sector gap is $g_\Pi=2m$, and $x=p/(2m)$.

Since $A^2=m^2\openone$, the first-order generator and the second-order effective Hamiltonian are
\begin{subequations}\label{eq:m-dirac}
\begin{align}
	S^{(1)}&=\frac{\beta\bm\alpha\cdot\bm p}{2m},\\
	H_{\mathrm{eff}}^{[2]}&=\beta\left(m+\frac{p^2}{2m}\right),
\end{align}
\end{subequations}
with truncation error $O(p^4/m^3)$.

The exact all-order Foldy--Wouthuysen transformation~\cite{foldy1950dirac} is
\begin{subequations}\label{eq:m-dirac-exact}
\begin{align}
	S&=\frac\theta2\beta\bm\alpha\cdot\hat{\bm p},\\
	H_{\mathrm{eff}}&=\beta E,
\end{align}
\end{subequations}
where $\theta=\arctan(p/m)$ and $\hat{\bm p}=\bm p/p$ for $p>0$, with the continuous limit $S=0$ at $p=0$ [Eq.~\eqref{eq:arctan-S}].

\paragraph*{Dual SWT: the ultrarelativistic expansion.}
For $p>0$, the dominant operator is $A=\bm\alpha\cdot\bm p$ and the perturbation is $B=\beta m$. We use the finest partition, whose blocks are the two spin-degenerate eigenspaces at $\pm p$. Here $B^{\parallel}=0$, the active gap set is $\mathcal G_A(B^{\perp})=\{2p\}$, the sector gap is $g_\Pi=2p$, and $x=m/(2p)$.

The first-order generator and the second-order effective Hamiltonian are
\begin{subequations}\label{eq:m-dirac-d}
\begin{align}
	\tilde S^{(1)}&=-\frac{m\beta\bm\alpha\cdot\bm p}{2p^2},\\
	\tilde H_{\mathrm{eff}}^{[2]}&=\bm\alpha\cdot\bm p
	\left(1+\frac{m^2}{2p^2}\right),
\end{align}
\end{subequations}
with truncation error $O(m^4/p^3)$.

The exact all-order result is the Cini--Touschek transformation~\cite{cini1958relativistic}:
\begin{subequations}\label{eq:m-dirac-d-exact}
\begin{align}
	\tilde S&=-\frac{\tilde\theta}2\beta\bm\alpha\cdot\hat{\bm p},\\
	\tilde H_{\mathrm{eff}}&=(\bm\alpha\cdot\hat{\bm p})E,
\end{align}
\end{subequations}
where $\tilde\theta=\arctan(m/p)=\pi/2-\theta$. Both exact forms have the spectrum $\pm E$ [Eqs.~\eqref{eq:arctan-S} and~\eqref{eq:arctan-Sd}]. Although the reference eigenvalues are degenerate, the Clifford algebra enforces a vanishing even part of the dual result: there is no spectral obstruction, $f=0$.

\subsection{Anderson impurity model}\label{sec:m-anderson}
\paragraph*{Physical setting.}
The two-electron Anderson impurity with one bath orbital interpolates between a local moment and a hybridized molecule~\cite{anderson1961localized,schrieffer1966relation}:
\begin{equation}\label{eq:H-anderson}
	H=\varepsilon_d n_d+Un_{d\uparrow}n_{d\downarrow}
	+V\sum_\sigma(c^\dagger_\sigma d_\sigma+\mathrm{H.c.}).
\end{equation}
Here $c_\sigma$ and $d_\sigma$ annihilate a bath and an impurity electron of spin $\sigma$, $n_{d\sigma}=d^\dagger_\sigma d_\sigma$ and $n_d=\sum_\sigma n_{d\sigma}$, $\varepsilon_d$ is the impurity level measured from the bath orbital, $U>0$ is the on-site repulsion, and $V>0$ is the hybridization amplitude. We work in the two-electron sector $n_c+n_d=2$, with $n_c=\sum_\sigma c^\dagger_\sigma c_\sigma$. In the graded form of Sec.~\ref{sec:dual}, the impurity terms form $H_0$, the hybridization term is the coupling, and $\Gamma=(-1)^{n_d}$. Throughout this subsection $V$ denotes the hybridization amplitude, not the coupling operator.

\paragraph*{SWT: virtual charge fluctuations.}
In the Kondo regime $\varepsilon_d<0<\varepsilon_d+U$, the dominant operator is $A=\varepsilon_d n_d+Un_{d\uparrow}n_{d\downarrow}$ and the perturbation is $B=V\sum_\sigma(c^\dagger_\sigma d_\sigma+\mathrm{H.c.})$. The charge sectors $n_d=0,1,2$ have dimensions $1,4,1$ and energies $0,\varepsilon_d,2\varepsilon_d+U$. We use the finest partition, whose blocks are the distinct eigenspaces of $A$. The $n_d=0$ and $n_d=2$ sectors merge when $2\varepsilon_d+U=0$. The hybridization changes $n_d$ by one and has operator norm $2V$ in the two-electron sector. Hence $B^{\parallel}=0$, and the active gap set is $\mathcal G_A(B^{\perp})=\{|\varepsilon_d|,\varepsilon_d+U\}$, with repeated values counted once. For three distinct charge energies the sector gap is $g_\Pi=\min\{|\varepsilon_d|,\varepsilon_d+U,|2\varepsilon_d+U|\}$, and $x=2V/g_\Pi$. At the degeneracy $2\varepsilon_d+U=0$ the last entry is omitted.

If only the charge fluctuations out of the local-moment sector are to be eliminated, the coarse-grained partition $\{n_d=1\}\oplus\{n_d=0,2\}$ has instead $g_\Pi=\min\{|\varepsilon_d|,\varepsilon_d+U\}$, which does not contain the spacing between the two charge states. This spacing is inactive at first order even in the finest partition.

Dividing each hopping process by its charge-transfer energy gives the same first-order generator for either partition. The first-order generator and the second-order effective Hamiltonian in the local-moment block $n_d=n_c=1$ are
\begin{subequations}\label{eq:m-anderson}
\begin{align}
	S^{(1)}&=V\sum_\sigma\left\{
	c^\dagger_\sigma d_\sigma
	\left[\frac{1-n_{d,-\sigma}}{-\varepsilon_d}
	+\frac{n_{d,-\sigma}}{-\varepsilon_d-U}\right]
	-\mathrm{H.c.}\right\},\\
	H_{\mathrm{eff}}^{[2]}\big|_{n_d=n_c=1}
	&=\varepsilon_d+J_K\left(\bm S_d\cdot\bm S_c-\frac14\right),
\end{align}
\end{subequations}
with truncation error $O(V^4/g_\Pi^3)$ and the Kondo exchange
\begin{equation}\label{eq:m-anderson-JK}
	J_K=2V^2\left(\frac1{|\varepsilon_d|}
	+\frac1{\varepsilon_d+U}\right)>0.
\end{equation}
Here $\bm S_d$ and $\bm S_c$ are the spin-$1/2$ operators of the singly occupied impurity and bath orbital. The occupation factors refer to the state before the hop. Virtual empty and doubly occupied impurity states both lower the singlet, producing antiferromagnetic Kondo exchange, while the triplet is unshifted~\cite{schrieffer1966relation}. The remainder is of fourth order at fixed nonzero charge-transfer energies because $\Gamma$ is constant on each chosen block.

\paragraph*{Dual SWT: a strongly hybridized molecule.}
For $|\varepsilon_d|,U\ll V$, the dominant operator is $A=V\sum_\sigma(c^\dagger_\sigma d_\sigma+\mathrm{H.c.})$ and the perturbation is $B=\varepsilon_d n_d+Un_{d\uparrow}n_{d\downarrow}$. The orbitals $(c_\sigma\pm d_\sigma)/\sqrt2$ have energies $\pm V$. In the two-electron sector we use the finest partition, whose blocks are the complete eigenspaces at $-2V,0,+2V$, with dimensions $1,4,1$, respectively. The sector gap is $g_\Pi=2V$, and $x=\max(|\varepsilon_d|,|2\varepsilon_d+U|)/(2V)$. Generically the active gap set is $\mathcal G_A(B^{\perp})=\{2V,4V\}$. The $2V$ component vanishes when $2\varepsilon_d+U=0$, and the $4V$ component vanishes when $U=0$. Using both gaps is always an exact superset construction for the first-order generator.

If $Q_\pm$ project onto the one-dimensional blocks at $\pm2V$, the first-order generator and the second-order effective Hamiltonian are
\begin{subequations}\label{eq:m-anderson-d}
\begin{align}
	\tilde S^{(1)}&=\frac5{16V^2}[A,B^{\perp}]
	-\frac1{64V^4}\operatorname{ad}_A^3B^{\perp},\\
	\tilde H_{\mathrm{eff}}^{[2]}&=A+B^{\parallel}
	+\frac1V\left[\frac{(2\varepsilon_d+U)^2}{16}
	+\frac{U^2}{64}\right](Q_+-Q_-),
\end{align}
\end{subequations}
with truncation error $O(\|B\|^3/V^2)$. The secular shift is $\varepsilon_d+U/4$ in each molecular level. In the zero-energy block it is $\varepsilon_d+U/2$ on the singlet and $\varepsilon_d$ on the three triplets. The second-order correction to this block vanishes because spin conservation forbids singlet--triplet mixing and parity alternation removes the remaining matrix elements. Here $\Gamma$ exchanges the molecular blocks, so odd orders beyond the secular shift are not generally forbidden.

The ground-state energy to this order is
\begin{equation}\label{eq:m-anderson-E0}
	E_0=-2V+\varepsilon_d+\frac U4
	-\frac1V\left[\frac{(2\varepsilon_d+U)^2}{16}
	+\frac{U^2}{64}\right].
\end{equation}
The two negative terms arise from virtual excitation to the blocks at $0$ and $+2V$, with gaps $2V$ and $4V$. For $V=1$, $\varepsilon_d=-0.3$, and $U=0.8$, this expression gives $E_0=-2.1125$, compared with the exact value $-2.11201185$.

\subsection{Fermi--Hubbard model}\label{sec:m-hubbard}

\paragraph*{Physical setting.}
The repulsive Fermi--Hubbard model describes the competition between electron hopping and the on-site repulsion~\cite{hubbard1963electron}:
\begin{equation}\label{eq:H-hubbard}
	H=-t\sum_{\langle ij\rangle\sigma}
	(c^\dagger_{i\sigma}c_{j\sigma}+\mathrm{H.c.})
	+U\sum_i n_{i\uparrow}n_{i\downarrow},
\end{equation}
where $c_{i\sigma}$ annihilates an electron of spin $\sigma$ on site $i$, $n_{i\sigma}=c^\dagger_{i\sigma}c_{i\sigma}$, $\langle ij\rangle$ runs over nearest-neighbor bonds, $t>0$ is the hopping amplitude, and $U>0$ is the on-site repulsion. On a finite bipartite lattice at fixed electron number, we write $D=\sum_i n_{i\uparrow}n_{i\downarrow}$ for the number of doubly occupied sites and $T$ for the hopping term. In the graded form of Sec.~\ref{sec:dual}, $H_0=UD$, $V=T$, and $\Gamma=(-1)^{N_a}$, where $N_a$ counts the electrons on one sublattice. The grading leaves the interaction invariant and reverses $T$.

\paragraph*{SWT: doublon elimination.}
The dominant operator is $A=UD$ and the perturbation is $B=T$. We use the finest partition, whose blocks are the eigenspaces of the doublon number $D$. We decompose $T=T_0+T_{+1}+T_{-1}$, where $[D,T_q]=qT_q$ and $T_{-1}=T_{+1}^\dagger$. Thus $B^{\parallel}=T_0$ and $B_\pm=T_{\pm1}$. Whenever doublon-changing hops are present, the active gap set is $\mathcal G_A(B^{\perp})=\{U\}$, the sector gap is $g_\Pi=U$, and $x=\|T\|/U$. The global norm can grow with lattice size, while each fixed order consists of local virtual processes.

The first-order generator and the second-order effective Hamiltonian are
\begin{subequations}\label{eq:m-hubbard}
\begin{align}
	S^{(1)}&=\frac{T_{+1}-T_{-1}}U,\\
	H_{\mathrm{eff}}^{[2]}&=UD+T_0+\frac{[T_{+1},T_{-1}]}U,
	\label{eq:m-hubbard-H}
\end{align}
\end{subequations}
with truncation error $O(t^3/U^2)$.

Projection onto $D=0$ gives the $t$--$J$ expansion, including the three-site terms away from half filling~\cite{macdonald1988expansion}. At half filling, $T_0$ vanishes in this block and
\begin{equation}\label{eq:m-hubbard-heis}
	H_{\mathrm{eff}}^{[2]}\big|_{D=0}
	=\frac{4t^2}{U}\sum_{\langle ij\rangle}
	\left(\bm S_i\cdot\bm S_j-\frac14\right),
\end{equation}
with truncation error $O(t^4/U^3)$ in this block. Each site is singly occupied, so $\Gamma$ is constant on this entire block and all odd effective orders vanish there. In other blocks the grading need not be constant, and third-order terms can survive.

\paragraph*{Dual SWT: weak interactions in a dimer.}
For two electrons on two sites, the dominant operator is $A=T$ and the perturbation is $B=UD$. We write $b_\sigma^\dagger=(c_{1\sigma}^\dagger+c_{2\sigma}^\dagger)/\sqrt2$ and $a_\sigma^\dagger=(c_{1\sigma}^\dagger-c_{2\sigma}^\dagger)/\sqrt2$ for the bonding and antibonding orbitals. We use the finest partition, which consists of the one-dimensional blocks spanned by the singlets $|bb\rangle=b_\uparrow^\dagger b_\downarrow^\dagger|0\rangle$ at $-2t$ and $|aa\rangle=a_\uparrow^\dagger a_\downarrow^\dagger|0\rangle$ at $+2t$, together with the complete four-dimensional eigenspace at zero.

We choose the relative phase of $|bb\rangle$ and $|aa\rangle$ so that $\langle aa|UD|bb\rangle=U/2$. Both molecular singlets have diagonal interaction $U/2$, and neither couples to the zero-energy block. Hence the active gap set is $\mathcal G_A(B^{\perp})=\{4t\}$, the sector gap is $g_\Pi=2t$, and $x=U/(2t)$. The smaller gaps to the zero-energy block are inactive. Restricting to the invariant span of $|bb\rangle,|aa\rangle$ instead gives $g_\Pi=4t$ and $x=U/(4t)$ for the unshifted perturbation.

On the full two-electron space, the first-order generator and the second-order effective Hamiltonian are
\begin{subequations}\label{eq:m-hubbard-d}
\begin{align}
	\tilde S^{(1)}&=\frac{[A,B^{\perp}]}{16t^2}
	=\frac U{8t}(|aa\rangle\langle bb|-|bb\rangle\langle aa|),\\
	\tilde H_{\mathrm{eff}}^{[2]}&=
	\left(-2t+\frac U2-\frac{U^2}{16t}\right)|bb\rangle\langle bb|
	+\left(2t+\frac U2+\frac{U^2}{16t}\right)|aa\rangle\langle aa|
	+UQ_s,
\end{align}
\end{subequations}
with truncation error $O(U^4/t^3)$. Here $Q_s$ projects onto the singlet in the zero-energy block, and the three triplets have energy zero.

The coupled two-state problem also gives the exact all-order generator and effective Hamiltonian,
\begin{subequations}\label{eq:m-hubbard-d-exact}
\begin{align}
	\tilde S&=\frac12\arctan\frac U{4t}
	(|aa\rangle\langle bb|-|bb\rangle\langle aa|),\\
	\tilde H_{\mathrm{eff}}&=
	\left(\frac U2-\sqrt{4t^2+U^2/4}\right)|bb\rangle\langle bb|
	+\left(\frac U2+\sqrt{4t^2+U^2/4}\right)|aa\rangle\langle aa|+UQ_s.
\end{align}
\end{subequations}
The generator vanishes on the zero-energy block. In the coupled pair, the secular term is the scalar $U/2$: it shifts the pair center, while the remaining corrections change the splitting and contain only even powers of $U$. This explains the fourth-order remainder of the second-order result.

\subsection{Su--Schrieffer--Heeger model}\label{sec:m-ssh}

\paragraph*{Physical setting.}
The Su--Schrieffer--Heeger model describes a particle hopping on a chain with alternating strong and weak bonds~\cite{su1979solitons},
\begin{equation}\label{eq:H-ssh}
	H=\sum_i\bigl(v|a_i\rangle\langle b_i|
	+w|b_i\rangle\langle a_{i+1}|+\mathrm{H.c.}\bigr),
\end{equation}
where $|a_i\rangle$ and $|b_i\rangle$ are the two sublattice sites of cell $i$, and $v>0$ and $w>0$ are the intra-cell and inter-cell hopping amplitudes. The two choices of dominant operator resolve the bonding and antibonding states of the corresponding isolated dimers.

Both hopping operators anticommute with the sublattice grading $\Gamma=P_a-P_b$, where $P_a$ and $P_b$ project onto the two sublattices. Thus ``SWT'' and ``dual SWT'' below denote the two exchanged choices of dominant operator, and the even--odd assumptions used for the parity-alternation theorem of the dual SWT do not apply to this decomposition. Here the canonical generator is $\Gamma$-even, and the effective Hamiltonian remains $\Gamma$-odd at every order.

\paragraph*{SWT: intra-cell dimers.}
For $w\ll v$ on a ring of $N_c\ge3$ cells, the dominant operator and the perturbation are
\begin{equation}
	A=v\sum_{i=1}^{N_c}(|a_i\rangle\langle b_i|+\mathrm{H.c.}),
	\qquad
	B=w\sum_{i=1}^{N_c}(|b_i\rangle\langle a_{i+1}|+\mathrm{H.c.}),
\end{equation}
with $a_{N_c+1}\equiv a_1$. We use the finest partition, whose blocks are the eigenspaces at $\pm v$, each of dimension $N_c$. The secular perturbation $B^{\parallel}$ produces hopping of magnitude $w/2$, with opposite signs in the two bands, while $B^{\perp}$ mixes the bands. The active gap set is $\mathcal G_A(B^{\perp})=\{2v\}$, the sector gap is $g_\Pi=2v$, and $\|B\|=w$ gives $x=w/(2v)$.

The single-gap formula, Eq.~\eqref{eq:single}, gives the first-order generator and the second-order effective Hamiltonian,
\begin{subequations}\label{eq:m-ssh}
\begin{align}
	S^{(1)}&=\frac{B_+-B_-}{2v}
	=\frac{[A,B^{\perp}]}{4v^2},\\
	H_{\mathrm{eff}}^{[2]}
	&=A+B^{\parallel}+\frac{[B_+,B_-]}{2v},
\end{align}
\end{subequations}
with truncation error $O(w^3/v^2)$, where $[A,B_\pm]=\pm2vB_\pm$. The bands of $H_{\mathrm{eff}}^{[2]}$ are
\begin{equation}\label{eq:m-ssh-bands}
	E_\pm^{[2]}(k)=\pm\left(v+w\cos k
	+\frac{w^2}{2v}\sin^2k\right).
\end{equation}

As an exact spectral benchmark, the bands obtained by Fourier transformation are $E_\pm(k)=\pm|v+we^{ik}|$. Their next term, $\mp[w^3/(2v^2)]\cos k\sin^2k$, is a cubic correction that survives because $B^{\parallel}\ne0$. In contrast, the filled lower band has, in the thermodynamic limit,
\begin{equation}
	\frac{E_0}{N_c}
	=-\frac1\pi\int_0^\pi|v+we^{ik}|\,dk
	=-v\sum_{n\ge0}\binom{1/2}{n}^{\!2}(w/v)^{2n},
	\qquad w\le v,
\end{equation}
so its second-order error is $O(w^4/v^3)$.

\paragraph*{Dual SWT: inter-cell dimers and an end state.}
For $v\ll w$, consider the open chain $a_1,b_1,\ldots,b_{N_c},a_{N_c+1}$ with $N_c\ge2$. The dominant operator and the perturbation are exchanged:
\begin{equation}
	A=w\sum_{i=1}^{N_c}(|b_i\rangle\langle a_{i+1}|+\mathrm{H.c.}),
	\qquad
	B=v\sum_{i=1}^{N_c}(|a_i\rangle\langle b_i|+\mathrm{H.c.}).
\end{equation}
We use the finest partition, whose blocks are the eigenspaces at $\pm w$, each of dimension $N_c$, and the one-dimensional kernel spanned by $|a_1\rangle$. The perturbation has intra-band matrix elements of magnitude $v/2$ and no secular matrix element on the kernel. The active gap set is $\mathcal G_A(B^{\perp})=\{w,2w\}$, which distinguishes end-to-band and inter-band transitions. The sector gap is $g_\Pi=w$, and $\|B\|=v$ gives $x=v/w$.

The first-order generator is obtained exactly by the two-gap polynomial, and the second-order effective Hamiltonian follows:
\begin{subequations}\label{eq:m-ssh-d}
\begin{align}
	\tilde S^{(1)}
	&=\frac5{4w^2}[A,B^{\perp}]
	-\frac1{4w^4}\operatorname{ad}_A^3B^{\perp},\\
	\tilde H_{\mathrm{eff}}^{[2]}
	&=A+B^{\parallel}
	+\frac12\mathcal P_A[\tilde S^{(1)},B^{\perp}],
\end{align}
\end{subequations}
with truncation error $O(v^3/w^2)$. To this order, its nonzero levels are
\begin{equation}\label{eq:m-ssh-d-bands}
	E_\pm(k)=\pm\left(w+v\cos k
	+\frac{v^2}{2w}\sin^2k\right)
	+O(v^3/w^2).
\end{equation}
For this open chain with an odd number of sites, the nonzero exact levels are $\pm|w+ve^{ik}|$ at $k=r\pi/(N_c+1)$, $r=1,\ldots,N_c$: this follows by diagonalizing the tridiagonal matrix $T^\dagger T$, whose diagonal and off-diagonal entries are $v^2+w^2$ and $vw$. Thus Eq.~\eqref{eq:m-ssh-d} also specifies the finite-chain levels through second order.

The remaining eigenvalue is exactly zero, because the bipartite chain has one more $a$ site than $b$ sites and $T$ has full column rank. The exact state is
\begin{equation}
	|\psi_0\rangle\ \propto\
	\sum_{i=1}^{N_c+1}(-v/w)^{i-1}|a_i\rangle,
\end{equation}
and $\langle a_1|\tilde H_{\mathrm{eff}}|a_1\rangle=0$ to all orders on the continued zero-mode block. The first-order dressing, $\tilde S^{(1)}|a_1\rangle=(v/w)|a_2\rangle$, therefore reproduces $|\psi_0\rangle=|a_1\rangle-(v/w)|a_2\rangle+O(v^2/w^2)$ up to normalization. The zero eigenvalue is an all-order result. For $v<w$ its localization length $\xi$ at the left end is given by $\xi^{-1}=\ln(w/v)$, in units of a cell. At $v=w$ it is extended, and for $v>w$ it localizes at the opposite end while remaining exactly at zero.

\section{Two-point Pad\'e approximant}\label{sec:pade}

A two-point Pad\'e approximant is a rational function $R(z)=P_L(z)/Q_M(z)$, with polynomials $P_L$ and $Q_M$ of degrees $L$ and $M$, that reproduces the leading terms of two expansions of the same function about two different points~\cite{baker1996pade,mccabe1976continued}. Here the two points are zero and infinite transverse field, where the SWT and the dual SWT provide the expansions of the ground-state energy.

For the collective spin model $H=\chi J_z^2+\Omega J_x$ of the main text (Sec.~\ref{sec:m-oat}), with $\chi>0$ and $\Omega>0$, the SWT and the dual SWT give the ground-state energy $E_0$ as expansions in $\Omega/\chi$ and in $\chi/\Omega$, starting from the ground states $|m=0\rangle$ of $\chi J_z^2$ and $|m_x=-j\rangle$ of $\Omega J_x$, respectively. To second order, the level $m=0$ of Eq.~\eqref{eq:m-oat-H} and the level $m_x=-j$ of Eq.~\eqref{eq:m-oat-dH} are
\begin{equation}\label{eq:pade-second}
	\begin{aligned}
		E_0^{\mathrm{SWT}}
		&=-\frac{c_j\Omega^2}{2\chi},\\
		E_0^{\mathrm{dual}}
		&=-j\Omega+\frac{j\chi}{2}
		-\frac{j(2j-1)\chi^2}{8\Omega},
	\end{aligned}
\end{equation}
with $c_j=j(j+1)$. These are the two dashed curves of Fig.~2 of the main text.

For $\Omega>0$ the Hamiltonian is an irreducible tridiagonal matrix in the $J_z$ basis and therefore has a simple spectrum, so $E_0(\Omega)$ is analytic, and the two series are its expansions about $\Omega=0$ and $\Omega=\infty$. Neither truncated series is controlled in the intermediate regime, where neither term of $H$ dominates. We therefore interpolate between them with a two-point Pad\'e approximant of $E_0(\Omega)$, constructed in the three steps below for $N=20$ ($j=10$), as in Fig.~2 of the main text, with $\chi=1$. The construction uses only the two perturbative series and exact properties of $E_0$, and no exact eigenvalue is fitted.

\paragraph*{Step 1: perturbative series.}
The two ground states are nondegenerate, so each is a one-dimensional block whose effective Hamiltonian is the exact level, and the SWT and dual SWT expansions are the Rayleigh--Schr\"odinger series of that level. We generate them through order $n_{\max}=20$,
\begin{equation}\label{eq:pade-series}
	E_0=\sum_{k=0}^{n_{\max}}a_k\Omega^k
	+O(\Omega^{n_{\max}+1}),
	\qquad
	E_0=\Omega\sum_{k=0}^{n_{\max}}b_k\Omega^{-k}
	+O\!\left(\Omega^{-n_{\max}}\right).
\end{equation}
The coefficients with $k\le2$ reproduce Eq.~\eqref{eq:pade-second}, and all odd $a_k$ vanish because the spectrum is even in $\Omega$.

\paragraph*{Step 2: matching the two series.}
We write
\begin{equation}\label{eq:pade-R}
	R(\Omega)=\frac{P_L(\Omega)}{Q_M(\Omega)},\qquad
	L=M+1,\qquad Q_M(0)=1,
\end{equation}
where the degree difference reproduces the linear asymptote $E_0\sim-j\Omega$. We require $R$ to reproduce the first $J$ coefficients of the weak-field series and the first $K$ coefficients of the strong-field series,
\begin{equation}\label{eq:pade-match}
	R(\Omega)=\sum_{k=0}^{J-1}a_k\Omega^k+O(\Omega^J),
	\qquad
	R(\Omega)=\Omega\sum_{k=0}^{K-1}b_k\Omega^{-k}
	+O(\Omega^{1-K}).
\end{equation}
These are linear equations for the coefficients of $P_L$ and $Q_M$, which are determined when $J+K=L+M+1=2M+2$. Since $n_{\max}+1=21$ coefficients are available from each series, the three largest totals, $J+K=38$, $40$, and $42$, give nine candidates.

\paragraph*{Step 3: selection.}
A rational function can have spurious poles. We therefore retain only the candidates that share the following exact properties of the ground-state energy for all $\Omega>0$:
\begin{equation}\label{eq:pade-props}
	-j\Omega\le E_0\le0,\qquad
	\frac{d}{d\Omega}\frac{E_0}{\Omega}
	=-\frac{\chi\langle J_z^2\rangle}{\Omega^2}\le0,
	\qquad
	\frac{d^2E_0}{d\Omega^2}\le0.
\end{equation}
The bounds follow from $\chi J_z^2\ge0$, $J_x\ge-j$, and the trial state $|m=0\rangle$. The derivative identity is the Hellmann--Feynman theorem. The energy $E_0$ is concave because it is the minimum of expectation values that are affine in $\Omega$.

Candidates with a pole on the positive real axis are rejected, and the remaining ones are tested against Eq.~\eqref{eq:pade-props} on a logarithmic grid covering the range of Fig.~2 of the main text. Seven candidates pass, and over this range they deviate from their pointwise median by at most $6.53\times10^{-5}$ in relative value. This spread measures the stability of the construction and is not a rigorous error bound. The Pad\'e curve of Fig.~2 of the main text is the candidate closest to the median, $[L/M]=[19/18]$ with $(J,K)=(21,17)$. All steps are carried out in multiprecision arithmetic, which controls the cancellations at these orders.

The two-point Pad\'e approximant is a numerical continuation of one observable, constrained by the two perturbative limits. It does not provide an effective Hamiltonian in the intermediate regime, and its agreement with the exact level there is not a convergence statement for either gSWT series. Likewise, the two vertical lines of Fig.~2 of the main text, at $\Omega/(N\chi)=N^{-2}$ and $N/2$, mark the sufficient spectral-separation condition $x=\tfrac12$ of the SWT and of the dual SWT. They are not convergence boundaries.

\bibliography{ref_supp}